\documentclass[12pt]{article}

\usepackage{amsfonts}
\usepackage{amssymb}
\usepackage{amsmath}
\usepackage{amsthm}
\usepackage{color}
\usepackage{graphicx}
\usepackage{verbatim}
\usepackage[longnamesfirst]{natbib}
\usepackage[dvipsnames]{xcolor}
\usepackage{caption}

\usepackage{makecell}
\usepackage{booktabs}  
\usepackage{subfig}

\newtheorem*{theorem*}{Theorem}

\newtheorem*{algorithm*}{Algorithm}

\newtheorem*{definition*}{Definition}

\usepackage{setspace} 
\usepackage[pagebackref,colorlinks=true,citecolor=blue]{hyperref}

\newcommand{\citeip}[1]{\citeauthor{#1}, \citeyear{#1}}

\usepackage{afterpage}
\usepackage{pdflscape}

\usepackage{textcomp}
\newcommand{\cents}{\text{\textcentoldstyle}}

\newcommand{\wdth}{0.65\textwidth}

\usepackage[bottom]{footmisc}
\usepackage{arydshln}

\begin{document}

\title{An Experimental Study of Decentralized Matching}
\author{
	Federico Echenique\thanks{%
		Department of Economics, UC Berkeley,
		fede@econ.berkeley.edu
	} 
	\and 
	Alejandro Robinson-Cort\'es\thanks{%
		Department of Economics, University of Exeter,
		a.robinson-cortes@exeter.ac.uk
	} 
	\and
	Leeat Yariv\thanks{%
		Department of Economics, Princeton University,
		lyariv@princeton.edu
	} 
	\thanks{
        We thank Drew Fudenberg, Jacob Goeree, Matthew Jackson, Gabriel Katz, Erik Madsen, Sergio Montero, Muriel Niederle, Charles Plott, Andrew Schotter, and Alistair Wilson for useful conversations and suggestions. We also thank the Editor and three anonymous reviewers for very helpful comments. Financial support from the National Science Foundation (SES 0963583), and the Gordon and Betty Moore Foundation (grant 1158) is gratefully acknowledged.
	}
}
\date{December 22, 2023}
\maketitle

\begin{abstract}
    We present an experimental study of decentralized two-sided matching markets with no transfers. Experimental participants are informed of everyone's preferences and can make arbitrary non-binding match offers that get finalized when a period of market inactivity has elapsed. Several insights emerge. First, stable outcomes are prevalent. Second, while centralized clearinghouses commonly aim at implementing extremal stable matchings, our decentralized markets most frequently culminate in the \textit{median} stable matching. Third, preferences’ cardinal representations impact the stable partners participants match with. Last, the dynamics underlying our results exhibit strategic sophistication, with agents successfully avoiding cycles of blocking pairs.
\end{abstract}

\begin{quote}
\textbf{Keywords:} Decentralized Matching, Experiments, Market Design.

\setcounter{page}{0}\thispagestyle{empty}
\end{quote}

\newpage

\section{Introduction}

This paper presents an experimental investigation of decentralized two-sided matching markets with no transfers. We consider two-sided markets, such as labor markets consisting of workers and firms, marriage markets comprising women and men, and so on. We study the outcomes emerging from free interactions between market participants. Would unconstrained decentralized interaction produce stable outcomes?\footnote{A stable outcome corresponds to a pairing of agents from both market sides such that no agent prefers to sever their partnership and no pair of agents prefer to pair with one another over remaining with their assigned partners.} When there are multiple stable matchings, would some have stronger drawing power? What dynamics would decentralized interactions follow? Our experiments are designed to answer these questions.

Matching theory offers predictions for the set of plausible market outcomes, namely the set of stable matchings, or the core, under two basic premises: all agents are completely informed of others' preferences, as well as their own; and agents can match with one another freely. Our experiments are designed to test the fundamental predictions of this model. While the assumption that market forces would guide matchings to stability has governed much of the literature, it has never been tested. In this regard, our approach is reminiscent of some of the original general equilibrium experiments \citep[e.g.,][]{smith1962experimental,plott1978experimental}, which examined whether markets reach an equilibrium without imposing constraints on the sequential actions that lead them there.

In the main treatments of our experiment each side of the market is composed of $8$ participants. Market participants are fully informed of their own and other participants' match payoffs, the payoffs derived from all possible pairings. Preferences are designed so that each market participant has either one, two, or three stable match partners. As part of our design, we utilize several cardinal utility representations for the same ordinal preference profiles, varying the utilitarian welfare of each side of the market as well as the marginal returns from matching with one partner as opposed to a more preferred one. 
At any point during a market's operation, agents on each side are free to make match offers to any agent on the other side of the market, up to one offer at a time, as well as accept any offer that arrives, up to one at a time. Importantly, agents can keep making and receiving offers while matched, so matches are tentative until the market ends, which occurs after $30$ seconds of inactivity.

Our experiments are designed to allow the exploitation of blocking opportunities, the main theoretical driving force behind stability. Nevertheless, there are reasons to suspect stability will be difficult to achieve through decentralized interactions, or at the very least require many offers, many rejections, and a long time. Indeed, the leading dynamics that the theoretical literature has offered assume sequential formation of blocking pairs \citep{knuth76}. \cite{roth1990random} proved that there is always \textit{at least one} path of such sequential blocking pairs leading to a stable matching. However, not every sequence of blocking pairs culminates in a stable matching. \cite{knuth76} showed that such paths can also generate indefinite cycles.\footnote{For example, suppose each market side has two agents---$f_1$ and $f_2$ on one, $c_1$ and $c_2$ on the other---and that all agents view all others as acceptable. Assume $f_i$'s preferred partner is $c_i$, while $c_i$'s preferred partner is $f_{3-i}$, $i=1,2$. Suppose, say, that $f_1$ and $c_1$ are the only matched participants initially. Then, $(f_2,c_1)$ is a blocking pair. If this match is formed, $(f_2,c_2)$ becomes a blocking pair, which, if formed, makes $(f_1,c_2)$ a blocking pair. Finally, if $f_1$ and $c_2$ match, then $(f_1,c_1)$ is a blocking pair. Upon matching, the market returns to its starting point.} In fact, the sequential formation of blocking pairs can have an almost chaotic behavior \citep{rudov2022fragile}. From a computational perspective, finding the set of stable matchings is a hard problem in general, see \cite{gusfield1989stable}, and even markets with $8$ individuals on each side entail $2 \times 8 \times 8 = 128$ match payoffs that participants would need to internalize in order to identify the market-wide stable matching.

In our data, \textit{stable outcomes appear in a predominance of cases}, with $88\%$ of markets reaching a stable matching. Despite the existence of cyclic ``traps,'' participants steer clear of cycles, and converge quickly to a stable matching, both in terms of offer volume, which averages $46$ offers per market, and in terms of market duration, which takes less than two minutes on average.

In general, stable outcomes exhibit a lattice structure, with two extremal stable matchings: one preferred by all the agents on one side of the market and one preferred by all the agents on the other side. These extremal stable matchings are computationally simple to identify via the \cite{gale1962college} Deferred Acceptance (DA) algorithm. Since the DA algorithm underlies many centralized clearinghouses in the field, centralized outcomes frequently correspond to extremal stable matchings for the reported preferences. As noted, the full set of stable matchings is computationally challenging to compute. In principle, however, in our markets with multiple stable matchings, decentralized interactions could yield any stable matching, including the \textit{median} stable matching, the matching ranked by all agents between the two extremal stable matchings \citep{teo2001gale}. Matching theory is mute regarding the selection of stable matchings, as well as on the effects of cardinal preference representations: stability accounts for only ordinal rankings of partners.

In our data, \textit{the median stable matching emerges as the modal outcome}. In fact, $77\%$ of pairings are between median stable match partners, and $80\%$ of markets converge to the market-wide median stable matching. We also find that the \textit{cardinal representation of preferences significantly influences the selection of stable matchings}. In particular, the side of the market that has ``more to lose'' by forgoing their most preferred matching tends to establish its most preferred matching more frequently. These findings are potentially useful for policy. When considering centralized interventions, our results highlight the potential importance of cardinal preference information, on top of ordinal rankings, for counterfactual decentralized outcomes. They also suggest that features of median stable matchings---utilitarian welfare, distributional attributes, and the like---may be relevant for assessing when centralized clearinghouses might be particularly beneficial.

The underlying dynamics of interactions are at the heart of the outcomes we observe. With field data, it is naturally challenging to track the complete protocol of market participants’ interactions.\footnote{One exception is \cite{hitsch10}, who use detailed data from an online dating site.} Our data allow a lens into decentralized dynamics and a comparison with some dynamics proposed by the theoretical matching literature, including the sequential formation of blocking pairs \citep[e.g.,][]{roth1990random,ackermann2011uncoordinated}, and the dynamics underlying the DA algorithm of \cite{gale1962college} \citep[e.g.,][]{dworczak21}.

Our last set of results pertains to the dynamic paths markets follow to reach stability. The speed at which stable pairings form decreases over markets' operation: a substantial fraction of agents find stable partners quickly, followed by a slow-down of market activity while the remaining agents take longer to find their stable matches. While, as expected, agents do not appear to see through the market-wide matching and make offers to stable partners immediately, \textit{participants exhibit  strategic sophistication and avoid blocking pairs that could trap them in a cycle}. Participants make more frequent offers to blocking partners, particularly to those who benefit more from the pairing. Furthermore, not all offers from blocking partners are accepted. These patterns help explain the convergence to the median stable matchings, as well as the avoidance of cycles and consequent quick convergence to stability that we observe. 

We also simulate some of the dynamics the theoretical literature has proposed to gain a comparison benchmark for our data. These simulations indicate that existing models do not organize our data well. Indeed, existing models are, by and large, non-strategic, and do not incorporate the history of past interactions. Our results open the door to new models, allowing strategic behavior as we observe.

In addition to our main treatments, we also report results from two auxiliary treatments that investigate our main results' robustness, in terms of market size and bargaining power. The first set of treatments involves larger markets, with $15$ participants, instead of $8$, on each side. The second set of treatments allows agents only on one side of the market to make offers. Qualitatively, the findings from our main treatments continue to hold: stable matchings, and in particular median stable matchings, are very frequent. Similarly, the dynamic patterns we identify in our main treatments are still present in larger markets, as well as when only side can make offers.

\section{Literature Review}

Our experimental design corresponds to the standard model of two-sided matching markets, see \cite{roth1990book}; which belongs to the realm of cooperative game theory. The model provides clear predictions: outcomes coincide with the core of the market, the set of stable matchings. In that respect, our experimental results provide a strong experimental validation of the theory underlying the stability notion.

Several authors propose particular dynamic decentralized processes by which one-to-one matchings are created; see, e.g., \cite{haeringer2011decentralized}, \cite{ferdowsian2020decentralized}, and \cite{pais2008incentives}. These papers usually impose some structure on the process by which offers are made and accepted. The main focus of the decentralized matching literature is on the identification of conditions under which stability is likely to arise in equilibrium. Complete information of the prevailing preferences, as in our experiments, allows for stability to emerge in equilibrium, while more stringent demands on preferences and equilibrium selection are required for stability to be the unique prediction. To the extent of our knowledge, the literature is silent on the selection of stable matchings when multiple ones exist in a market.

While the experimental literature on matching markets has grown rapidly in recent years \citep{hakimov2021experiments}, there are only a few studies of decentralized markets. \cite{nalbantian1995matching} analyze several procedures for matching with transferable utility, decentralized matching among them, where agents have private payoff information. \cite{nalbantian1995matching} include private negotiations between potential match partners. Offers in our treatment are private as well; only accepted offers become public, but they are non-binding. \cite{kagel2000dynamics} analyze the transition from decentralized matching to centralized clearinghouses when market features lead to inefficient matching through unraveling. \cite{haruvy2007equilibrium} study the effects of repetition on a two-stage decentralized matching protocol. \cite{agranov2022paying} allow for transfers in small decentralized markets that follow protocols similar to ours, where information about preferences is either complete or incomplete. Transfers and incomplete information make stability elusive, particularly when preferences are submodular. \cite{pais2012decentralized} study the effects of information and costly-offer frictions on outcomes and find that enough frictions may make stability difficult to achieve. Finally, \cite{niederle2009market} also look at an incomplete information setting in which one side of the market (the firms) makes offers to the other side (the workers) over three experimental periods. They study the effects of offer structure on the information that gets used in the final matching and consequent market efficiency.\footnote{There is a growing experimental literature studying centralized matching systems, e.g., \cite{bergstrom2013choosing}, \cite{harrison1996expectations}, \cite{chen2006school}, \cite{pais2008school}, \cite{echenique2016clearinghouses}, \cite{featherstone2011some}, and \cite{featherstone2011school}.}

There is also a methodological link between the current paper and some of the experimental work studying financial markets and general equilibrium predictions in the lab, see for instance \cite{smith1962experimental}, \cite{plott1978experimental}, or the survey in Chapter 6 of \cite{roth1995handbook}. As in our paper, the underlying predictions of general equilibrium theory pertain to outcomes, and by and large shine through in experiments; this despite the precise dynamics leading to these outcomes not having been imposed by the experimenters.

\section{Theoretical Preliminaries}\label{theory}

We start by reviewing the one-to-one two-sided matching model, and the theoretical results that are pertinent to our paper.

Let $F$ and $C$ be disjoint, finite sets.  We call the elements of $F$ ``foods'' and the elements of $C$ ``colors.'' We use the language of foods and colors in our experimental design, but these sets can stand for firms and workers in labor markets, men and women in heterosexual marriage markets, etc. A \emph{matching} is a function $\mu :F\cup C\rightarrow F\cup C$ such that for all $f\in F$ and $c\in C$,  

\begin{enumerate}
	\item $\mu \left( c\right) \in F\cup \left \{ c\right \} $,	
	\item $\mu \left( f\right) \in C\cup \left \{ f\right \} $,	
	\item $f=\mu \left( c\right) $ if and only if $c=\mu \left( f\right) $.
\end{enumerate}

Whenever $a\in F\cup C$ is unmatched under $\mu$, we write $\mu (a)=a$; when $f$ and $c$ are matched under $\mu$, then $c=\mu (f)$ (and $f=\mu(c)$).

A \emph{preference relation} is a linear order (a complete, transitive, and antisymmetric binary relation). In particular, we assume preferences are \emph{strict}. A preference relation for a food $f\in F$, denoted $P(f)$, is understood to be over the set $C\cup \left \{ f\right \} $, with $f$ representing the possibility of being unmatched. Similarly, for $c\in C$, $P(c)$ denotes a preference relation over $F\cup \left \{c\right \}$. For simplicity and consistency with our experimental design, we ignore individual rationality by assuming that each food (color) prefers any color (food) over remaining unmatched. A \emph{preference profile} is a list $P$ of preference relations for foods and colors, i.e.,\ 
\begin{equation*}
P=\left( \left( P(f)\right) _{f\in F},\left( P(c)\right) _{c\in C}\right) .
\end{equation*}%

Denote by $R(f)$ the weak version of $P(f)$. That is, $c^{\prime }\mathbin{R(f)}c$ if either $c^{\prime }=c$ or $c^{\prime }P(f)c$. The definition of $R(c)$, the weak version of $P(c)$, is analogous.

Fix a preference profile $P$. We say that a pair $(c,f)$ \emph{blocks} $\mu$ if $c\neq \mu (f)$, $c\mathbin{P(f)}\mu (f)$, and $f\mathbin{P(c)}\mu (c)$. In words, $(c,f)$ is a blocking pair if $c$ and $f$ prefer to be matched to one another over their assigned matches under $\mu$. In the absence of individual rationality, a matching is \emph{stable} if there is no pair that blocks it. Denote by $S(P)$ the set of all stable matchings.

\begin{theorem*}[\citealp{gale1962college}]
    $S(P)$\textit{\ is non
		empty, and there are two matchings }$\mu _{F}$\textit{\ and }$\mu _{C}$%
	\textit{\ in }$S(P)$\textit{\ such that, for all }$f\in F$\textit{, }$c\in C$\textit{,
		and }$\mu \in S(P)$\textit{, }%
	\begin{align*}
	& \mu _{F}(f)\mathbin{R(f)}\mu (f)\mathbin{R(f)}\mu _{C}(f), \\
	& \mu _{C}(c)\mathbin{R(c)}\mu (c)\mathbin{R(c)}\mu _{F}(c).
	\end{align*}
\end{theorem*}

The matchings $\mu _{F}$ and $\mu _{C}$ coincide when the market has a unique stable matching. The matching $\mu_{F}$ is called \emph{food optimal}, while $\mu _{C}$ is called \emph{color optimal}, and both are known as the \textit{extremal stable matchings}. The matching $\mu_{F}$ is preferred by all foods to any other stable matching, and all colors prefer any stable matching to $\mu _{F}$. Analogously for $\mu _{C}$. The proof of the Gale-Shapley Theorem is constructive, and uses the Deferred Acceptance (DA) algorithm to identify one of the extreme matchings, $\mu_{F}$ or $\mu_{C}$. Beyond its theoretical role in establishing existence, DA is the algorithm often used in centralized markets. For instance, the National Resident Matching Program uses a variation of it \citep{roth1999redesign}.

The set of \textit{stable partners} of an agent $a\in F\cup C$ is the set of agents that are matched to $a$ under some stable matching, i.e., $\left \{ \mu ^{\prime}(a)\mid \mu ^{\prime }\in S(P)\right \} $.\footnote{Stable partners are also referred to as ``achievable partners'' in the literature \citep[e.g.,][]{roth84}.} Likewise, a pair $(f,c)\in F\times C$ is said to be a \textit{stable pair}, or a \textit{stable match}, if $f$ and $c$ are matched under some stable matching.\footnote{Note the distinction between a \textit{match} and a \textit{matching}. A match refers to a pair $(f,c)$ who are matched. A matching refers to the function $\mu$ describing all the matches in a market.} Therefore, in a stable matching, all matches are stable, and all agents are matched to a stable partner. Nonetheless, in unstable matchings, it might be reasonable for some agents who are not stable partners to be matched: from an individual perspective, it makes sense to form matches in which no agent has a blocking partner, regardless of whether the match is stable. Similarly, while in markets with a unique stable matching every agent has a unique stable partner, the number of stable partners may differ among agents, and, in particular, may be less than the number of stable matchings when not unique.

Consider a preference profile $P$ for which $S(P)$ has an odd number $K$ of matchings, and denote the partners of agent $a$ in each of these matchings by $a_1,\dots,a_K$ (which may not all be distinct). A \emph{median stable matching} is a matching $\mu \in S(P)$ such that, for all agents $a\in F\cup C$, $\mu (a)$ is $a$'s median partner among $a$'s stable partners under $P(a)$. That is, $\mu(a)$ occupies the $\frac{K+1}{2}$-th place in $a$'s preference among $a_1,\dots,a_K$. We refer to $\mu(a)$ as the \textit{median stable partner} of $a$.

In general, median stable matchings are guaranteed to exist, see \cite{teo2001gale}---in fact, they also exist when $K$ is even. Median stable matchings present a compromise between the two sides of the market. Interestingly, there are no known simple algorithms that generate median stable matchings. Certainly, one can search for all stable matchings of a market and then identify a median one. From a computational perspective, however, this can potentially be quite demanding as the problem of finding all stable matchings is computationally hard (see \citeip{gusfield1989stable}, for general references; \citeip{irving1986complexity}, show that determining the number of all stable matchings is generally \#P-complete; while \citeip{cheng2008generalized}, shows that finding the median stable matching is hard). These results contrast with the problem of finding a color- or food-optimal stable matching, which can be done in polynomial time by using DA.

The notion of stability, as well as the ranking of the different stable matchings, are ordinal in nature. In particular, the theory does not allow for refined predictions on the basis of \emph{how much} agents prefer certain partners to others.

\section{Experimental Design\label{Design}}

Our experimental design corresponds to a decentralized one-to-one, two-sided market with no transfers. The two sides are termed colors and foods, and contain the same number of participants each. In each round, each participant is randomly assigned a role: ``red,'' ``blue,'' etc.\ if a color;  ``apple,'' ``banana,'' etc.\ if a food. A participant can match with one and only one participant from the other side of the market, each match resulting in a potentially different monetary payoff. All participants observe all potential payoffs from a numerical matrix on the experimental interface. If a participant is unmatched, they earn a payoff of $0$.\footnote{The instructions and the set of payoff matrices we used are available at the following link: \href{https://sites.google.com/view/decentralized-matching}{https://sites.google.com/view/decentralized-matching}.}

We implemented three \emph{main treatments} with 8 participants on each side of the market, and two additional \emph{auxiliary treatments} intended to check for robustness with respect to bargaining power and market size. In our main treatments, over the course of the experiment, participants are free to propose a match to anyone on the other side of the market. At any point in time, participants observe all current matches through a panel of the experimental interface. Importantly, participants can make an offer while (tentatively) matched, and offers can be made to any member of the opposing side of the market, including participants who are already matched. If a matched agent accepts a new offer, their existing match is undone. When receiving an offer, a participant has 10 seconds to respond. Each market ends after 30 seconds have transpired without any new offer.

We ran our main treatments across five experimental sessions, each consisting of 2 practice rounds and 10 real rounds.\footnote{Four of the sessions included 32 participants, so we ran two markets simultaneously, for a total of 20 non-practice experimental rounds per session. In the remaining session, we only had 16 participants for a total of 10 non-practice experimental rounds. Hence, in total, we ran 90 experimental rounds across five sessions. Due to software malfunctions, 5 markets were not presented as intended to participants. We drop these from the data. In total, we have 85 rounds of experimental market data for our main treatments. Including data from the dropped rounds does not alter our results qualitatively.\label{foot_main_sessions}} Participants took part in a different market each round. Markets were designed with two objectives in mind. First, in order to see whether cardinal representations of preferences matter, we implemented multiple cardinal representations of the same ordinal markets. Specifically, for any ordinal preference---say, red prefers apple to banana---there are many ways by which these preferences can be presented cardinally. For example, red receiving \$50 and \$10, or \$5 and \$4, from matching with apple and banana, respectively, would both correspond to the same ordinal ranking. Our design entailed 6 underlying ordinal descriptions of markets with 17 different cardinal representations. Second, in order to study the endogenous selection of stable matchings, we designed our experimental markets so that all participants had either one, two, or three possible stable partners. When all agents have only one stable partner, the market as a whole has a unique stable matching. For each fixed number of stable partners (ranging from $1$ to $3$), we use several markets differing in market participants' ordinal and cardinal preferences.\footnote{The mapping from rounds to payoff matrices varied; in some sessions we used the same markets across rounds, and in others we shuffled them. We see no strong evidence of learning in either.} 

The top panel of \autoref{tab_final_1} summarizes our three main experimental treatments. The following is a general description of the match payoffs we used.

\begin{table}[tbph!]
\centering\footnotesize
\caption{Description of treatments
\label{tab_final_1}}
\begin{tabular}{ c c c c c c c c c}
 &  &  &  &  &  &  &  &   \\\hline\hline
\textit{\scriptsize\makecell[b]{ Agents \\ per side }} & 
\textit{\scriptsize\makecell[b]{ Proposing \\ sides }} & 
\textit{\scriptsize\makecell[b]{\#Exp. \\ Mkts.}} & 
\textit{\scriptsize\makecell[b]{\#Ordinal \\ Mkts.}} & 
\textit{\scriptsize\makecell[b]{\#Cardinal \\ Mkts.}} & 
\textit{\scriptsize\makecell[b]{\#Sessions}} & 
\textit{\scriptsize\makecell[b]{\#Particp's}} & 
\textit{\scriptsize\makecell[b]{\#Stable \\ matchings}} & 
\textit{\scriptsize\makecell[b]{Avg.\ \\ \#stable \\ partners}} \\\hline
 &  &  &  &  &  &  &  &  \\
\multicolumn{9}{l}{\hspace{0.0cm}\textit{Main treatments}} \\
 &  &  &  &  &  &  & &  \\
\multicolumn{9}{l}{\hspace{0.2cm}\textit{Unique stable matching}*} \\
 &  &  &  &  &  &  & &  \\
\hspace{0.2cm} 8 & 2 & 30 & 4 & 7 & 5 & 144 & 1 & 1.00  \\
 &  &  &  &  &  &  & &  \\
\multicolumn{9}{l}{\hspace{0.2cm}\textit{Two embedded 4-by-4 markets$^\dagger$}} \\
 &  &  &  &  &  &  & &  \\
\hspace{0.2cm} 8 & 2 & 35 & 1 & 6 & 5 & 144 & $2\times2$ & 1.75  \\
 &  &  &  &  &  &  & &  \\
\multicolumn{9}{l}{\hspace{0.2cm}\textit{5 stable matchings \& 3 stable partners}} \\
 &  &  &  &  &  &  & &  \\
\hspace{0.2cm} 8 & 2 & 20 & 1 & 4 & 3 & 80 & 5 & 3.00  \\
 &  &  &  &  &  &  &  & \\\hline
 &  &  &  &  &  &  &  & \\
\multicolumn{9}{l}{\hspace{0.0cm}\textit{Auxiliary treatments}} \\
 &  &  &  &  &  &  & &  \\
\multicolumn{9}{l}{\hspace{0.2cm}\textit{Unilateral offers}} \\
 &  &  &  &  &  &  &  & \\
\hspace{0.0cm} 8 & 1 & 9 & 1 & 2 & 2 & 80 & 1 & 1.00  \\
\hspace{0.0cm} 8 & 1 & 19 & 1 & 4 & 2 & 80 & $2\times2$ & 1.75  \\
\hspace{0.0cm} 8 & 1 & 15 & 1 & 3 & 3 & 112 & 5 & 3.00  \\
 &  &  &  &  &  &  &  &  \\
\multicolumn{9}{l}{\hspace{0.2cm}\textit{Large markets}$^\ddagger$} \\
 &  &  &  &  &  &  &  &  \\
\hspace{0.0cm} 15 & 2 & 8 & 3 & 3 & 3 & 90 & 1 & 1.00  \\
\hspace{0.0cm} 15 & 2 & 4 & 2 & 2 & 2 & 60 & 3 & 2.93  \\
 &  &  &  &  &  &  & &  \\\hline
&  &  &  &  &  &  & &  \\
\multicolumn{2}{l}{Total}  & 140 & 11 & 22 & 12 & 346 &  &  \\\hline\hline
\multicolumn{9}{p{14.5cm}}{\scriptsize\textit{Notes.} The table reports all experimental markets (rounds within a session) across our three main treatments (\textit{unique stable matching}, \textit{four-by-four}, and \textit{five stable matchings}), and two auxiliary treatments (\textit{unilateral offers} and \textit{large markets}). For each treatment, the table reports: the number of agents on each side of the market, the number of sides that can make match offers, the number of experimental markets ran in this treatment, and the ordinal and cardinal representations used across them, the number of sessions in which at least one market in the corresponding treatment was played, the number of participants who played in at least one market in this treatment, and the number of stable matchings and the average number of stable partners in the markets in the treatment. The last row reports the totals across all treatments.\newline
$^*$One of the ordinal markets (out of 4) had a salient egalitarian unstable matching, which was used in 4 experimental markets (out of 30). Eight experimental markets (out of 30) were run with two alternative ordinal payoff matrices: four of them had one-sided aligned preferences, and the other four had two-sided aligned preferences.\newline
$^\dagger$In each of the two embedded markets, there were eight agents, four on each side. Three agents on each side had two stable partners, and one agent had one, for an average of 1.75 stable partners per agent.\newline
$^\ddagger$In the large markets with a unique stable matching, five (out of 8) had one-sided aligned preferences. In the large markets with multiple stable matchings, in one of the markets every agent had three stable partners, while in the other one some agents had two stable partners and most had three (avg $=$ 2.87).\newline\vspace{1cm}
}
\end{tabular}
\end{table}

\paragraph{Unique stable matching.}We used $4$ different ordinal markets that have a unique stable matching: assortative preference markets, where participants on each side of the market agree on the ranking of participants on the other market side; markets with assortative preferences on one side, where only the members of one side of the market are in agreement; a market including a fully egalitarian matching, providing all agents the same payoff, which is unstable; and a ``generic'' market with a unique stable matching without agents agreeing on the ranking of others on either side. Each of the first three markets was implemented via one cardinal representation. The last, generic market was implemented via four cardinal representations. In two of these we varied how aligned the interests were across the market: that is, if $\mu $ is the stable matching, we computed the correlation of the vectors $(u_{f}(\mu ))_{f\in F}$ and $(u_{\mu (f)}(\mu ))_{f\in F}$, where $u_{a}(\mu )$ denotes the payoff of agent $a\in F\cup C$ in the matching $\mu$. We created one market in which the correlation was $-0.9$ and one where it was $0.9$. We used two additional cardinal representations: one in which, for each agent, the difference in utilities between matching with the agent's $k$'th and $k+1$'th choices was $20\cents $ and one in which these marginal differences were $70\cents$. Altogether, we used $7$ different cardinal markets with a unique stable matching.

\paragraph{Two embedded four-by-four markets.}We used one ordinal market in which each agent had two possible stable match partners. These were constructed so that there were two $4\times 4$ embedded markets, where any agent within a submarket preferred to match with anyone from that submarket over anyone from the other. We varied the overall utilitarian efficiency of each matching, the utilitarian efficiency of foods relative to colors from each matching, the distribution within each matching,\footnote{Since egalitarian motives appear frequently in experiments, we were concerned that some form of altruism would be driving our results. We therefore designed payoffs so that in some treatments, fully egalitarian matchings were unstable (see the description of our markets with a unique stable matching above). Furthermore, we included cardinal representations in which certain \emph{stable} matchings were more egalitarian than others.} as well as the marginal loss for either side of the market from switching from their more preferred stable matchings to their less preferred ones (higher for foods or for colors). Overall, we used $6$ cardinal markets of this sort.

\paragraph{Five stable matchings and three stable partners.}We used one ordinal market represented cardinally in $4$ ways: one in which the marginal differences between utilities derived from matching with one's $k$'th and $k+1$'th most preferred partners was $20\cents$, one in which it was $70\cents$, one in which for foods it was $20\cents$ and for colors $70\cents$, and one in which these differences were $20\cents $ for both market sides, but colors' payoffs were all shifted up by $\$1$. In these markets, while each individual has precisely three distinct stable partners, there are five different market-wide stable matchings.\bigskip

In addition to our three main treatments, with 8 participants on each side who could all make match offers, we implemented two auxiliary treatments that differed in either the relative bargaining power each market side had or the market size.\footnote{We ran four sessions with unilateral markets. In three sessions there were 32 participants, so we ran two markets at the same time, each consisting of 16 participants, 8 per side. In the remaining session we only had 16 participants. However, we exclude the first session from the analysis, which had 32 participants, due to a software glitch that affected how the termination rule was implemented. While none of our results change qualitatively if we include this session in the analysis, some markets terminated before 30 seconds of inactivity and some after in this session. This glitch was fixed between the first and the second session. Additionally, for a similar reason as with some markets in our main treatments (see footnote \ref{foot_main_sessions}), some markets were not presented as intended to participants, so we drop them from the analysis. In total, this adds up to 43 experimental markets in our unilateral treatments. We ran three sessions with large markets, each with 30 participants, 15 on each side. Since large markets took longer to converge than our main and unilateral treatments, we ran fewer rounds. In two sessions, we ran five non-practice rounds, and in one, we ran two, for a total of 12 experimental markets in our large treatment.} The bottom panel of \autoref{tab_final_1} summarizes our two auxiliary experimental treatments. 

\paragraph{Unilateral offers.}We ran treatments with $8\times 8$ markets and payoffs as in our main treatments where only foods could make offers. Otherwise, markets operated as in our main treatments.

\paragraph{Large markets.}We also ran several treatments with larger markets, containing 15 participants on each side. We concentrated on markets with either a unique stable matching or $3$ stable matchings. We used $3$ distinct markets with a unique stable matching: assortative preferences on one side, assortative preferences with a fully egalitarian unstable matching, and preferences that were not assortative on either side. We used two markets with $3$ stable matchings (in one market, all agents had three stable partners; in the other one, most agents had three stable partners, while some had two).\bigskip

All sessions took place at the California Social Science Experimental Laboratory (CASSEL), using a modification of the multi-stage software. All participants were UCLA undergraduates and each participant participated in only one session. The average payment per participant was $\$40$ in our main treatments, $\$33$ in the treatments in which only foods were able to make offers, and $\$24$ in the large $15\times 15$ market treatments.\footnote{%
    Standard deviations were $\$3$, $\$13$, and $\$9$, respectively.} %
All of these were combined with a $\$5$ show-up fee. 

\section{Market Outcomes\label{outcomes}}

Three main findings emerge from our experiments regarding final outcomes. First, across our treatments, \emph{most outcomes are stable}. Furthermore, market outcomes that are unstable are close to stable; they are close in the sense that both the number of blocking pairs, as well as the unrealized payoff gains from not forming them, are small. Second, in our treatments with three stable partners, \emph{most agents are matched to their median stable partner}. Even when we ``handicap'' one side of the market, so that it cannot make any offers, we continue to see the median as the modal outcome. Last, \emph{cardinal representations of preferences affect the particular stable matchings that get selected}. Specifically, higher cardinal incentives to colors make the color optimal matching more likely to be selected; similarly for foods. 

In all our treatments, learning across rounds did not appear to have significant effects on either outcomes or behavior. Therefore, we present aggregate results across rounds. We return to the within-round dynamics underlying our final outcomes in \autoref{market_dynamics}.

\subsection{Stability in Experimental Markets\label{stabilitysection}}

Virtually all agents match through our markets' operations, and a large fraction of markets culminate in a stable matching, as shown in \autoref{tab_final_2}. The table summarizes the overall outcomes in our main treatments. Over 99\% of agents are matched when markets terminate. Furthermore, $88\%$ of markets are fully stable: no agent in the whole market has a blocking partner. Markets with three stable partners exhibit slightly fewer market-wide stable outcomes, $75\%$. As a measure of stability at the match or pair level, we check the proportion of matched pairs in which no member was part of a blocking pair. In over 95\% of matches, no member has a blocking partner, with little variation across markets. The treatment with the lowest proportion of blocking partners is the one in which every agent has three stable partners. Even in this case, in 90\% of matches, no agent can form a blocking pair.\footnote{The use of color and food labels in our markets did not seem to have any effect. For example, if one considers banana and mango to be associated with yellow, apple and cherry with red, and kiwi and pear with green, there is no significant increase in the corresponding matches relative to any other classification.} 

\begin{table}[tbp!]
\centering\footnotesize
\caption{Outcomes in main treatments
\label{tab_final_2}}
\begin{tabular}{lcccc}
 &  &  & & \\\hline\hline
 &  &  & & \\
 &
\textit{\footnotesize\makecell[b]{Unique stable \\ matching}} & 
\textit{\footnotesize\makecell[b]{Two embedded \\ 4-by-4 markets}} & 
\textit{\footnotesize\makecell[b]{5 stable matchings \\  \& 3 stable partners}} & 
\textit{\footnotesize\makecell[b]{All main \\ treatments}} \\\hline
&  &  &  &  \\
\textit{\footnotesize\makecell[l]{\#Mkts.\ with stable \\ matching / \#Mkts}} & 27 / 30 & 33 / 35 & 15 / 20 & 75 / 85 \\
 &  &  & &  \\
\textit{\footnotesize\makecell[l]{\% Mkts.\ with stable \\ matching}} & 90.00 & 94.29 & 75.00 & 88.24 \\
&  &  & &  \\
\textit{\footnotesize\makecell[l]{ Avg.\ \% pairs w/o \\ blocking partners (BPs) }} & 95.32 & 98.65 & 90.00 & 95.44 \\
&  &  & &  \\
\textit{\footnotesize\makecell[l]{ Avg.\ \% pairs w/o BPs $\mid$ \\ unstable matching }} & 53.24 & 76.39 & 60.00 & 61.25 \\
&  &  & &  \\
\textit{\footnotesize\makecell[l]{ Avg.\ \% agents with \\ $\geq1$ BP }} & 2.50 & 0.71 & 5.00 & 2.35 \\
&  &  & &  \\
\textit{\footnotesize\makecell[l]{ Avg.\ \% agents with $\geq1$ \\ BP $\mid$ unstable matching }} & 25.00 & 12.50 & 20.00 & 20.00\\
&  &  & &  \\
\textit{\footnotesize\makecell[l]{ Avg.\ \# of BPs per \\ agent $\mid$ $\geq1$ BP }} & 1.35 & 1.00 & 1.23 & 1.22 \\
&  &  & &  \\
\textit{\footnotesize\makecell[l]{ Avg.\ \% unmatched \\ agents }} & 0.42 & 0.36 & 0.00 & 0.29 \\
&  &  & &  \\
\textit{\footnotesize\makecell[l]{ Avg.\ \% unmatched \\ agents $\mid$ $\geq1$ BP }} & 4.76 & 25.00 & 0.00 & 6.43 \\
&  &  & &  \\\hline\hline
\multicolumn{5}{p{14.5cm}}{\scriptsize\textit{Notes.} The table reports the following final outcomes for each of our \textit{main treatments}: (i) number and (ii) percent of final matchings that are stable; avg.\ number of final pairs (matches) in which no agent has a blocking partner across (iii) all markets and (iv) markets with an unstable final matching; average percent of agents with at least one blocking partner across (v) all markets and (vi) markets with an unstable final matching; (vii) average number of blocking partners per agent across agents with at least one blocking partner; average number of unmatched agents (viii) among all agents, and (ix) among agents who have at least one blocking partner.}
\end{tabular}
\end{table}

\begin{figure}[tbph!]
	\centering
	\includegraphics[width=0.49\textwidth]{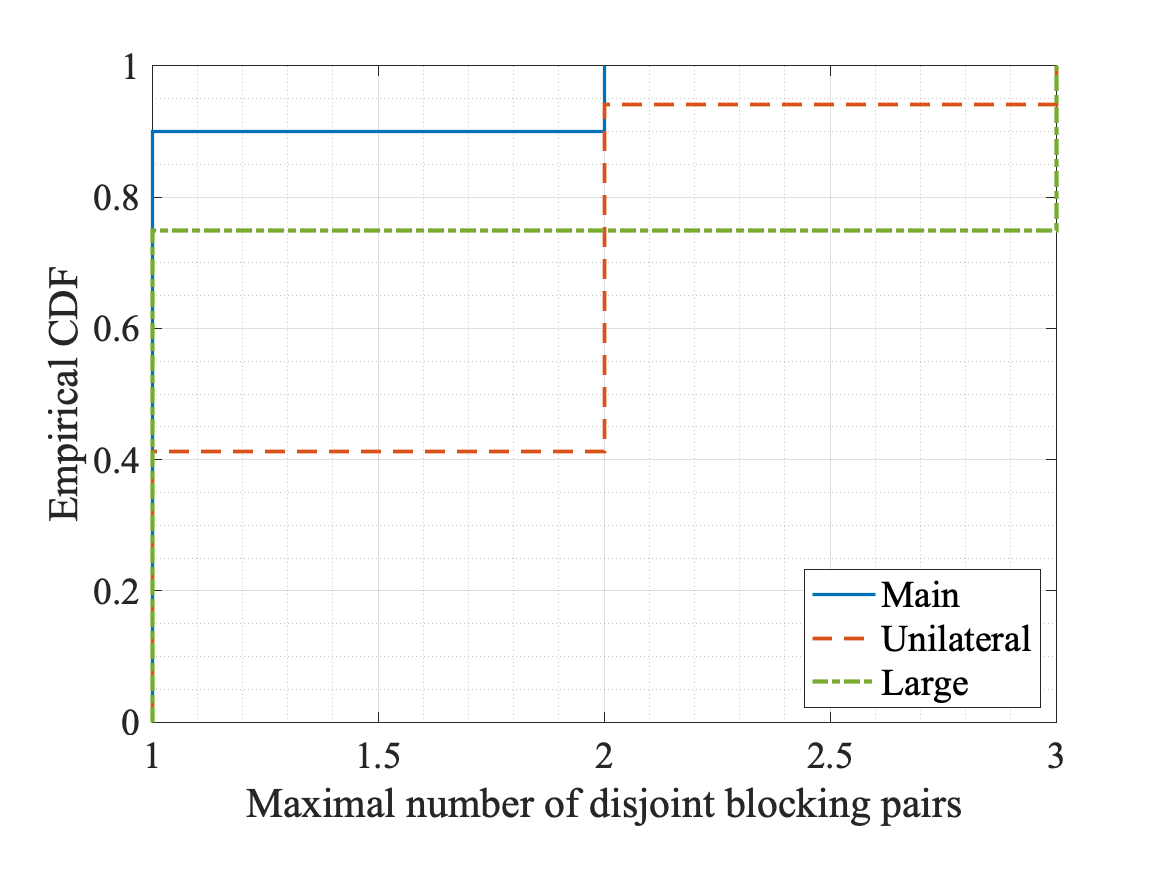}
	\includegraphics[width=0.49\textwidth]{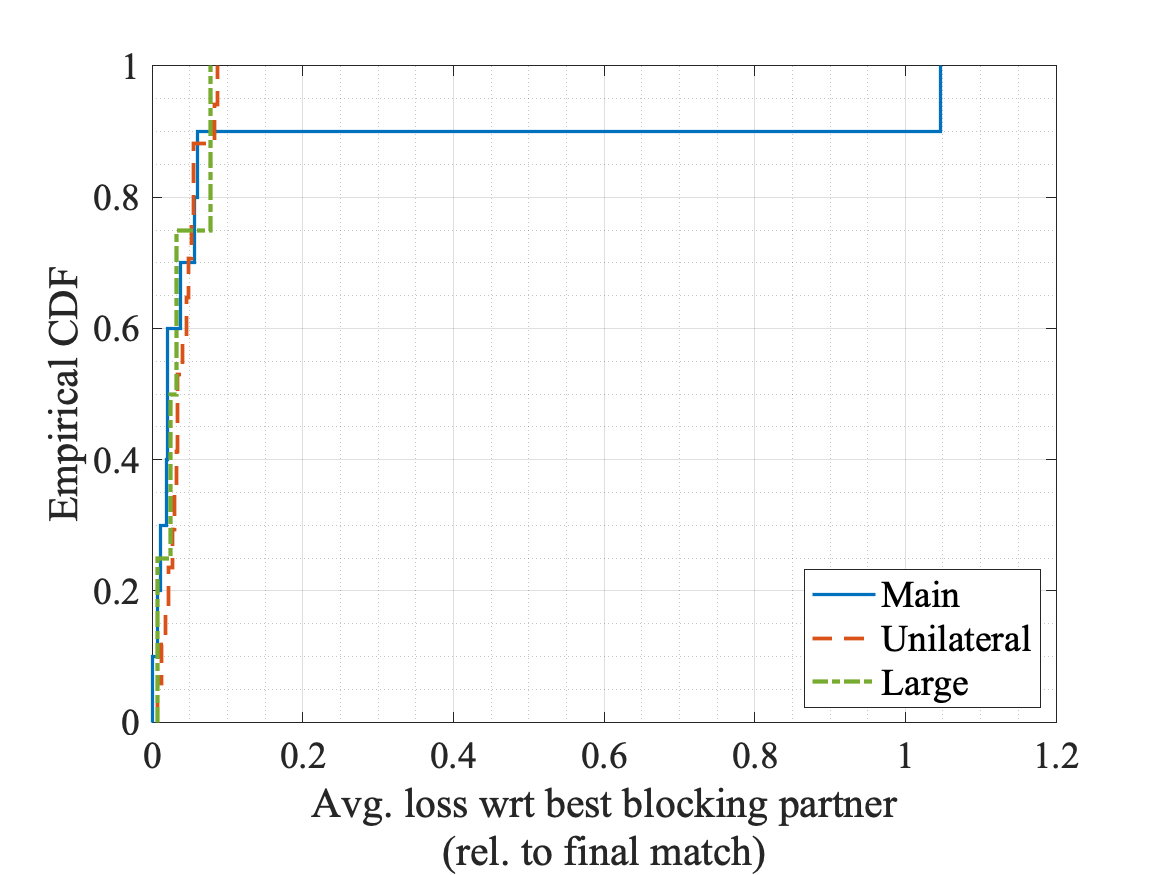}
	\caption{Distance to stability in unstable markets
    \label{fig_final_1}}
\end{figure}

Not all our markets reached full stability, but markets that were not fully stable were close to stable. Even in unstable markets, \autoref{tab_final_2} indicates that, in the majority of matched pairs (61\%), no agent has a blocking partner. Indeed, only 20\% of agents have a blocking partner when markets finalize (compared to 2\% of agents across all markets). The left panel of \autoref{fig_final_1} presents the empirical cumulative distribution functions (CDFs) of the maximal number of disjoint blocking pairs across all treatments, main and auxiliary.\footnote{In our main treatments, when no agents are matched, the maximal number of disjoint blocking pairs is $8$, the number of potential disjoint pairs. When a stable matching is in place, there are $0$ blocking pairs.} The figure contains the distribution of blocking pairs pertaining to \textit{markets in which outcomes were not fully stable}.\footnote{Including all markets in the figure would not allow to visualize the fine-grained distribution of blocking pairs, as the prevalence of stable matchings implies a large spike at zero blocking pairs.} In our main treatments, most outcomes are close to stable: most markets culminating in unstable outcomes have only one or two disjoint blocking pairs.

We can also use payoffs to measure distance to stability. For each agent, we compare the payoff they received from their final match, with the payoff they would have received had they matched with their most preferred blocking partner at the end of the market. That is, we compute the maximum immediate losses agents incur from not exploiting blocking opportunities. Agents in markets that did not culminate in a stable matching lost, on average, 12.8\% of their final market payoff, equivalent to around 50$\cents$. By contrast, the same average across all markets is 1.5\%, equivalent to around 6$\cents$. The right panel of \autoref{fig_final_1} reports the CDFs of this measure across all unstable markets. Even in unstable markets, most agents had very small payoff losses due to instability. Indeed, the median loss is equivalent to 2.1\% of the final payoff, which comes down to around 9$\cents$.\footnote{In the Online Appendix, we plot CDFs for alternative measures of distance to stability across the distinct treatments: total number of blocking pairs, and the average immediate loss with respect to the best blocking partner (i) in absolute terms, (ii) relative to average market payoffs, and (iii) relative to the average payoff across random matchings (which with high likelihood are unstable). Roughly speaking, all measures point to the same conclusion: the majority of agents in markets that did not reach full stability had few blocking partners and did not incur substantial immediate payoff losses.}

In terms of the duration of market interactions, convergence to the stable matching was rapid. On average, in our main treatments the final matching was reached after $1.62$ minutes (sd = 1.00). While markets terminated 30 seconds after the last offer was made, the last offer usually took place after the final matching was reached (i.e., the last offers are generally not accepted and do not change the matching). On average, the last offer took place after 2.28 minutes (sd = 1.35), and markets terminated 30 seconds afterwards, with 89\% of markets terminating before the five minute mark. We return to the dynamics underlying these observations in \autoref{market_dynamics}.

\subsection{The Emergence of Median Stable Matches\label{selectionsection}}

The median stable matching has strong drawing power, and cardinal representations of preferences affect outcomes, as shown in \autoref{tab_final_3}. The top panel of the table reports the distribution of final matchings and matched pairs across markets in our main treatments in which each agent has three stable match partners. A key difference between the markets is the level, and marginal differences, of payoffs within each side of the market. We use the notation of ``$x$-$y$ marginals'' to denote a market in which the marginal difference in utilities between one partner and the next-best partner was $x$ cents for foods and $y$ cents for colors. For example, in markets with 20-70 marginals the payoff differences to the next-best partner was $20\cents$ for foods and $70\cents$ for colors. In the 20-20 marginals market with $100$ color shift (labeled as 20-20$_{+100}$ in the table), a $100\cents $ (or $\$1$) was added to the color payoffs of the 20-20 market.

\begin{table}
\centering\footnotesize
\vspace{-0.5cm}
\caption{Selection of stable matching and cardinal effects 
\label{tab_final_3}}
\begin{tabular}{llllll}
 &  &  &  &  &   \\\hline\hline
\textit{Marginal Utility} &  &  &  &  &  \\
\textit{Differences (foods--colors)} \hspace{0.5cm} & 
\multicolumn{1}{p{1.55cm}}{\textit{20--20}} & 
\multicolumn{1}{p{1.56cm}}{\textit{20--20$_{+100}$}} & 
\multicolumn{1}{p{1.55cm}}{\textit{20--70}} & 
\multicolumn{1}{p{1.55cm}}{\textit{70--70}}  &
\multicolumn{1}{p{1cm}}{\textit{All}} \\\hline
&  &  &  &  &   \\
\multicolumn{6}{l}{\textit{Main: markets with three stable partners}} \\
&  &  &  &  &   \\
\hspace{0.2cm} \textit{\# markets} & 5 & 5 & 5 & 5 & 20 \\
&  &  &  &  &   \\
\multicolumn{6}{l}{\hspace{0.2cm} \textit{Market-level outcomes (matchings)}}  \\
&  &  &  &  &   \\
\hspace{0.2cm} \% \textit{stable} & 60 & 40 & 100 & 100 & 75\\
\hspace{0.2cm} \% \textit{median} $\mid$ \textit{stable} & 67 & 100 & 60 & 100 & 80\\
\hspace{0.2cm} \% \textit{non-extremal} $\mid$ \textit{stable} & 100 & 100 & 100 & 100 & 100\\
\hspace{0.2cm} \% \textit{food-optimal} $\mid$ \textit{stable} & 0 & 0 & 0 & 0 & 0 \\
\hspace{0.2cm} \% \textit{color-optimal} $\mid$ \textit{stable} & 0 & 0 & 0 & 0 & 0 \\
&  &  &  &  &   \\
\multicolumn{6}{l}{\hspace{0.2cm} \textit{Individual-level outcomes (matches)}}  \\
&  &  &  &  &   \\
\hspace{0.2cm} \% \textit{stable} & 95 & 93 & 100 & 100 & 97\\
\hspace{0.2cm} \% \textit{median} $\mid$ \textit{stable} & 64 & 63 & 80 & 100 & 77\\
\hspace{0.2cm} \% \textit{food-optimal} $\mid$ \textit{stable} & 11 & 0 & 0 & 0 & 3\\
\hspace{0.2cm} \% \textit{color-optimal} $\mid$ \textit{stable} & 24 & 37 & 20 & 0 & 20\\
&  &  &  &  &   \\\hline
&  &  &  &  &   \\
\multicolumn{6}{l}{\textit{Unilateral offers (foods propose): markets with three stable partners}} \\
&  &  &  &  &   \\
\hspace{0.2cm} \textit{\# markets} & 0 & 5 & 5 & 5 & 15\\
&  &  &  &  &   \\
\multicolumn{6}{l}{\hspace{0.2cm} \textit{Market-level outcomes (matchings)}}  \\
&  &  &  &  &   \\
\hspace{0.2cm} \% \textit{stable} & -- & 20 & 20 & 60 & 33\\
\hspace{0.2cm} \% \textit{median} $\mid$ \textit{stable} & -- & 0 & 100 & 100 & 80\\
\hspace{0.2cm} \% \textit{non-extremal} $\mid$ \textit{stable} & -- & 100 & 100 & 100 & 100\\
\hspace{0.2cm} \% \textit{food-optimal} $\mid$ \textit{stable} & -- & 0 & 0 & 0 & 0\\
\hspace{0.2cm} \% \textit{color-optimal} $\mid$ \textit{stable} & -- & 0 & 0 & 0 & 0\\
&  &  &  &  &   \\
\multicolumn{6}{l}{\hspace{0.2cm} \textit{Individual-level outcomes (matches)}}  \\
&  &  &  &  &   \\
\hspace{0.2cm} \% \textit{stable} & -- & 93 & 88 & 95 & 92\\
\hspace{0.2cm} \% \textit{median} $\mid$ \textit{stable} & -- & 49 & 77 & 83 & 70\\
\hspace{0.2cm} \% \textit{food-optimal} $\mid$ \textit{stable} & -- & 14 & 17 & 17 & 16\\
\hspace{0.2cm} \% \textit{color-optimal} $\mid$ \textit{stable} & -- & 37 & 6 & 0 & 14\\
&  &  &  &  &   \\\hline\hline
\multicolumn{6}{p{14cm}}{\scriptsize\textit{Notes.} The table reports (i) \% of markets which final matching corresponds to the median, non-extremal, food-optimal, or color-optimal stable matching, and (ii) the percentage of final matches that are food-optimal, color-optimal, or median stable matches (i.e., part of a stable matching). The table reports results for all experimental markets with five stable matchings, in the \textit{main treatments} (top panel) and \textit{unilateral offers} (bottom panel). The results are disaggregated into cardinal treatments, according to the marginal utility of each side (payoff difference from less preferred to more preferred partner). The second column corresponds to markets in which utility differences are 20\cents\, on each side, with the payoffs of colors shifted upwards by \$1.}
\end{tabular}
\end{table}

The median stable matching is the modal outcome: $80\%$ of stable matchings correspond to the median stable matching, entailing \emph{all} participants being matched to their median stable partner. Moreover, \emph{none} of these markets converged to an extremal stable matching, food-optimal or color-optimal. Recall that the markets in which each agent has three stable partners have five market-wide stable matchings. Hence, there are two stable matchings that are neither the median nor extremal. We emphasize that, in these markets, a median matching had to be ``discovered'' by participants: it is quite hard to infer the set of stable matchings from looking at the payoff tables (an $8\times 8$ table with $128$ numerical entries); almost impossible without using a computer program. We return to the dynamics underlying the convergence to the median in \autoref{market_dynamics}.

Shifting attention to individual match-level outcomes, we observe similar patterns. $77\%$ of the final matched pairs that were stable correspond to median stable partners. From the perspective of individual behavior, agents care about stable partners, not market-wide matchings. It is remarkable that the vast majority of stable matches---which comprise 97\% of all final matched pairs---correspond to the median. That is, in the vast majority of cases, agents ended up in stable matches in which neither of the two parties were matched to their most preferred stable partner: no side of the market systematically got their way. Of the remaining stable matches, the majority were color-optimal stable matches (20\% out of 23\%), and the rest were food-optimal matches.\footnote{The preferences in these markets, both ordinal and cardinal, were not symmetric across the two sides, which explains the imbalance.} 

Stability depends on agents' ordinal preferences alone, not on their cardinal representation. Of course, our experimental participants receive different monetary payments depending on who they match with, and one might expect the different monetary magnitudes to play a role. It is interesting, then, that in our data, different cardinal representations seem to affect more \emph{which} stable matching gets selected than whether a stable matching is achieved, as shown in \autoref{tab_final_3}. Participants on the side of the market (foods or colors) having ``more to lose'' tend to partner with their most preferred match at higher frequencies (significant at any conventional levels). We return to the effects of cardinal preference representations on outcomes in \autoref{cardinalIncentives}.

\section{Robustness to Bargaining Power and Market Size}

We investigate the robustness of our results along two dimensions: the bargaining power afforded to each side of the market, and the market's size. As for bargaining power, in our \emph{unilateral design}, only one side of the market is allowed to make offers. Specifically, only foods are able to make offers to colors. The effect of market size is studied by means of our \emph{large-market design}, which resembles our main treatments but with 15 agents on each market side. \autoref{tab_final_4} summarizes our findings.

\begin{table}[p!]
\centering\footnotesize
\caption{Outcomes in unilateral offers and large markets 
\label{tab_final_4}}
\begin{tabular}{lcc|c}
 &  &  \multicolumn{2}{c}{}  \\\hline\hline
 &  &  &  \\
& \textit{\footnotesize\makecell[b]{Unilateral offers}} & 
\textit{\footnotesize\makecell[b]{Large markets}} & 
\textit{\footnotesize\makecell[b]{Main}} \\\hline
 &  &  &  \\
\textit{\footnotesize\makecell[l]{\#Mkts.\ with stable \\ matching / \#Mkts}} & 26 / 43 & 8 / 12 & 75 / 85 \\
 &  &  &  \\
\textit{\footnotesize\makecell[l]{\% Mkts.\ with stable \\ matching}} & 60.47 & 66.67 & 88.24 \\
 &  &  &  \\
\textit{\footnotesize\makecell[l]{ Avg.\ \% pairs w/o \\ blocking partners (BPs) }} & 83.30 & 93.33 & 95.44 \\
 &  &  &  \\
\textit{\footnotesize\makecell[l]{ Avg.\ \% pairs w/o BPs $\mid$ \\ unstable matching }} & 57.76 & 80.00 & 61.25 \\
 &  &  &  \\
\textit{\footnotesize\makecell[l]{ Avg.\ \% agents with \\ $\geq1$ BP }} & 9.60 & 3.61 & 2.35 \\
 &  &  &  \\
\textit{\footnotesize\makecell[l]{ Avg.\ \% agents with $\geq1$ \\ BP $\mid$ unstable matching }} & 24.27 & 10.83 & 20.00 \\
 &  &  &  \\
\textit{\footnotesize\makecell[l]{ Avg.\ \# of BPs per \\ agent $\mid$ $\geq1$ BP }} & 1.19 & 1.04 & 1.22 \\
 &  &  &  \\
\textit{\footnotesize\makecell[l]{ Avg.\ \% unmatched \\ agents }} & 0.87 & 0.00 & 0.29 \\
 &  &  &  \\
\textit{\footnotesize\makecell[l]{ Avg.\ \% unmatched \\ agents $\mid$ $\geq1$ BP }} & 3.00 & 0.00 & 6.43 \\
 &  &  &  \\\hline\hline
\multicolumn{4}{p{11cm}}{\scriptsize\textit{Notes.} The table reports the following final outcomes for markets with \textit{unilateral offers} and in \textit{large markets}: (i) number and (ii) percent of final matchings that are stable; avg.\ number of final pairs (matches) in which no agent has a blocking partner across (iii) all markets and (iv) markets with an unstable final matching; average percent of agents with at least one blocking partner across (v) all markets and (vi) markets with an unstable final matching; (vii) average number of blocking partners per agent across agents with at least one blocking partner; average number of unmatched agents (viii) among all agents, and (ix) among agents who have at least one blocking partner. For ease of comparison, we replicate the rightmost column of \autoref{tab_final_2}, corresponding to markets in our \textit{main treatments}, in the  last column.}
\end{tabular}
\end{table}

\paragraph{Unilateral Offers.} A natural possible explanation for the prevalence of the median stable matches is that, in our experiments, participants on both sides of the market could make offers. In a sense, they had equal bargaining power. In contrast, in Gale and Shapley's DA, only one side makes offers while the other side decides which offers to accept. DA produces the optimal matching for one side, so it is possible that allowing both sides of the market to make offers is at the root of the frequent median matches we observe.\footnote{In \autoref{market_dynamics}, we contrast the dynamics we observe in the experiment with a two-sided version of DA \citep{dworczak21}, and still observe that median stable matches are more prone to occur in our experiment than when both sides propose in DA.} This conjecture is interesting, particularly in view of the fact that several real-world matching markets that operate in a decentralized manner allow one side greater, if not sole responsibility for making offers: e.g., the job market for academics, the marriage market in certain cultures, etc.\ To explore the effects of bargaining or proposal power, in our \emph{unilateral offers} treatments, only foods were allowed to propose matches.  While stability rates are not as high as in our main treatments, we find that outcomes are mostly stable and also correspond to median outcomes.

With unilateral offers, outcomes are qualitatively similar to those in our main treatments, as seen in the bottom panel of \autoref{tab_final_3} and the first and third columns of \autoref{tab_final_4}. In these treatments, $99\%$ of participants are matched. In 83\% of the final matches, no agent is in a blocking pair, compared with 95\% in the markets in which both sides can make offers. At the market level, $60\%$ of the markets culminate in a stable matching, compared with 88\% in the main treatments (see \autoref{tab_final_2}).\footnote{The corresponding percent of final matches in which no agent is in a blocking pair (final matchings that are stable) for markets with a unique stable matching, two embedded four-by-four markets, and with five stable matchings, in unilateral markets, are $94\%$ ($78\%$), $91\%$ ($74\%$), and $67\%$ ($33\%$).} As \autoref{fig_final_1} shows, the observed unstable matchings are close to stable, although not as close as in our main treatments. Across the markets that fail to reach stability, on average, around a quarter of agents has at least one blocking partner, compared to a fifth in the main treatments. Similar to the main treatments, agents who have blocking partners, have $1.2$ such partners on average. The convergence time in unilateral markets is similar to our main treatments: on average, it takes $1.64$ minutes to reach the final matching (sd = 1.04), the last offer takes place after $1.69$ minutes (sd = 1.10), and markets terminate 30 seconds afterwards, with $95\%$ of markets terminating before 5 minutes.

Regarding selection, two important observations emerge. First, in markets with four stable matchings (two per each embedded submarket), the stable matchings favored by foods, the proposers, are much more frequent. In our main treatments, when both sides can make offers, $27\%$ of markets converge to the food-optimal stable matching, $21\%$ to the color-optimal, and the rest to a non-extremal stable matching.\footnote{In these markets, non-extremal stable matchings are the ones in which one of the $4\times 4$ submarkets is at the food-optimal stable matching, and the other submarket is at the color-optimal one.} By contrast, in markets in which only foods can make offers, $86\%$ of the markets converge to the food-optimal stable matching, and none to the color-optimal stable matching (the rest to one of the two non-extremal ones). The same finding emerges when looking at individual matches across all markets, not only those reaching stability. When both sides can make offers, among all the agents who have two stable partners ($75\%$), $54\%$ of final matches are food-optimal, and the rest are color-optimal. With unilateral offers, a higher proportion of the final matches are food-optimal ($90\%$). 

For the markets with two embedded four-by-four markets, results are consistent with extremal outcomes, but this is no longer true in markets with three stable partners. Indeed, for markets in which each agent has three stable partners, median outcomes are again the most common. As the bottom panel of \autoref{tab_final_3} shows, $70\%$ of final matches that are stable correspond to median stable matches when only foods are able to make offers, compared to $77\%$ when both sides can do so. Similarly, among markets that reach full stability, $80\%$ reach the median stable matching, same as when both sides can make offers. At the individual level, there is an increase in the number of matches that are optimal for foods, the proposing side, from $3\%$ to $16\%$, but this comes mainly at the expense of matches that are optimal for colors, the receiving side. Nonetheless, cardinal incentives matter for the distribution of outcomes as they do in our main treatments: when foods have more to lose from forgoing higher-ranked partners, color-optimal matches are less common.

\paragraph{Market Size.} The outcomes in our large-market treatments are similar to those in our main treatments, as seen in the second and third columns of \autoref{tab_final_4}. Our large markets involve 30 participants, which is quite large as experimental markets go. They do not, of course, come near the size of some real-world matching markets---such as, for example, the medical residents market---but it is a comfort that duplicating the size of our the markets in our main treatments does not upset our main results. 

In our large-market treatments, in 93\% of the final matches, no agent was part of a blocking pair. In terms of market-wide outcomes, while $67\%$ of matchings are fully stable, the number of blocking pairs within unstable markets is, again, very small, see \autoref{fig_final_1}. Our large experimental markets exhibit at most $3$ disjoint blocking pairs, and in 75\% of cases just one. Indeed, as \autoref{tab_final_4} reports, even in markets that fail to reach full stability, the vast majority of agents have no blocking partners (89\%, as compared to 96\% across all markets, unstable and stable). Agents who have blocking partners have, on average, just one; and there are no agents left unmatched. Interestingly, the seemingly more complex markets with three stable matchings always culminate in a stable matching in our large $15\times 15$ markets. 

As one might expect, convergence is slower in large markets. On average, it takes $4.47$ minutes for large markets to reach a final matching (sd = $3.16$). The last offer takes place after $4.86$ minutes (sd = $3.06$), and markets terminate 30 seconds afterwards, with half terminating before the five-minute mark.

As for the selected stable match partners, the results are similar to those in our main treatments, although arguably more extreme. In the large market treatments, \textit{every} agent who has three stable partners matches with their median stable partner. Hence, the market-wide stable matchings are exclusively median stable matchings.

\section{Cardinal Incentives and Social Preferences\label{cardinalIncentives}}

A matching delivers cardinal (monetary) payoffs to market participants, and these payoffs may be more or less fair: they may be more or less equal across participants. The degree of fairness of a matching can reasonably be expected to influence whether it is chosen: a taste for egalitarian outcomes is well-documented in experimental economics (the literature has suggested different types of social preferences, for surveys see, e.g., Chapter 4 in \citeip{kagel2020handbook}, and \citeip{fehr2000fairness}). Fairness considerations may, in particular, be important given our finding that median matches are very common.

Our design entailed several markets with identical ordinal preference profiles, but different cardinal utility representations. Focusing on such variations, we have already seen that the side that faces steeper cardinal incentives is more likely to achieve its optimal stable matching. We now inspect the impact of payoff \textit{distributions}. We show that equality of payoffs across match partners makes a matching more likely, \emph{as long as it is stable.}

We first analyze markets with a salient egalitarian, but unstable, matching. In these markets, one unstable matching entails identical payoffs to all market participants, and comparable utilitarian welfare to that generated by the unique stable matching.\footnote{By specifying identical payoffs to all market participants, we avoid the challenge of identifying the individuals that are relevant for participants' social preferences.} Participants consistently fail to select the unstable egalitarian matching. In our main treatments, none of these markets end up in the unstable, albeit more equal, matching outcome. 

Our larger markets, in which more payoff variation could be introduced, generate similar findings. For example, we ran one $15\times 15$ market in which there was an unstable matching under which all agents received exactly $\$4$; there was a unique stable matching in which the average payoff was also $\$4$, which was much more unequal.\footnote{Its Gini index was 26. For a country's income distribution, this is in the Scandinavian range, but it appears starkly unequal compared to perfect egalitarianism with no utilitarian efficiency loss.} While the final outcome in these markets is not always stable---in some instances a few blocking pairs remain---participants clearly avoid the egalitarian unstable matching.

Second, we use the markets with two stable matchings to assess the effects of cardinal utilities on the selection of food- and color-optimal stable matchings.\footnote{Recall that each of these markets was constructed by embedding two smaller $4\times 4$ markets, which allows us to gain more payoff variations. In these markets, $99\%$ of final matches are stable. In addition, the embedding of the ``submarkets'' was effective---indeed, there are only $6\%$ of cross-offers and $1\%$ of cross-matches (the acceptance probability of cross-offers is 0.05, compared with 0.42 of non-cross offers).} Markets are more likely to converge to the extremal stable matching, food- or color-optimal, which has the lowest payoff variation. That is, when the dispersion of payoffs is relatively high in, say, the color-optimal matching, markets tend to achieve the food-optimal matching, and vice-versa. Specifically, we compute the \emph{coefficient of variation} of agents' payoffs in a given matching: the standard deviation of payoffs divided by their mean, which is a ``scale free'' measure of the dispersion in payoffs across agents. We then compute the ratio of the coefficient of variation at the food-optimal stable matching over that at the color-optimal stable matching. Markets with a high ratio are those in which the payoff variation in the food-optimal stable matching is high, relative to that in the color-optimal stable matching. When the ratio is above the median ratio in our data, the food-optimal matching obtains $11.7\%$ of the time, and the color-optimal $35.3\%$. When it is below the median, they obtain $43.8\%$ and $6.3\%$, respectively. The difference between these values is significant at any conventional level of confidence. The implication is that when the variance in payoffs at the color-optimal matching is relatively high, we tend to get more food-optimal outcomes, and vice-versa.

To summarize, our results suggest that egalitarian, or fairness, considerations play a role in selecting outcomes, but they are not so strong as to trump stability.\footnote{Our results cannot be directly compared with the experimental bargaining literature that has documented a preference for fairness. The nature of bargaining in our experiments is substantially different from other experiments, such as the dictator game, where the compromise between the two sides is obvious. In fact, as noted, identifying the set of stable matchings in our markets is arguably not trivial.}

\section{Market Dynamics\label{market_dynamics}}

We now turn to the dynamics underlying the final outcomes discussed till now. Several insights emerge from our analysis of the observed dynamics. First, markets progress quickly early on, with many agents finding stable matches in a short amount of time, and slow down as they reach the final matching. Second, agents' behavior exhibits \textit{strategic sophistication}. Specifically, when making proposals, participants target highly preferred partners, but also take receivers' payoffs into consideration: they are more likely to make proposals to blocking partners that gain substantial payoff benefits from the match. Furthermore, participants are successful at avoiding cycles of blocking pairs. While it is challenging to determine exactly how participants manage to avoid cycles, we observe that agents incorporate past interactions into their decisions and reject some offers from blocking partners. Finally, our experiments challenge the dynamic models that have been proposed in the literature. We use simulation exercises to show that \textit{observed dynamics in our data differ from those generated by some of the standard dynamic models in the literature} that, for the most part, are non-strategic and do not incorporate past interactions.

\subsection{Evolution towards Stable and Median Outcomes}\label{section:evolution}

\begin{figure}[h!]
	\centering
	
    \includegraphics[width=0.49\textwidth]{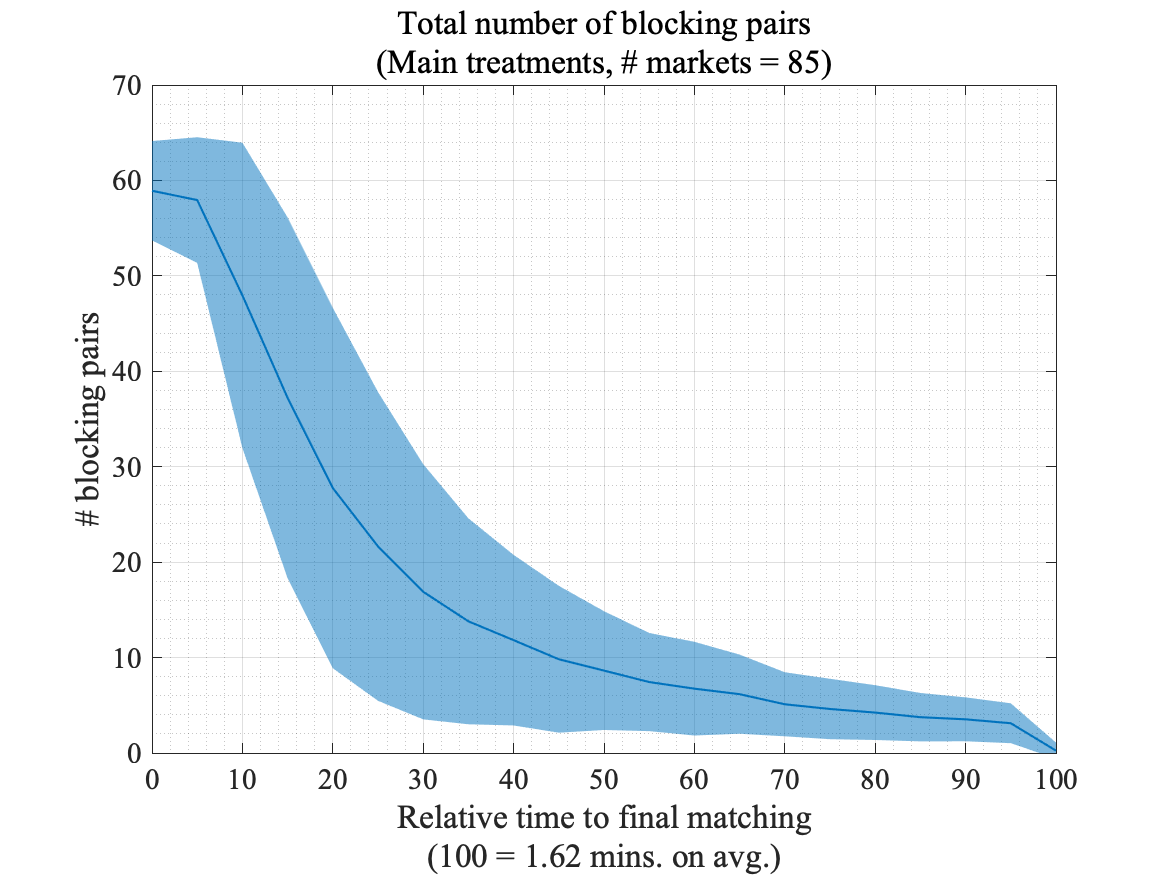}
    \includegraphics[width=0.49\textwidth]{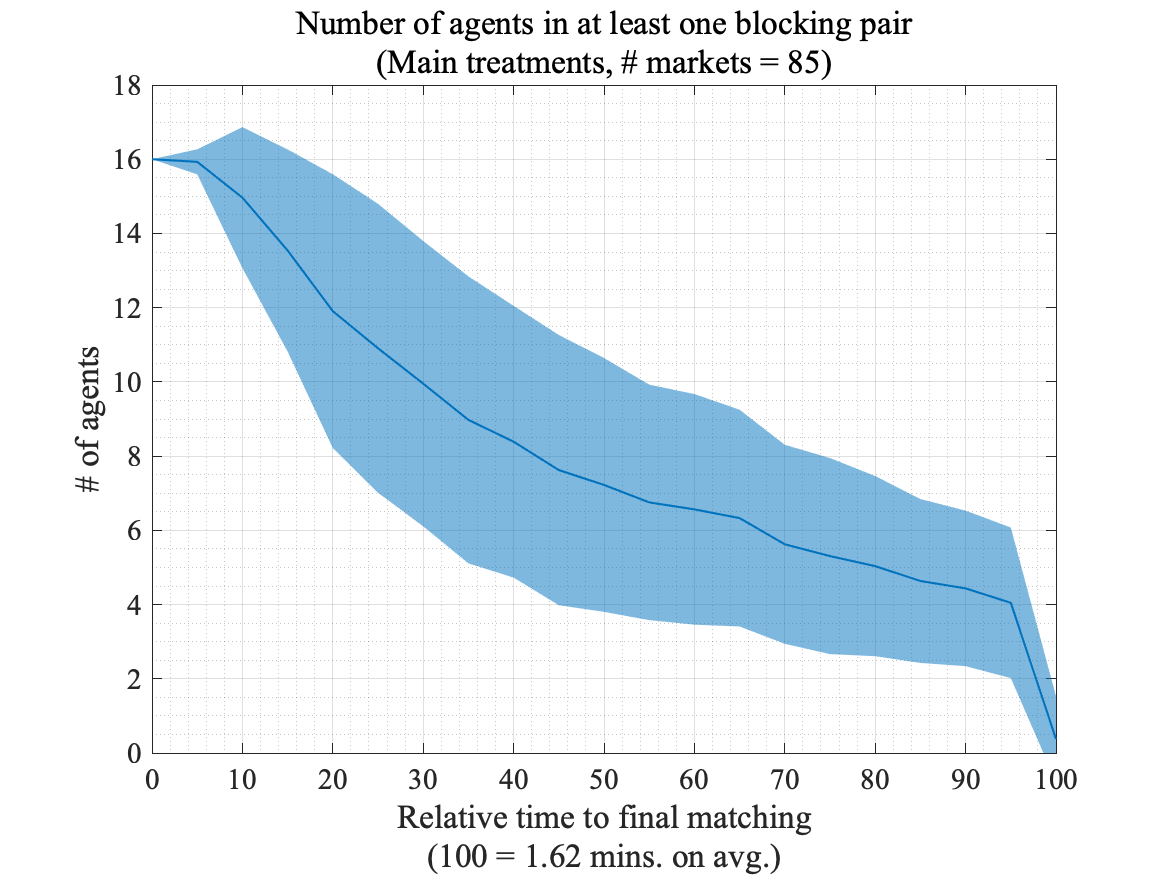}
    
    \vspace{0.75cm}

    \includegraphics[width=0.49\textwidth]{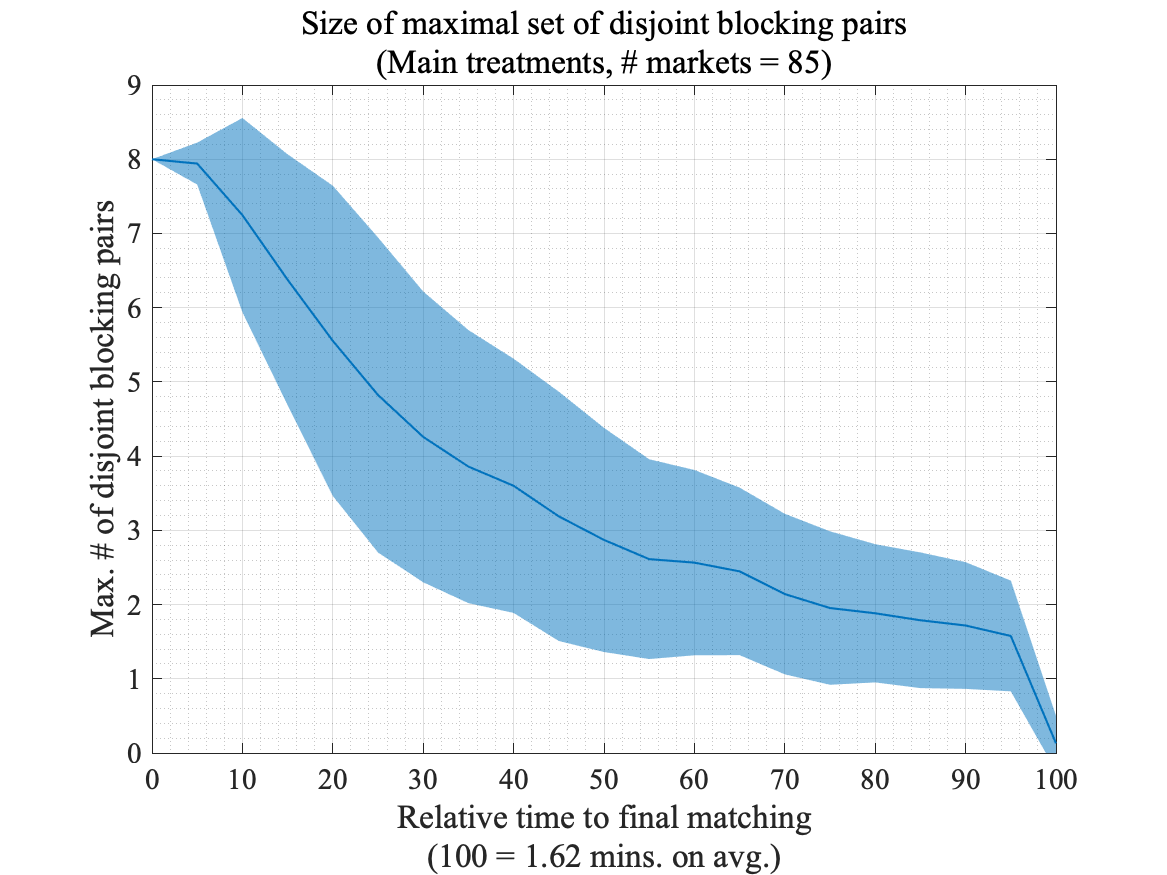}
    \includegraphics[width=0.49\textwidth]{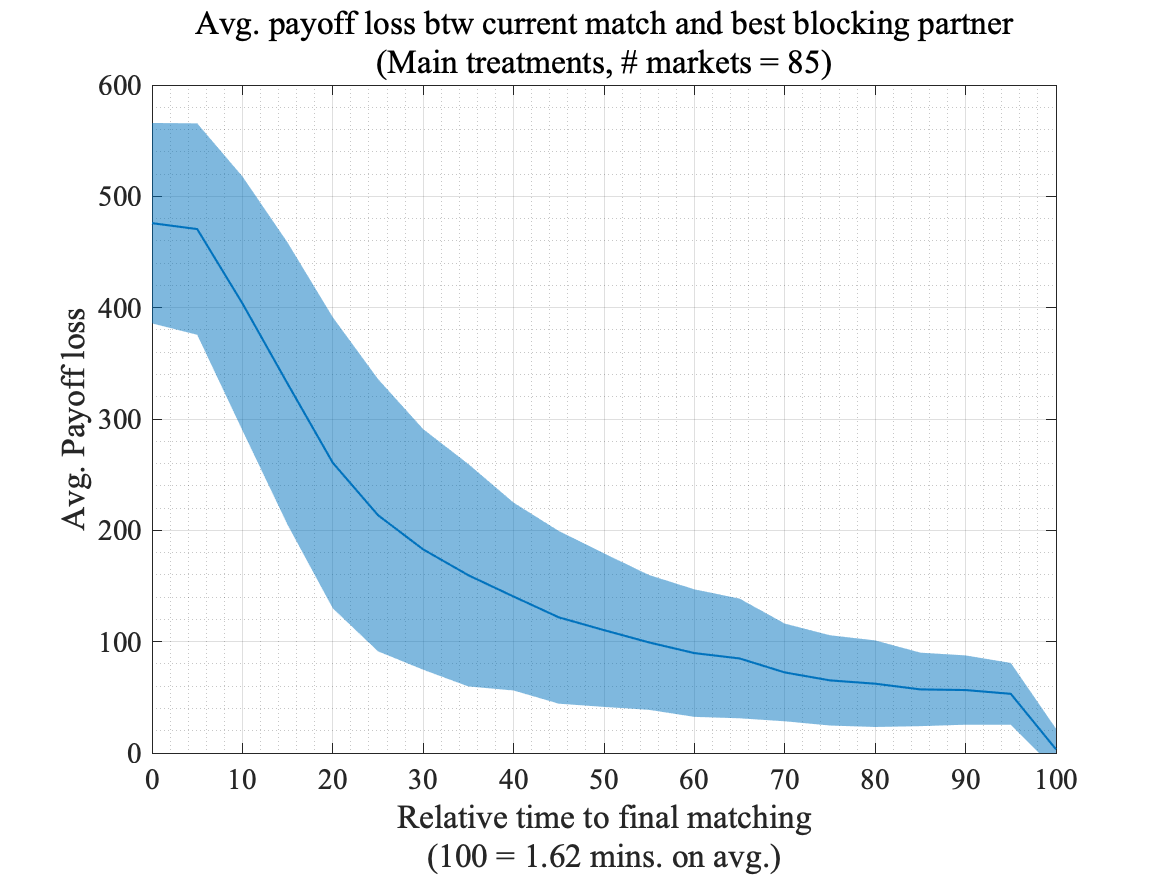}

    \vspace{0.25cm}
    
    \caption{Distance to stable matching over time
    \label{Dist2SMoverTime}}
\end{figure}

Markets reach stability gradually, with the vast majority of blocking opportunities, especially the most profitable ones, vanishing during the initial stages of markets, as shown in \autoref{Dist2SMoverTime}. On the $x$-axis, we plot the normalized time within a market, where 0 indicates the start of the market and 100 the time at which the final matching is reached (on average, $1.62$ minutes across our main treatments). On the $y$-axis, we plot the total number of blocking pairs (top-left panel), the number of agents in at least one blocking pair (top-right panel), the maximum number of disjoint blocking pairs (bottom-left panel), and the average immediate payoff loss with respect to the best blocking partner of each agent (bottom-right panel). The solid lines correspond to averages across all markets in our main treatments, and the shaded regions to a standard deviation above and below the average at every point. On average, after 20\% of time has elapsed, half of all blocking opportunities vanish, and the average gains per agent from forming blocking pairs have are cut in half. Nonetheless, over half the agents still have at least one potential blocking pair. As markets approach the final matching, the four measures dip towards zero. This dip reflects the fact that the vast majority of markets in our main treatments converge to stability: all four measures equal zero when outcomes are stable.\footnote{%
    A similar picture emerges when differentiating markets according to their number of stable matchings; as well as for our auxiliary treatments entailing larger markets and unilateral offers. See the Online Appendix for more details.}

\renewcommand{\wdth}{0.35\textwidth}
\afterpage{
\begin{landscape}
\begin{figure}[h!]
    \centering
    \subfloat[Main: unique]
    {\includegraphics[width=\wdth]{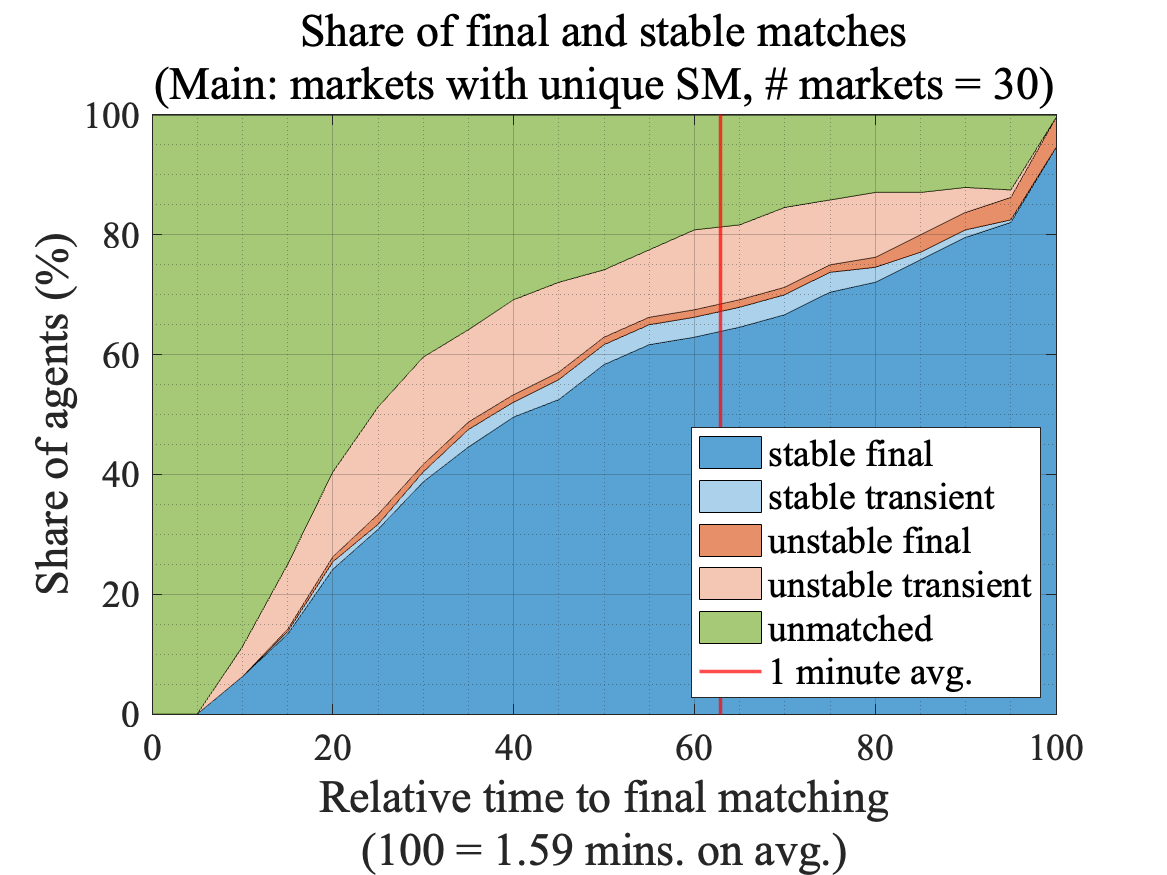}}
    \subfloat[Main: four-by-four]
    {\includegraphics[width=\wdth]{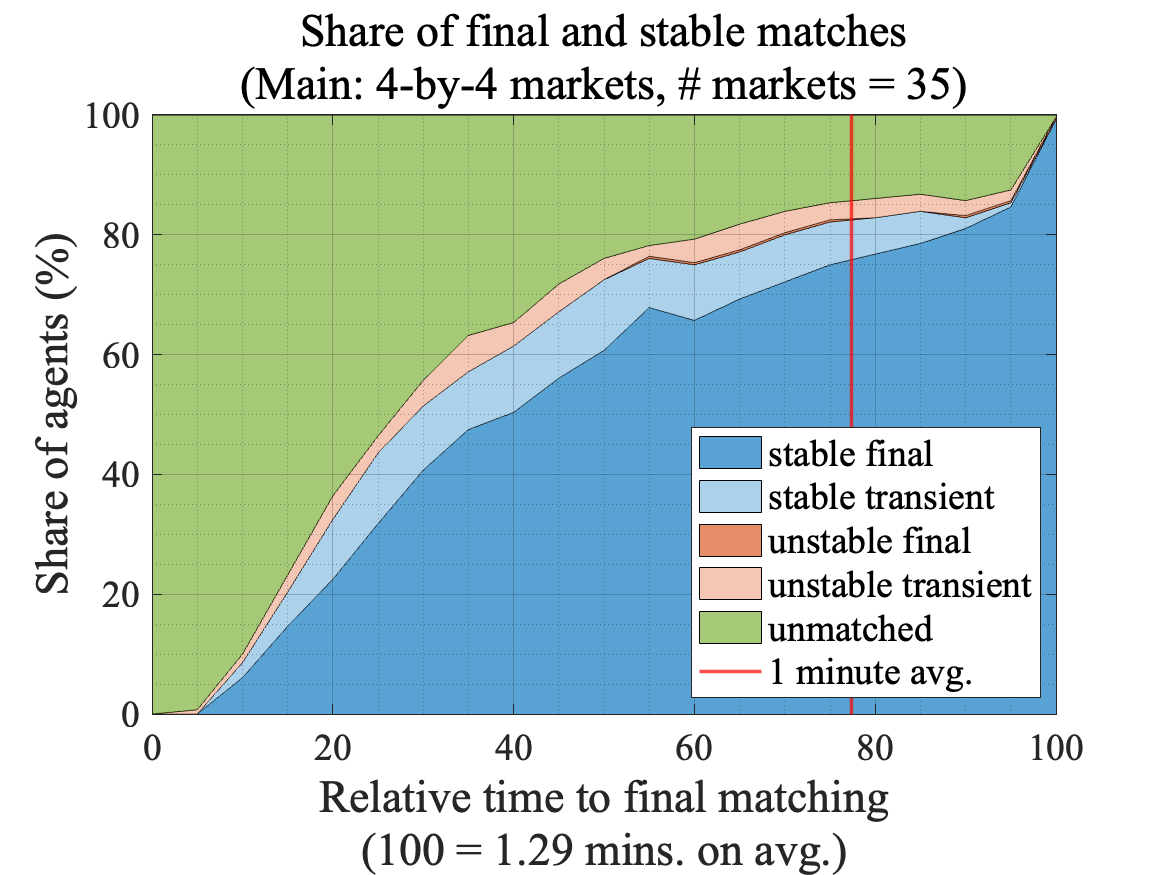}}  
    \subfloat[Main: 5 SMs]{\includegraphics[width=\wdth]{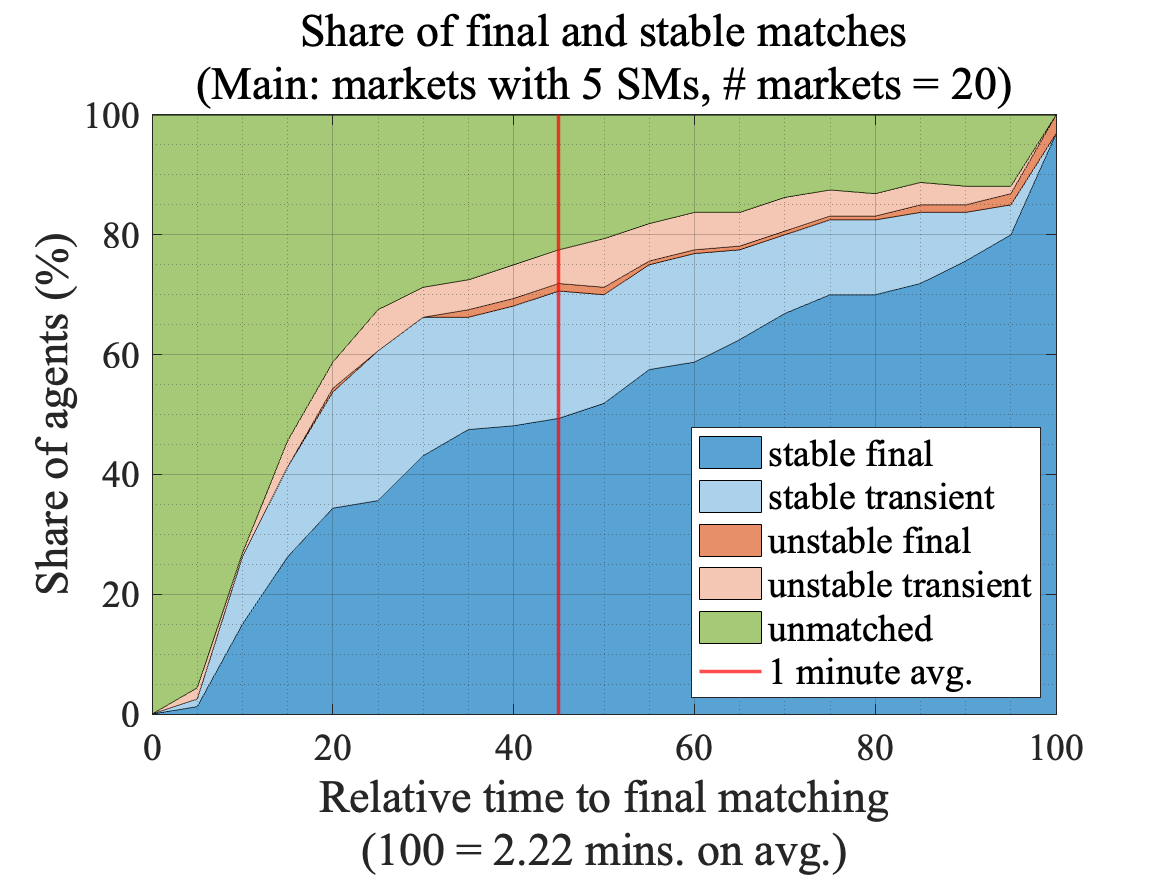}}  
    \subfloat[Main: 5 SMs]{\includegraphics[width=\wdth]{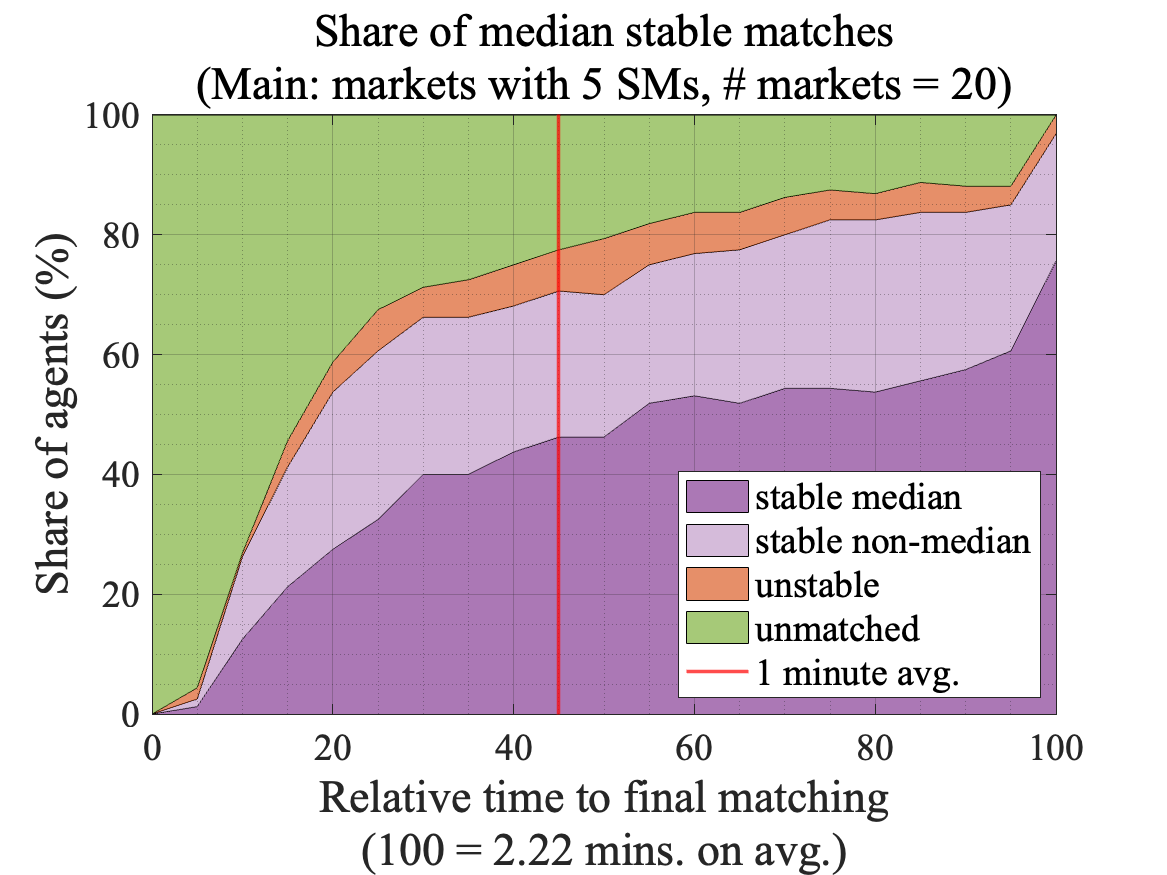}}

    \subfloat[Unilateral: unique]
    {\includegraphics[width=\wdth]{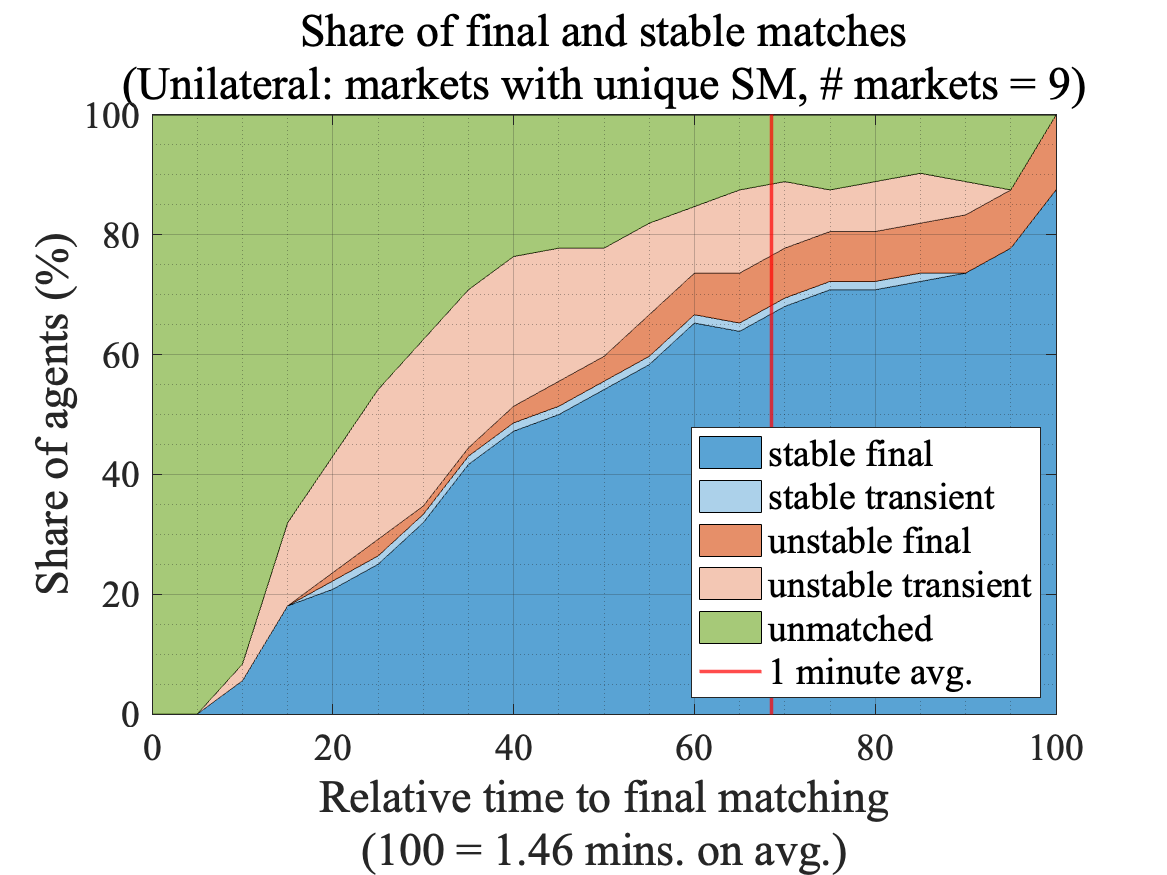}}
    \subfloat[Unilateral: four-by-four]
    {\includegraphics[width=\wdth]{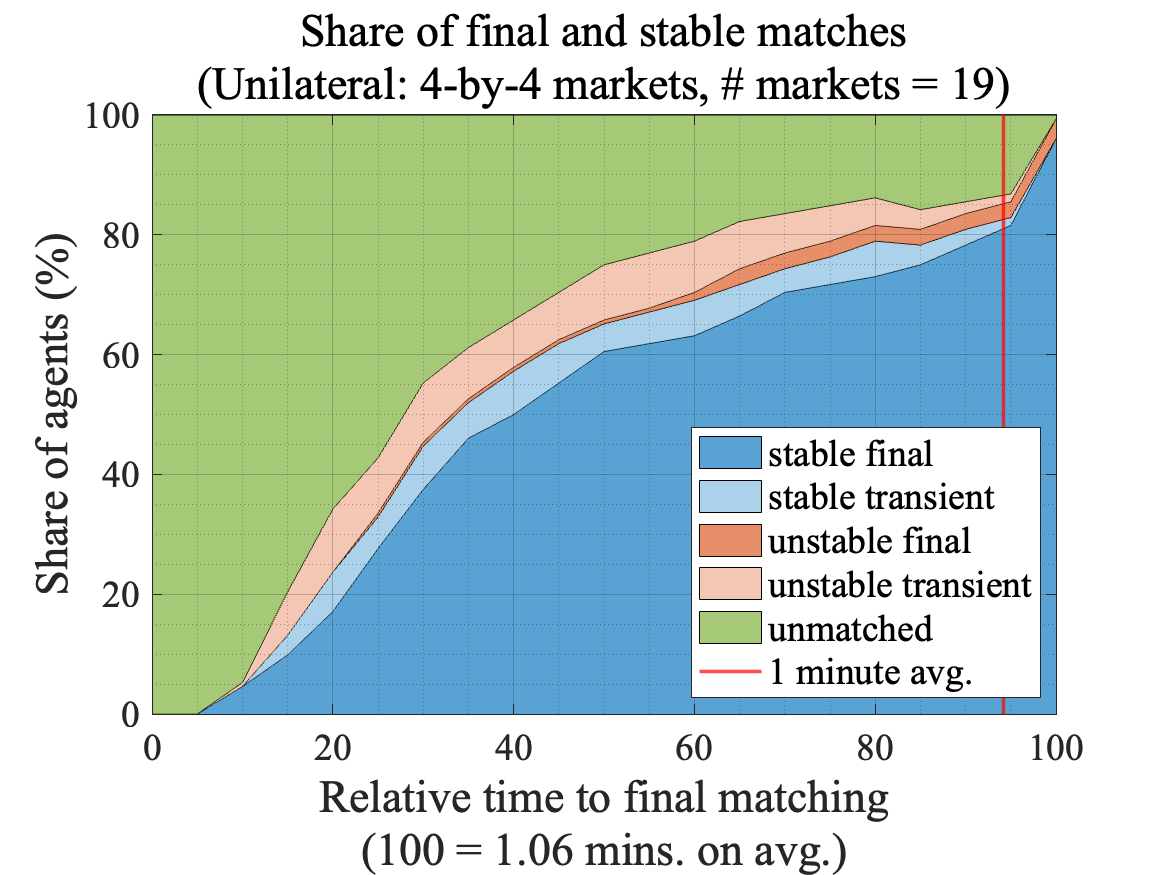}}  
    \subfloat[Unilateral: 5 SMs]{\includegraphics[width=\wdth]{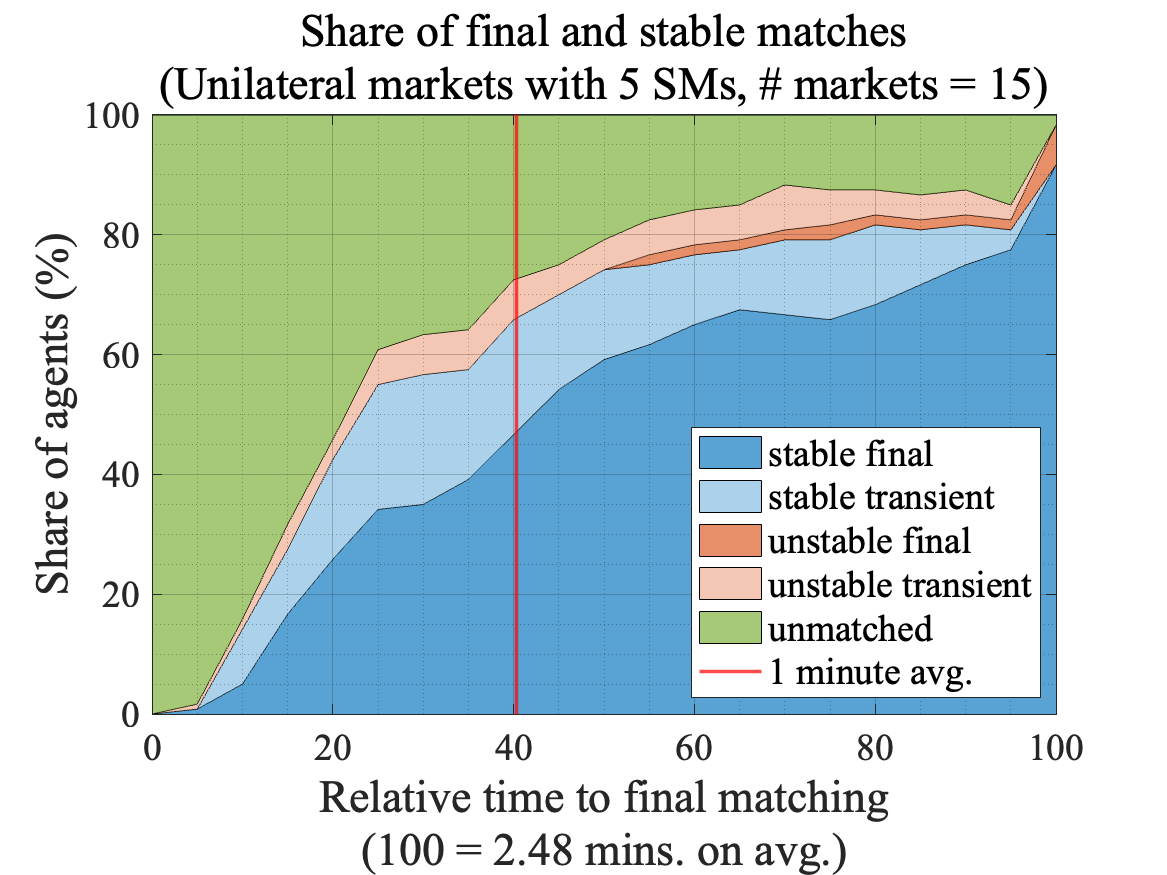}}  
    \subfloat[Unilateral: 5 SMs]{\includegraphics[width=\wdth]{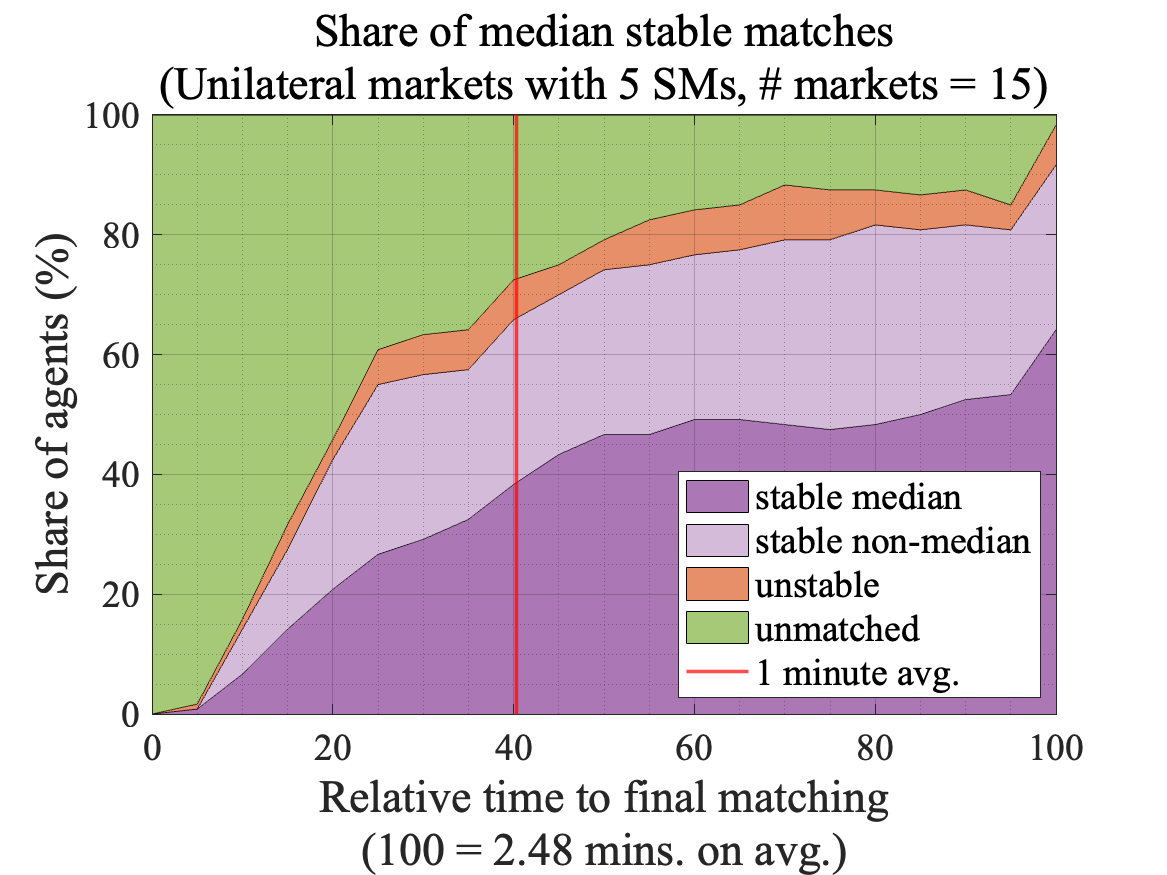}}

    \subfloat[Large: unique]
    {\includegraphics[width=\wdth]{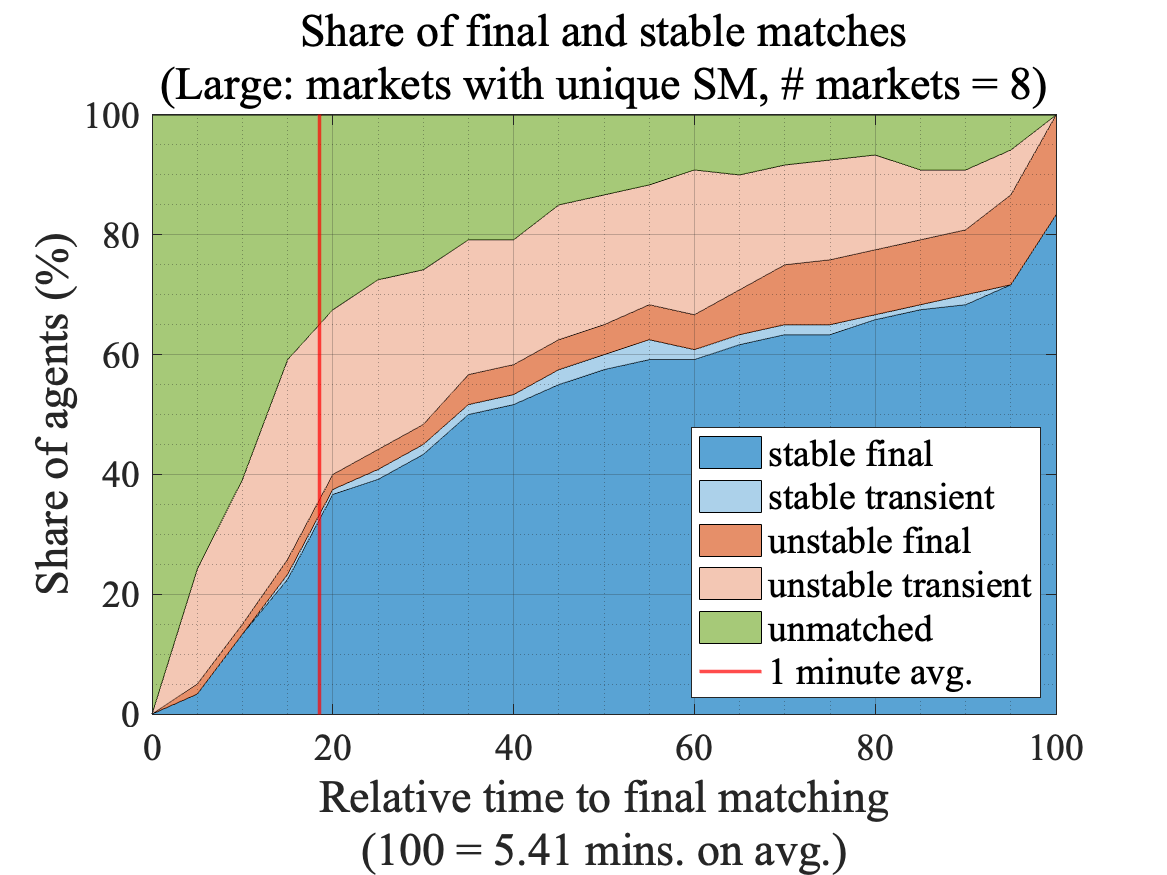}}
    \subfloat[Large: 3 SPs]{\includegraphics[width=\wdth]{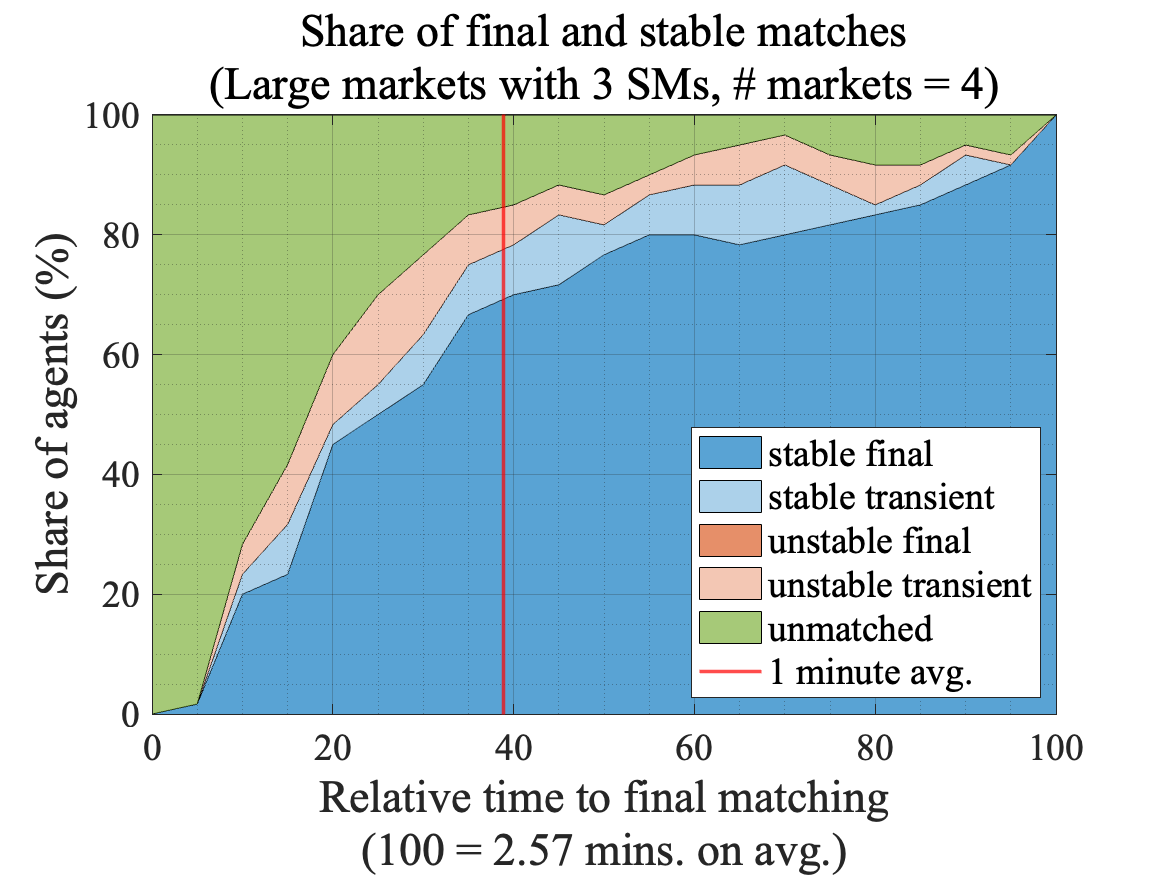}}  
    \subfloat[Large: 3 SPs]{\includegraphics[width=\wdth]{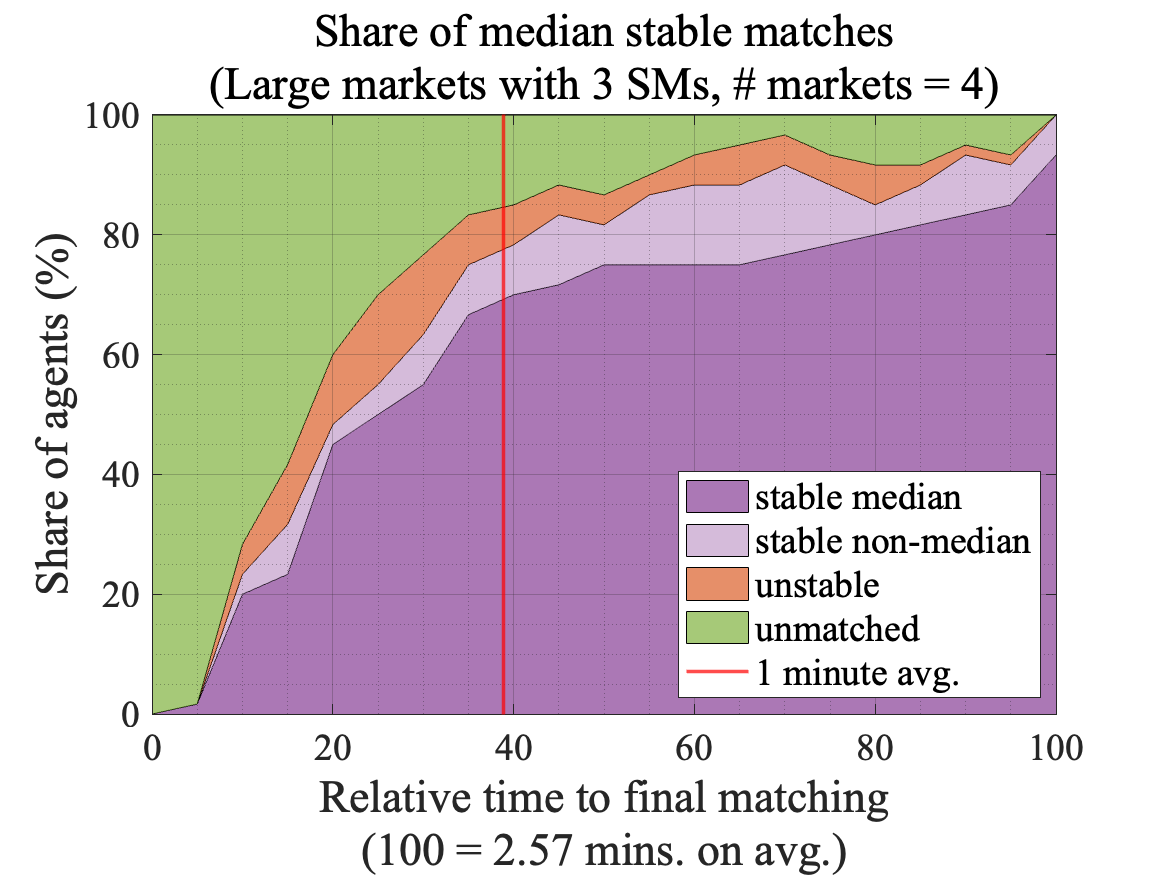}}
    
    \caption{Progression of final, stable, and median stable matches over time
    \label{fig:progression_over_time_all}}
\end{figure}
\end{landscape}
}

The pattern of stable match formation echoes the patterns depicted in \autoref{Dist2SMoverTime}: a significant fraction of agents find their stable match partners early on. \autoref{fig:progression_over_time_all} depicts the evolution of the shares corresponding to agents matched to a stable partner (in blue), matched to an unstable partner (in orange), and remaining unmatched (in green).\footnote{A similar picture emerges when replacing the $x$-axis with the percentage of offers made so far. See the Online Appendix for more details.} Markets in our three main treatments are shown in the top row, panels (a)--(c). The plots further differentiate between \textit{final} and \textit{transient} partners. Two agents are \textit{final partners} (or, equivalently, form a \textit{final match}) if matched to one another at the market's culmination; a pair of matched agents is \textit{transient} if their match is not final.\footnote{Under this definition, a final match \textit{may} break and then form again; all transient matches eventually break.} The shares of agents matched to a final (transient) stable (unstable) partner are shown in dark (light) blue (orange). The figure shows the path by which agents find stable and final partners from the early stages of market activity. For instance, when 30\% of time has elapsed, over 40\% of agents are matched with a stable partner in our main treatments, with the vast majority constituting final partners. This fraction is even higher in markets in which agents have more than one stable partner: over 50\% in four-by-four markets, and over 65\% in markets with five stable matchings. However, we also observe a larger fraction of transient stable pairs in these markets, indicating that agents match with more than one stable partner throughout the course of the market.\footnote{\autoref{Dist2SMoverTime} and \autoref{fig:progression_over_time_all} are reminiscent of the dynamics observed by \cite{roth1997} in the entry-level market for clinical psychologists. \cite{roth1997} document more intense activity at early stages of the market, with a majority of agents finding their final match early on, and a few agents remaining unmatched toward the final stages of the market.} The second row of \autoref{fig:progression_over_time_all} depicts analogous dynamics in our unilateral markets, see panels (e)--(g). Patterns appear remarkably similar to those observed in our main treatments, with agents in unilateral markets also finding stable and final partners at early market stages, and markets converging at similar speeds. The figure also illustrates the similarity in convergence speeds across our main treatments and our unilateral treatment. In particular, whether or not both sides can make offers, convergence speeds are ranked similarly by the number of stable matchings. Last, the bottom row of \autoref{fig:progression_over_time_all} shows the same shares for our large markets, see panels (i) and (j). While large markets take longer to converge, especially those with a unique stable matching, many agents still find stable and final partners early on.\footnote{We also analyzed the first proposal made by agents in our experimental markets. The starkest differences are observed in large markets, where agents are less likely to target their first proposal at their highest-ranked partner or at a stable partner. See the Online Appendix for more details.}

The formation of median stable matchings is also rapid initially, and then slows down, as seen in the rightmost panels of \autoref{fig:progression_over_time_all}, which plot the shares of agents who are matched to their median stable partner (in dark purple), to a non-median stable partner (in light purple), and to an unstable partner (in orange). For example, when 30\% of the time has elapsed in our main treatment with five stable matchings, 40\% of agents are matched with their median stable partner, and this figure rises to 77\% by the time the final matching has been reached.

\subsection{Market Activity across Treatments}

The time and volume of activity it takes for markets to reach a final matching is small, as shown in \autoref{table:dynamics:volume:all}. In our main treatments, with $8$ agents on each market side, the number of distinct (directional) offers is $8 \times 8 = 64$. Regardless of the number of stable matchings, the average number of offers observed till an experimental market terminates is lower than $64$. The average number of matches is also limited, and the number of partners each participant has ranges between $2$ and $3$, depending on market features. Market ``complexity'' affects the volume of interactions, where markets with more stable matchings exhibit more offers and more repeat matchings. Notably, the activity required under DA offers a poor predictor of behavior in our data. In terms of offer volume alone, under DA, there are no repeat matches by definition, and the number of offers the algorithm entails is lower than the number observed in our data, although of the same order of magnitude.

\begin{table}[t!]
\centering\scriptsize
\caption{Volume of market activity across main and auxiliary treatments
\label{table:dynamics:volume:all}
\centering}
\begin{tabular}{lccc|cc|ccc}
&  \multicolumn{8}{c}{}  \\\hline\hline
 &  &  &  &  &  &  &  &  \\
& 
\multicolumn{3}{c|}{\scriptsize \textit{Unique stable matching}}  & 
\multicolumn{2}{c|}{\scriptsize \textit{Four-by-four}}  &  
\multicolumn{3}{c}{\scriptsize \textit{3 Stable Partners}}  \\
 &  &  &  &  &  &  &  &  \\
 & 
{\scriptsize\textit{Main}} & 
{\scriptsize\textit{Unilateral}} & 
{\scriptsize\textit{Large}}  & 
{\scriptsize\textit{Main}} & 
{\scriptsize\textit{Unilateral}} & 
{\scriptsize\textit{Main}} & 
{\scriptsize\textit{Unilateral}} & 
{\scriptsize\textit{Large}}  \\\hline
 &  &  &  &  &  &  &  &  \\
\# \textit{Mkts.} & 30 & 9 & 8 & 35 & 19 & 20 & 15 & 4 \\
 &  &  &  &  &  &  &  &  \\
\# \textit{offers} & 44.8 & 18.1 & 198.9 & 39.7 & 16.8 & 59.2 & 29.9 & 92.8 \\
\# \textit{offers per agent (avg.)} & 2.8 & 1.1 & 6.6 & 2.5 & 1.0 & 3.7 & 1.9 & 3.1 \\
\textit{offers per minute (avg.)} & 22.3 & 12.6 & 33.3 & 22.5 & 17.3 & 21.3 & 12.0 & 32.7 \\
\textit{time to last offer (mins.)} & 2.2 & 1.5 & 5.8 & 2.0 & 1.0 & 2.9 & 2.6 & 3.0 \\
 &  &  &  &  &  &  &  &  \\
\% \textit{accepted offers} & 41.3 & 71.4 & 38.6 & 46.5 & 76.0 & 40.4 & 56.2 & 39.5 \\
 &  &  &  &  &  &  &  &  \\
\# \textit{matches} & 15.9 & 12.8 & 69.2 & 15.9 & 12.5 & 24.6 & 15.9 & 37.0 \\
\# \textit{matches per agent (avg.)} & 2.0 & 1.6 & 4.6 & 2.0 & 1.6 & 3.1 & 2.0 & 2.5 \\
\textit{matches per minute (avg.)} & 11.6 & 9.1 & 14.3 & 14.9 & 12.5 & 11.5 & 7.2 & 15.5 \\
\textit{time to final matching (mins.)} & 1.6 & 1.5 & 5.4 & 1.3 & 1.1 & 2.2 & 2.5 & 2.6 \\
 &  &  &  &  &  &  &  &  \\
\% \textit{matches formed are repeated} & 14.0 & 5.6 & 27.6 & 19.3 & 7.7 & 27.2 & 11.5 & 23.3 \\
\% \textit{matches formed break} & 21.4 & 6.3 & 34.7 & 10.4 & 12.0 & 17.4 & 11.2 & 9.3 \\
 &  &  &  &  &  &  &  &  \\
\# \textit{offers in food-proposing DA} & 21.2 & 16.0 & 87.8 & 12.0 & 12.0 & 13.0 & 13.0 & 36.0 \\
\# \textit{offers in color-proposing DA} & 18.3 & 15.0 & 86.1 & 10.0 & 10.0 & 14.0 & 14.0 & 36.0 \\
 &  &  &  &  &  &  &  &  \\
\# \textit{matches in food-proposing DA} & 11.0 & 13.0 & 18.4 & 10.0 & 10.0 & 11.0 & 11.0 & 24.0 \\
\# \textit{matches in color-proposing DA} & 10.3 & 11.0 & 31.0 & 8.0 & 8.0 & 10.0 & 10.0 & 23.0 \\
 &  &  &  &  &  &  &  &  \\\hline\hline
\multicolumn{9}{p{15.5cm}}{\scriptsize\textit{Notes.} The table reports averages across all our experimental markets in each of our main and auxiliary treatments of: 
number of offers made by both sides, 
average number of offers made per agent, 
average number of offers made per minute,
time (minutes) of the last offer made, 
percentage of offers that are accepted,
number of matches (same as accepted offers), 
average number of matches (not necessarily distinct) per agent, 
average number of matches formed per minute, 
time (minutes) to reach the final matching (last accepted offer), 
\% of matches that are repeated (proposer and receiver had already been matched previously), 
\% of matches that break (a match breaks if any of the two agents is matched to someone else at a later point in time),
and number of offers made and matches formed in the food/color-proposing versions of the DA algorithm.}
\end{tabular}
\end{table}

We also see some differences across treatments. Within all our treatments, convergence time and offer volume are arguably low. Nonetheless, we observe a lower offer volume---roughly a half, on average---in our unilateral treatments relative to our main treatments. This finding reflects, in part, the higher acceptance rates observed in our unilateral markets: on average, more than half of the offers in our unilateral markets are accepted. In addition,  the activity volume we observe in large markets does not align with the number of stable matchings. On average, large markets with a unique stable matching exhibit close to 200 offers and 70 matches, compared to 93 offers and 37 matches in markets with three stable partners. 

\subsection{Offers and Acceptances through Time}

Proposal and acceptances exhibit different features as markets progress, illustrated in \autoref{tab:desc_dynamics_time}. The table reports average summary statistics regarding offers made and accepted at distinct stages in our main treatments. We split every market into three segments of equal length relative to the time at which the last offer is made: the early stage (first third), middle stage (second third), and later stage (last third).\footnote{Our auxiliary treatments exhibit similar patterns, see the Online Appendix for details.} 

The top panel of \autoref{tab:desc_dynamics_time} echoes our observations up to now. Offers, as well as their acceptance rates, occur at greater intensities early on in our markets and consistently decline as markets converge. The second panel shows that, while the majority of offers are made to blocking partners, this is more common during early stages, with $80\%$ of offers directed at blocking partners in the first tercile, and only $36\%$ directed at blocking partners in the last tercile. In part, this pattern is mechanical in nature: as markets converge toward stability, fewer blocking pairs are available.\footnote{We also considered how the ratio between the number of matches formed among blocking pairs and the size of the maximal disjoint set of blocking pairs evolves as the market progresses. This analysis parallels the message of \autoref{Dist2SMoverTime}. Initially, there are many outstanding blocking pairs, generating a large volume of offers. Towards the market's termination, activity tapers down at the same time as the number of blocking pairs diminishes.} Nevertheless, the rate at which blocking partners accept offers increases over time, from $61\%$ in the first tercile to $71\%$ in the last tercile. Importantly, offers from blocking partners are rejected at substantial rates throughout, a point we soon return to. In markets with $8$ agents on each side, there are $64$ distinct pairs. Given the volume of offers made and accepted, repeat matches are inherent. Indeed, the table illustrates that repeat offers and matches become more frequent as markets progress. 

\newcommand{\cwidth}{\hspace{0.1cm}\,}
\newcommand{\spone}{\hspace{0.25cm}}
\newcommand{\sptwo}{\hspace{0.25cm}}
\begin{table}[p!]
\centering\footnotesize
\caption{Proposing and accepting behavior across time
\label{tab:desc_dynamics_time}}
\begin{tabular}{lcccc}
 &  &  &  &  \\\hline\hline
 &  &  &  &  \\
 & 
\cwidth \textit{1st/3} \cwidth& 
\cwidth \textit{2nd/3} \cwidth& 
\cwidth \textit{3rd/3} \cwidth& 
\textit{All} \\\hline
&  &  &  &  \\ 
\spone\% \textit{offers} & 54.4 & 28.0 & 17.6 & 100.0 \\
\spone\% \textit{accepted} & (49.9) & (39.3) & (25.4) & (43.2) \\
 &  &  &  &  \\ 
\spone\# \textit{offers per minute (avg.)} & 37.7 & 10.1 & 4.0 & 22.1 \\
\spone\# \textit{matches per minute (avg.)} & 17.3 & 4.6 & 2.0 & 13.0 \\
 &  &  &  &  \\ 
\spone\% \textit{matches formed are repeated} & 8.7 & 31.5 & 54.6 & 19.3 \\
\spone\% \textit{matches formed break} & 20.0 & 15.3 & 3.2 & 15.9 \\
\spone\% \textit{break $\mid$ final} & 10.2 & 6.6 & 0.8 & 6.7 \\
\spone\% \textit{break $\mid$ stable} & 13.3 & 9.1 & 2.0 & 10.0 \\
&  &  &  &  \\ 
\multicolumn{5}{l}{\textit{Characteristics of offers}} \\
\multicolumn{5}{l}{\textit{(\% accepted)}} \\
 &  &  &  &  \\ 
\spone\% \textit{to blocking partners} & 80.0 & 50.9 & 35.7 & 64.8 \\
 & (60.7) & (69.7) & (71.3) & (63.4) \\
\spone\% \textit{are only-proposer beneficial\cwidth} & 17.9 & 42.3 & 56.5 & 31.2 \\
 & (7.2) & (5.5) & (2.8) & (5.0) \\
\spone\% \textit{are repeated} & 12.5 & 49.7 & 69.1 & 32.7 \\
 & (35.1) & (33.0) & (31.4) & (28.8) \\
\spone\% \textit{to previous match} & 6.9 & 26.2 & 37.0 & 17.8 \\
 & (65.2) & (47.9) & (50.6) & (52.7) \\
&  &  &  &  \\ 
\multicolumn{5}{l}{\textit{In markets with three stable partners:}} \\
 &  &  &  &  \\ 
\sptwo\% \textit{to best stable partner} & 32.3 & 34.7 & 39.8 & 33.5 \\
 & (36.6) & (21.3) & (10.2) & (27.2) \\
\sptwo\% \textit{to median stable partner} & 25.8 & 28.0 & 20.3 & 26.5 \\
 & (76.9) & (69.5) & (67.7) & (74.6) \\
\sptwo\% \textit{to worst stable partner} & 6.9 & 7.6 & 7.2 & 7.4 \\
 & (87.0) & (94.0) & (60.0) & (82.4) \\
&  &  &  &  \\\hline\hline
\multicolumn{5}{p{12cm}}{\scriptsize\textit{Notes.} The table reports averages across markets in main treatments (``thirds'' are relative to the time of the last offer) of: 
avg.\ number of offers made and accepted per minute,
\% of repeated matches,
\% of matches that break,
and \% of offers made and accepted (\% accepted shown in parentheses): 
to blocking partners, that are only beneficial to the proposer, repeated, and to a previous match.
For markets that have five stable matchings and three stable partners, the table also reports the average \% of offers made and accepted (\% accepted shown in parentheses) to the proposer's best, median, and worst stable partner.
}
\end{tabular}
\end{table}

The bottom panel of \autoref{tab:desc_dynamics_time} speaks to dynamics and the selection of stable matchings. In markets where agents have three stable partners, the share of offers agents make to their most-preferred stable partner is higher than the share of offers they make to their median stable partner, which in turn is higher than the share they make to their least-preferred stable partner. Acceptance rates follow the opposite pattern: offers made to best stable partners, who are the worst stable partners for receivers, are less likely to be accepted than those made to median stable partners, which in turn are less likely to be accepted than those made to worst stable partners, who are the best stable partners for receivers. From this perspective, matches with median stable partners can be seen as a form of compromise between the two market sides. This form of the ``median as a compromise'' is also reflected in convergence times. On average, markets converging to the median stable matching converge faster than those converging to a non-median stable matching, which in turn converge faster than markets that fail to reach a stable matching.\footnote{The corresponding average convergence times are 2.07, 2.21, and 2.60 minutes, respectively. While illustrative, these differences are not statistically significant. We also do not see any clear relationship between the volume of offers foods make in any particular market and the likelihood the market culminates in the food-preferred stable matching. On average, foods make 50.9\% of offers in our main treatment with 5 stable matchings and three stable partners. See the Online Appendix for more differences between markets accounting for final matching reached.}

\subsection{Comparison with Dynamic Models}\label{comparison_dynamic}

We now evaluate several theoretical models of market dynamics that serve as benchmarks for the dynamics we observe.\footnote{%
    The theoretical literature on stabilization dynamics in matching markets is limited. There is an active literature on dynamic matching markets, see \cite{baccara2022dynamic}, and our literature review. However, relatively little attention has been dedicated to how agents reach stable outcomes, analogous to the literature on learning in game theory, or t\^{a}tonnement dynamics in competitive markets.}
First, we consider the dynamics underlying Gale and Shapley's DA algorithm, in which proposers ``go down their preference lists'' when making offers and receivers accept beneficial offers myopically. We consider two versions of these dynamics: a version of DA in which proposers are chosen uniformly at random from both sides (2RDA), and a version with Compensation Chains (DACC) by \cite{dworczak21}, where agents dropped by partners who had made them offers initially get priority in proposing themselves. DACC is guaranteed to converge to a stable matching, while 2RDA is not. Second, we consider the dynamics implied by the sequential formation of blocking pairs. We also consider two versions of these dynamics: the Random Paths to Stability (RPS) model of \cite{roth1990random}, whereby blocking pairs are formed sequentially and uniformly at random, and the Random Best Response (RBR) dynamics proposed by \cite{ackermann2011uncoordinated}, which allows (myopic) optimization in a process resembling RPS. Both RPS and RBR are guaranteed to converge to a stable matching. In their standard versions, the four dynamic models we consider only depend on ordinal preferences. We also analyze alternative versions of these algorithms in which we allow the probabilities with which agents are chosen as proposers, or the probabilities different blocking pairs form, to depend on the cardinality of the match payoffs. We provide a detailed description of each algorithm, and discuss these alternative versions, in the Online Appendix.

To contrast the proposing behavior in our main treatments with the ``going down the list'' DA-type dynamics, we look more closely at \textit{when} and to \textit{whom} agents make offers. We call an agent \textit{active} if there exists an agent they prefer to their current match whom they have not made an offer to previously. Across our main treatments, most proposers (84\%) are active. In terms of \textit{whom} offers are made to, we classify offers into three overlapping categories: \textit{downward}, \textit{Gale-Shapley}, and \textit{skip someone}. An offer is \textit{downward} if the proposer has not made an offer to an agent whom they prefer less than the receiver. Intuitively, offers are downward if a proposer is going down their preference ranking when making offers.\footnote{The condition for a downward offer relates to offers made beforehand, not to offers that have been received or accepted: an offer made to an agent the proposer prefers to their current match may still be downward.} While the vast majority of the offers we observe are downward (74\%), it is not uncommon for agents to go up their ranking. An offer is said to be \textit{Gale-Shapley} if the receiver is the proposer's most preferred agent among those the proposer has not made an offer to (all offers are Gale-Shapley in the DA algorithm). A minority of offers are Gale-Shapley (43\%), implying that agents either skip potential partners or make repeat offers throughout the market. Accordingly, an offer is said to \textit{skip someone} if there exists an agent the proposer has not made an offer to, whom they prefer to the receiver. Around 32\% of offers skip someone.

In sum, while agents tend to go down their preference lists when making offers, they generally do not follow the order of offers prescribed by the DA algorithm: they frequently make repeat offers and skip agents on their preference list. In this regard, our proposing behavior deviates from the dynamics prescribed by DA.

In relation to blocking pair dynamics, while the majority of offers are made to blocking pairs (65\%), this number is far from 100\%. And, as already noted, a large fraction of offers made to blocking pairs is not accepted (37\%). These two observations go against the dynamics prescribed by both RPS and RBR, in which all offers target blocking partners and are accepted. Moreover, as we soon describe, both RPS and RBR yield different patterns than those we observe in our experimental markets with five stable matchings.

\subsection{Market Simulations and Cycles of Blocking Pairs\label{simulation}}

We now compare our experimental markets with simulations from the dynamic models we have discussed, paying particular attention to the role of cycles in the formation of blocking pairs \citep{knuth76}. By comparing simulated dynamics with the observed behavior in our experiments, we evaluate the extent to which agents avoid cycles of blocking pairs in our experiments. We show that participants appear more intentional in forming blocking pairs than some of the theoretical models allow for, and thus avoid falling prey to cycles predicted by these models. 

\afterpage{
\begin{landscape}
\begin{table}[htbp!]
\centering\scriptsize
\caption{Market dynamics and simulations
\label{table:mktdynandsimulations}\centering}
\begin{tabular}{lccccc|ccccc|ccccc}
 \multicolumn{16}{c}{}  \\\hline\hline
 &  &  &  &  &  &  &  &  &  &  &  &  &  &  &  \\
 & \multicolumn{5}{c|}{\textit{Unique stable matching}}
 & \multicolumn{5}{c|}{\textit{Two embedded 4-by-4 markets}}
 & \multicolumn{5}{c}{\textit{5 stable matchings and 3 stable partners}} \\
 &  &  &  &  &  &  &  &  &  &  &  &  &  &  &  \\
& {\scriptsize\textit{Exp't}} & 
{\scriptsize\textit{2RDA}} & 
{\scriptsize\textit{DACC}} & 
{\scriptsize\textit{RPS}} & 
{\scriptsize\textit{RBR}} & 
{\scriptsize\textit{Exp't}} & 
{\scriptsize\textit{2RDA}} & 
{\scriptsize\textit{DACC}} & 
{\scriptsize\textit{RPS}} & 
{\scriptsize\textit{RBR}} & 
{\scriptsize\textit{Exp't}} & 
{\scriptsize\textit{2RDA}} & 
{\scriptsize\textit{DACC}} & 
{\scriptsize\textit{RPS}} & 
{\scriptsize\textit{RBR}} \\\hline
 &  &  &  &  &  &  &  &  &  &  &  &  &  &  &  \\
\multicolumn{3}{l}{\textit{Market activity}} &  &  &  &  &  &  &  &  &  &  &  &  &  \\
 &  &  &  &  &  &  &  &  &  &  &  &  &  &  &  \\
\# \textit{offers} & 44.8 & 36.2 & 35.3 & 36.4 & 19.8 & 39.7 & 25.4 & 23.8 & 28.1 & 15.3 & 59.2 & 54.2 & 48.4 & 682.0 & 145.0 \\
\# \textit{matches} & 15.9 & 17.2 & 17.3 & 36.4 & 19.8 & 15.9 & 15.9 & 15.1 & 28.1 & 15.3 & 24.6 & 30.7 & 27.9 & 682.0 & 145.0 \\
\% \textit{accepted offers} & 41.3 & 51.6 & 53.5 & 100.0 & 100.0 & 46.5 & 62.3 & 63.3 & 100.0 & 100.0 & 40.4 & 56.3 & 57.3 & 100.0 & 100.0 \\
 &  &  &  &  &  &  &  &  &  &  &  &  &  &  &  \\
\# \textit{offers to blocking pairs} & 27.2 & 17.2 & 17.3 & 36.4 & 19.8 & 22.8 & 15.9 & 15.1 & 28.1 & 15.3 & 37.3 & 30.7 & 27.9 & 682.0 & 145.0 \\
\% \textit{offers to blocking pairs} & 65.5 & 51.6 & 53.5 & 100.0 & 100.0 & 65.0 & 62.4 & 63.4 & 100.0 & 100.0 & 63.5 & 56.3 & 57.3 & 100.0 & 100.0 \\
 &  &  &  &  &  &  &  &  &  &  &  &  &  &  &  \\
\# \textit{accepted $\mid$ to BP} & 15.2 & 17.2 & 17.3 & 36.4 & 19.8 & 15.1 & 15.9 & 15.1 & 28.1 & 15.3 & 22.6 & 30.7 & 27.9 & 682.0 & 145.0 \\
\% \textit{accepted $\mid$ to BP} & 60.5 & 100.0 & 100.0 & 100.0 & 100.0 & 68.0 & 99.8 & 99.8 & 100.0 & 100.0 & 59.5 & 100.0 & 100.0 & 100.0 & 100.0 \\
 &  &  &  &  &  &  &  &  &  &  &  &  &  &  &  \\
\# \textit{matches formed are repeated} & 2.7 & 0.9 & 1.0 & 9.1 & 2.8 & 3.8 & 1.8 & 1.4 & 5.8 & 1.9 & 8.2 & 4.5 & 3.8 & 624.4 & 108.9 \\
\% \textit{matches formed are repeated} & 14.0 & 4.7 & 5.1 & 20.2 & 10.9 & 19.3 & 9.3 & 7.9 & 17.1 & 9.6 & 27.2 & 12.9 & 12.1 & 79.6 & 58.7 \\
 &  &  &  &  &  &  &  &  &  &  &  &  &  &  &  \\
\# \textit{repeated matchings} & 1.2 & 0.0 & 0.0 & 0.4 & 0.1 & 1.5 & 0.0 & 0.0 & 0.3 & 0.1 & 1.8 & 0.0 & 0.0 & 17.1 & 2.6 \\
\% \textit{repeated matchings} & 6.0 & 0.0 & 0.0 & 0.6 & 0.3 & 8.0 & 0.1 & 0.0 & 0.6 & 0.2 & 6.2 & 0.1 & 0.0 & 1.3 & 1.0 \\
 &  &  &  &  &  &  &  &  &  &  &  &  &  &  &  \\
\# \textit{match-level cycles} & 2.5 & 1.4 & 1.6 & 13.5 & 3.8 & 4.2 & 2.6 & 1.9 & 8.3 & 2.6 & 10.4 & 6.9 & 5.7 & 8.24e21 & 2.61e7 \\
\textit{avg.\ match-level cycle length} & 3.3 & 3.3 & 3.3 & 4.4 & 3.4 & 3.2 & 3.3 & 3.2 & 3.9 & 3.2 & 3.5 & 3.9 & 3.5 & 96.3 & 12.2 \\
 &  &  &  &  &  &  &  &  &  &  &  &  &  &  &  \\
\% \textit{offers to stable pairs} & 35.8 & 31.1 & 32.3 & 35.6 & 53.1 & 65.2 & 66.5 & 67.9 & 55.2 & 78.7 & 67.5 & 63.0 & 63.8 & 62.4 & 73.2 \\
\% \textit{accepted $\mid$ to SP} & 73.5 & 89.8 & 90.5 & 100.0 & 100.0 & 63.1 & 73.5 & 73.9 & 100.0 & 100.0 & 51.4 & 66.7 & 67.5 & 100.0 & 100.0 \\
 &  &  &  &  &  &  &  &  &  &  &  &  &  &  &  \\
\% \textit{proposer is active} & 88.1 & 100.0 & 100.0 & 98.9 & 99.4 & 78.7 & 100.0 & 100.0 & 98.3 & 99.3 & 84.7 & 100.0 & 100.0 & 65.9 & 91.8 \\
\% \textit{offer is downward} & 76.7 & 100.0 & 99.9 & 61.0 & 81.7 & 77.8 & 100.0 & 100.0 & 64.7 & 86.1 & 61.1 & 100.0 & 99.8 & 21.5 & 44.6 \\
\% \textit{offer is Gale-Shapley} & 42.8 & 100.0 & 97.7 & 19.6 & 31.4 & 48.6 & 99.8 & 98.4 & 27.4 & 39.7 & 31.8 & 100.0 & 94.8 & 9.0 & 17.9 \\
\% \textit{offer skips someone} & 35.6 & 0.0 & 1.8 & 77.5 & 66.6 & 23.4 & 0.2 & 1.4 & 67.8 & 57.5 & 41.0 & 0.0 & 3.9 & 39.5 & 59.3 \\
 &  &  &  &  &  &  &  &  &  &  &  &  &  &  &  \\
\multicolumn{3}{l}{\textit{Stability of outcomes}} &  &  &  &  &  &  &  &  &  &  &  &  &  \\
 &  &  &  &  &  &  &  &  &  &  &  &  &  &  &  \\
\% \textit{final matching is stable} & 90.0 & 89.0 & 100.0 & 100.0 & 100.0 & 94.3 & 83.3 & 100.0 & 100.0 & 100.0 & 75.0 & 46.9 & 100.0 & 100.0 & 100.0 \\
\% \textit{final pairs are stable} & 94.7 & 98.4 & 100.0 & 100.0 & 100.0 & 99.6 & 96.5 & 100.0 & 100.0 & 100.0 & 96.9 & 93.8 & 100.0 & 100.0 & 100.0 \\
 &  &  &  &  &  &  &  &  &  &  &  &  &  &  &  \\\hline\hline
\multicolumn{16}{p{22.25cm}}{\scriptsize\textit{Notes.} The table reports average activity and outcome measures across markets in our three main treatments. Within each of the three panels, the first column reports the experimental data (Exp't), and the second through fifth columns report simulations using Two-Sided Random Deferred Acceptance (2RDA), \cite{dworczak21}'s Deferred Acceptance with Compensation Chains (DACC), \cite{roth1990random}'s Random Paths to Stability (RPS), and \cite{ackermann2011uncoordinated}'s Random Best Response (RBR). For each algorithm and market, we ran 10,000 simulations. In 2RDA and DACC, proposers are chosen uniformly at random (with the corresponding adjustment in DACC). In RPS, blocking pairs are chosen uniformly at random. In RBR, an agent is chosen uniformly at random among those who have at least one blocking partner.}
\end{tabular}
\end{table}
\end{landscape}
}

\autoref{table:mktdynandsimulations} reports averages of several summary statistics across 10,000 simulations of each market in our main treatments, with each of the four algorithms: 2RDA, DACC, RPS, and RBR. With one key exception, our experimental markets take more offers to converge than the four algorithms, and exhibit lower acceptance rates, yielding a roughly similar number of ultimate matches. The one exception is in markets with five stable matchings, when compared with the RPS and RBR algorithms. As shown in the last two columns of \autoref{table:mktdynandsimulations}, in these markets, both algorithms take a high number of offers to converge to a stable matching---682 and 145, respectively---which  is much higher than what we observe in the experiment (avg~=~$59$, max~=~$109$). As we show below, the reason for this difference is due to both RPS and RBR getting ``stuck'' in cycles of blocking pairs. Participants in our experiments avoid getting stuck. 

As observed by \cite{knuth76}, when markets evolve through the sequential formation of blocking pairs, markets may cycle between matchings. \autoref{table:mktdynandsimulations} indicates that, on average, our experimental markets with five stable matchings return to a previous matching 1.8 times. This figure is somewhat higher than in our other treatments (1.2 with a unique stable matching, and 1.5 in four-by-four markets). When simulating our markets with five stable matchings with the RPS and RBR dynamics, markets return to a previous matching, on average, 17.1 and 2.6 times, respectively. These figures are echoed when looking at the matched-pair level. In our markets with five stable matchings, on average, 27.2\% of matches are repeated. Under RPS and RBR, 79.6\% and 58.7\% of matches are repeated, respectively. As in the experiment, both RPS and RBR are more likely to cycle across matchings in these markets than in our markets with fewer stable matchings. However, the differences across treatments are starker than in the experiment. On average, RPS and RBR return less than once to a previous matching in our markets with a unique stable matching and our four-by-four markets. 

To glean more insight into our experimental participants' engagement in cycles, we now consider repeated matches in more detail. A \textit{match-level cycle} is a maximal sequence of matches, $(f_k,c_k)_{k=1}^K$, in an experimental or simulated market, that starts and ends at the same pair, and in which every two consecutive matches are distinct and share exactly one agent. In particular, $(f_1,c_1)=(f_K,c_K)$, and $f_{k+1}=f_k$ or $c_{k+1}=c_k$, for all $k=1,\dots,K-1$. The \textit{length} of a match-level cycle is the number of matches that it contains. For example, if we observe that food $f$ matches with color $c$, then severs this match to form one with another color, say $c'$, and then severs this second match to match with $c$ again, then this sequence of matches constitutes a match-level cycle of length 3: $(f,c)\to(f,c')\to(f,c)$. Intuitively, match-level cycles describe the paths agents follow when they return to previous partners.\footnote{\label{foot_match_level_cycle}We include a formal definition of a match-level cycle in the Online Appendix, along with examples and a discussion. In particular, the definition allows for there to be more than a single cycle between two identical pairs. For example, in the sequence of matches $(f,c)$, $(f,c')$, $(f',c)$, and $(f,c)$, there are two match-level cycles of length 3: $(f,c)\to(f,c')\to(f,c)$, and $(f,c)\to(f',c)\to(f,c)$. See the Online Appendix for further details.}

\autoref{table:mktdynandsimulations} reports the average number of match-level cycles and their average lengths in our experimental markets, as well as in their simulated counterparts. The average number of match-level cycles in the experiment is below RPS across all our treatments, and below RBR in our markets with a unique stable matching and with five stable matchings. 

Zooming in on our markets with five stable matchings highlights the differences between the dynamics implied by random formation of blocking pairs and the dynamics in our data \textit{at the individual level}. The average number of match-level cycles in our experimental markets with five stable matchings is higher than in our other treatments, but of a similar order of magnitude: 10.4 vs.\ 2.5 in markets with a unique stable matching, and 4.2 in four-by-four markets. By contrast, when simulating our markets with RPS and RBR, the difference between the number of match-level cycles in markets with five stable matchings and markets with fewer stable matchings explodes. Thus, participants in our experiment do not appear to be pursuing blocking pairs randomly, and, consequently, generate matches that lead them back to previous partners less than what the RPS and RBR dynamics imply.\footnote{As mentioned, we simulated different versions of these two algorithms where blocking pairs (in the case of RPS) and proposers (in the case of RBR) are chosen with a probability that is increasing in the profitability of the blocking pair formed. These calibrated algorithms bring the number of offers, repeated matches, and match-level cycles to the same order of magnitude of what we observe in the experiment. While this suggests a channel through which cardinal preferences may play a role in avoiding cycles, using the same calibration in our other treatments with fewer stable partners results in very fast convergence and a mismatch with the data. See the Online Appendix for details.} 

The simulation results in \autoref{table:mktdynandsimulations} also show how the individual dynamics we observe in the experiment are more sophisticated than those implied by the ``going down the list'' DA-type dynamics. While 2RDA and DACC are very effective at avoiding cycles---never returning back to previous matchings, and generating few repeat matches and match-level cycles---in our experiment, the fraction of offers made to blocking partners is higher, and acceptance rates are lower, than in both of these algorithms. This indicates two features of the data. First, participants in our experiment do not appear to avoid cycles by making offers according to DA-type dynamics (``going down their lists''). Second, the behavior of our experimental participants appears more strategically sophisticated than what DA-type dynamics prescribe: participants take into account the incentives of receivers more often than DA, and are not as myopic as DA-type dynamics prescribe when determining whether to accept offers.   

\begin{table}[p]
\centering\footnotesize
\caption{Simulated dynamics in markets with five stable matchings
\label{tab:median-simulations}\centering}
\begin{tabular}{lccccc}
 &  &  &  &  &  \\\hline\hline
 &  &  &  &  &  \\
 & 
{\scriptsize\textit{Experiment}} & 
{\scriptsize\textit{2RDA}} & 
{\scriptsize\textit{DACC}} & 
{\scriptsize\textit{RPS}} & 
{\scriptsize\textit{RBR}} \\\hline
 &  &  &  &  &  \\
\multicolumn{3}{l}{\textit{Target of offers}} &  &  &  \\
 &  &  &  &  &  \\
\% \textit{offers to best stable partner} & 33.5 & 30.3 & 32.1 & 19.5 & 22.7 \\
\% \textit{accepted $\mid$ to best SP} & 27.2 & 55.6 & 57.0 & 100.0 & 100.0 \\
 &  &  &  &  &  \\
\% \textit{offers to median stable partner} & 26.5 & 23.4 & 23.7 & 23.4 & 27.8 \\
\% \textit{accepted $\mid$ to median SP} & 74.6 & 69.1 & 71.6 & 100.0 & 100.0 \\
 &  &  &  &  &  \\
\% \textit{offers to worst stable partner} & 7.4 & 9.4 & 8.0 & 19.5 & 22.7 \\
\% \textit{accepted $\mid$ to worst SP} & 82.4 & 95.6 & 95.2 & 100.0 & 100.0 \\
 &  &  &  &  &  \\
\multicolumn{3}{l}{\textit{Market-level outcomes (matchings)}} &  &  &  \\
 &  &  &  &  &  \\
\% \textit{median $\mid$ stable} & 80.0 & 64.1 & 50.4 & 39.4 & 46.2 \\
\% \textit{non-extremal $\mid$ stable} & 100.0 & 89.1 & 75.1 & 77.5 & 79.1 \\
\% \textit{food-optimal $\mid$ stable} & 0.0 & 5.5 & 19.4 & 10.8 & 8.1 \\
\% \textit{color-optimal $\mid$ stable} & 0.0 & 5.3 & 5.4 & 11.6 & 12.9 \\
 &  &  &  &  &  \\
\multicolumn{3}{l}{\textit{Individual-level outcomes (matches)}} &  &  &  \\
 &  &  &  &  &  \\
\% \textit{median $\mid$ stable} & 76.8 & 52.9 & 62.8 & 58.5 & 62.6 \\
\% \textit{fruit-optimal $\mid$ stable} & 2.9 & 22.0 & 19.4 & 10.8 & 8.1 \\
\% \textit{color-optimal $\mid$ stable} & 20.4 & 24.7 & 17.8 & 30.7 & 29.3 \\
 &  &  &  &  &  \\\hline\hline
\multicolumn{6}{p{12.5cm}}{\scriptsize\textit{Notes.} The table reports average activity and outcome measures across markets with five stable matchings. The first column reports the experimental data, and the second through fifth columns report simulations using Two-Sided Random Deferred Acceptance (2RDA), \cite{dworczak21}'s Deferred Acceptance with Compensation Chains, \cite{roth1990random}'s Random Paths to Stability (RPS), and \cite{ackermann2011uncoordinated}'s Random Best Response (RBR). For each algorithm and market, we ran 10,000 simulations. In 2RDA and DACC, proposers are chosen uniformly at random (with the corresponding adjustment in DACC). In RPS, blocking pairs are chosen uniformly at random. In RBR, an agent is chosen uniformly at random among those who have at least one blocking partner.}
\end{tabular}
\end{table}

Finally, \autoref{tab:median-simulations} reports features of the dynamics leading to different stable partners across simulations of the four algorithms in our markets with five stable matchings. Both 2RDA and DACC predict similar shares of offers made to proposers' best, median, and worst stable partners. However, the  acceptance rates in our simulations differ from the ones we observe in the experiment. In the algorithms, offers made to proposers' best stable partners (i.e., coming from the receivers' worst stable partners) are more likely to be accepted than in the experiment. Hence, while in both 2RDA and DACC agents match with their median stable partners with relatively high frequency--- $53\%$ and $63\%$ on average, respectively---the frequency at which they do so is lower than what we observe experimentally ($77\%$). In RPS and RBR, the frequency with which proposers make offers to all their stable partners is similar, and acceptance rates are $100\%$ by definition. Hence, while these algorithms converge to median outcomes at times, they do so less frequently than our experimental participants.\footnote{Strictly speaking, in RPS there are no ``proposers,'' since the algorithm indicates only which blocking pair should be formed. When calculating the average, we compute an expectation, treating each relevant blocking agent as proposer with equal probability.}
 
\subsection{Individual Dynamics\label{individual_dynamics}}

We now analyze the main determinants driving market participants to make and accept offers at the individual level. 

\begin{table}
\centering\scriptsize
\caption{Proposal conditional logit estimations
\label{final_table_7}}
\begin{tabular}{lcccccc}
 &  &  &  &  &  &  \\\hline\hline
  &  &  &  &  &  &  \\
 & (1) & (2) & (3) & (4) & (5) & (6) \\\hline
 &  &  &  &  &  &  \\
\textit{Proposer's PA $> 0$}                    &       0.159***&       0.099***&       0.165***&       0.085   &       0.107*  &       0.194***\\
                                   		&     (0.033)   &     (0.031)   &     (0.031)   &     (0.057)   &     (0.061)   &     (0.062)   \\
\textit{Receiver's rank (in prop's list)}       &      -0.077***&      -0.068***&      -0.060***&      -0.086***&      -0.083***&      -0.058***\\
                                   		&     (0.001)   &     (0.002)   &     (0.004)   &     (0.002)   &     (0.004)   &     (0.007)   \\
\textit{Receiver is matched}                    &       0.080***&       0.032   &       0.029   &       0.081***&       0.014   &       0.014   \\
                                   		&     (0.031)   &     (0.019)   &     (0.021)   &     (0.022)   &     (0.024)   &     (0.029)   \\
\textit{Are blocking pair (BP)}                 &       0.148***&       0.159***&       0.145***&       0.192***&       0.168***&       0.169***\\
                                   		&     (0.021)   &     (0.018)   &     (0.022)   &     (0.022)   &     (0.028)   &     (0.030)   \\
\textit{$\max\{\text{Rec's PA},0\}$}            &       0.028***&       0.014***&       0.020***&       0.039***&       0.021***&       0.031***\\
                                   		&     (0.007)   &     (0.005)   &     (0.005)   &     (0.006)   &     (0.006)   &     (0.007)   \\
\textit{$\min\{\text{Rec's PA},0\}$}            &       0.044***&       0.028***&       0.025***&       0.047***&       0.025   &       0.018   \\
                                   		&     (0.010)   &     (0.009)   &     (0.009)   &     (0.014)   &     (0.015)   &     (0.019)   \\
\textit{Matched previously}                     &               &       0.036***&       0.062***&               &       0.013   &       0.053***\\
                                   		&               &     (0.011)   &     (0.012)   &               &     (0.027)   &     (0.017)   \\
\textit{\# previous offers (prop.\ to rec.)}    &               &       0.019***&       0.027***&               &      -0.037   &       0.007   \\
                                   		&               &     (0.004)   &     (0.007)   &               &     (0.028)   &     (0.010)   \\
\textit{Are stable partners (SP)}               &               &       0.088***&       0.093***&               &       0.061***&       0.093***\\
                                   		&               &     (0.007)   &     (0.008)   &               &     (0.012)   &     (0.016)   \\
\textit{Offer is downward}                      &               &               &       0.091***&               &               &       0.007   \\
                                   		&               &               &     (0.026)   &               &               &     (0.025)   \\
\textit{Offer is Gale-Shapley}                  &               &               &       0.108***&               &               &       0.116***\\
                                   		&               &               &     (0.037)   &               &               &     (0.024)   \\
\textit{Offer skips someone}                    &               &               &      -0.009   &               &               &      -0.046   \\
                                   		&               &               &     (0.045)   &               &               &     (0.030)   \\
  &  &  &  &  &  &  \\\hline
   &  &  &  &  &  &  \\
Treatment                          		&        Main   &        Main   &        Main   &  Unilateral   &  Unilateral   &  Unilateral   \\
 &  &  &  &  &  &  \\
Observations                                  	&      29,760   &      29,760   &      29,760   &       7,246   &       7,246   &       7,246   \\
 &  &  &  &  &  &  \\
Adj. $R^2$                            		&       0.390   &       0.423   &       0.445   &       0.458   &       0.470   &       0.500   \\
 &  &  &  &  &  &  \\
$MSE$ \textit{(sample)}                    	&       7.730   &       7.439   &       7.187   &       6.992   &       6.823   &       6.391   \\
$MSE$ \textit{(2-fold $\times$ valid)}          &       7.751   &       7.472   &       7.218   &       7.028   &       6.906   &       6.461   \\
$MSE$ \textit{(future $|$ present)}            	&       7.315   &       7.155   &       6.755   &       5.958   &       5.813   &       5.016   \\
$MSE$ \textit{(present $|$ future)}             &       8.241   &       8.003   &       7.848   &       8.571   &       8.397   &       7.980   \\
 &  &  &  &  &  &  \\
$\%CorrMaxCP$ \textit{(sample)}             		&      56.559   &      56.279   &      59.010   &      61.075   &      63.118   &      63.548   \\
$\%CorrMaxCP$ \textit{(2-fold $\times$ valid)}        	&      56.228   &      56.023   &      59.035   &      61.075   &      59.140   &      62.903   \\
$\%CorrMaxCP$ \textit{(future $|$ present)}      	&      60.114   &      58.457   &      62.686   &      73.816   &      67.409   &      76.323   \\
$\%CorrMaxCP$ \textit{(present $|$ future)}      	&      52.629   &      53.413   &      54.889   &      49.737   &      50.438   &      53.415   \\
 &  &  &  &  &  &  \\
$Avg\,\mathbb{P}(OK\,Pred)$ \textit{(sample)}         		&      40.570   &      43.409   &      45.696   &      44.482   &      45.611   &      49.729   \\
$Avg\,\mathbb{P}(OK\,Pred)$ \textit{(2-fold $\times$ valid)}    &      40.496   &      43.306   &      45.612   &      44.377   &      45.343   &      49.453   \\
$Avg\,\mathbb{P}(OK\,Pred)$ \textit{(future $|$ present)}  	&      40.646   &      43.607   &      46.492   &      47.509   &      49.172   &      55.456   \\
$Avg\,\mathbb{P}(OK\,Pred)$ \textit{(present $|$ future)}  	&      41.194   &      42.575   &      44.461   &      43.180   &      43.813   &      45.624   \\
 &  &  &  &  &  &  \\\hline\hline
\multicolumn{7}{p{14.5cm}}{\scriptsize\textit{Notes.} The table reports average marginal effects of conditional logits. The response variable indicates the receiver of every offer in the data. Standard errors are clustered at the participant level. *, **, and *** indicate significance at the 90\%, 95\%, and 99\% confidence levels, respectively. The table also reports the mean-squared error (\textit{MSE}) of the predicted choice probability, percentage of choices in which the predicted probability of the alternative chosen in the data is the greatest among all alternatives ($\%CorrMaxCP$), and the average probability of correctly predicting the data ($Avg\,\mathbb{P}(OK\,Pred)$). Each is computed in the estimation sample and out of the sample using random two-fold cross-validation, predicting the final five rounds with the first five rounds, and the first five rounds using the final five. See the Online Appendix for more details.}
\end{tabular}
\end{table}

\paragraph{Determinants of Offer Targets.} At an individual level, offers are driven by yield---the chance that the offer will be accepted---and payoff, as well as by past offers and matches, as \autoref{final_table_7} shows. The table reports the results of multiple conditional logits explaining offer targets (who are offers made to). We use the following regressors. For any offer, the proposer's or receiver's \textit{Payoff Advantage (PA)} is the change in payoff to the proposer, or receiver, if the offer is accepted. We include the proposer PA as a dummy, indicating whether it is positive (i.e., the proposer finds the match profitable given the payoff from their current match). We also differentiate receivers' gains from losses by considering receivers' positive and negative PA separately. In addition, we include receivers' rank in proposers' rank-order list (higher rank means less preferred), whether a receiver is currently matched, whether the pair formed by the proposer and the receiver is a blocking pair, whether the pair has been matched previously, the number of offers the proposer has made to the receiver previously, whether the pair formed by the proposer and the receiver is a stable pair, and whether the offer is downward, Gale-Shapley, or skips someone (as defined above). We estimate several specifications, and test their fit in-sample and out-of-sample using the mean-squared error, the percentage of choices in which the alternative chosen in the data has the maximum predicted choice probability, and the average probability of correctly predicting the data (see the Online Appendix for details on how we compute these measures). The first three columns report average marginal effects across all our main treatments, while the final three columns focus on our unilateral markets.

The estimates in \autoref{final_table_7} confirm some of the observations already made. First, proposers not only take their own preferences into account, but also those of the receivers. The probability of making an offer to a blocking partner is around 0.15 higher than to someone who is not. Likewise, proposers are more likely to target receivers that find it more profitable to match with them (a sort of ``second order reasoning'' in game theoretic terms). Second, proposers are more likely to make offers to receivers whom they have already proposed to in the past and, especially, to receivers whom they have already been matched with. Third, proposers are more likely to make downward or Gale-Shapley offers. Last, and perhaps surprisingly, proposers are more likely to make offers to receivers who are stable partners, even after including all the regressors discussed above. As already noted, it is not trivial to identify stable partners in our experimental markets through visual inspection. The draw of stable partners indicates that stability may be intrinsically attractive.\footnote{We also conducted similar analysis including cardinal payoff information for both proposers and receivers. Results are qualitatively identical. See the Online Appendix for further details.} 

\paragraph {Determinants of Offer Responses.} Receivers are more likely to accept offers that are more profitable in monetary terms, disregarding proposers' payoffs, and are more selective the more offers they receive over time, as \autoref{final_table_8} illustrates. The table reports the results of binary logits explaining which offers are accepted with similar regressors to those in \autoref{final_table_7}. 

\begin{table}
\centering\scriptsize
\caption{Acceptance binary logit estimations
\label{final_table_8}}
\begin{tabular}{lcccccc}
 &  &  &  &  &  &  \\\hline\hline
  &  &  &  &  &  &  \\
 & (1) & (2) & (3) & (4) & (5) & (6) \\\hline
 &  &  &  &  &  &  \\
\textit{Receiver's PA $> 0$}                    &       0.562***&       0.562***&       0.495***&       0.356***&       0.355***&       0.417***\\
                                   		&     (0.044)   &     (0.043)   &     (0.034)   &     (0.113)   &     (0.113)   &     (0.121)   \\
\textit{Receiver is matched}                    &       0.224***&       0.224***&       0.200***&       0.082   &       0.082   &       0.048   \\
                                   		&     (0.037)   &     (0.037)   &     (0.032)   &     (0.062)   &     (0.062)   &     (0.066)   \\
\textit{$\max\{\text{Rec's PA},0\}$}            &       0.034***&       0.034***&       0.031***&       0.029   &       0.029   &       0.013   \\
                                   		&     (0.010)   &     (0.010)   &     (0.008)   &     (0.019)   &     (0.019)   &     (0.019)   \\
\textit{$\min\{\text{Rec's PA},0\}$}            &       0.034   &       0.034   &      -0.003   &       0.354***&       0.355***&       0.415** \\
                                   		&     (0.031)   &     (0.031)   &     (0.019)   &     (0.130)   &     (0.130)   &     (0.174)   \\
\textit{Proposer's rank (in rec's list)}        &      -0.038***&      -0.038***&      -0.015***&      -0.067***&      -0.067***&      -0.052***\\
                                   		&     (0.005)   &     (0.005)   &     (0.004)   &     (0.008)   &     (0.008)   &     (0.009)   \\
\textit{Proposer's PA $> 0$}                    &               &      -0.008   &      -0.062*  &               &      -0.035   &      -0.018   \\
                                   		&               &     (0.040)   &     (0.037)   &               &     (0.151)   &     (0.131)   \\
\textit{Matched previously}                     &               &               &       0.128***&               &               &       0.276***\\
                                   		&               &               &     (0.020)   &               &               &     (0.085)   \\
\textit{\# previous offers (prop.\ to rec.)}	&               &               &      -0.038***&               &               &      -0.072***\\
                                   		&               &               &     (0.014)   &               &               &     (0.023)   \\
\textit{\# previous offers (total to rec.)}     &               &               &      -0.016***&               &               &       0.004   \\
                                   		&               &               &     (0.005)   &               &               &     (0.008)   \\
\textit{Are stable partners (SP)}               &               &               &       0.124***&               &               &       0.074** \\
                                   		&               &               &     (0.012)   &               &               &     (0.032)   \\
 &  &  &  &  &  &  \\\hline
 &  &  &  &  &  &  \\
Treatment                          		&        Main   &        Main   &        Main   &  Unilateral   &  Unilateral   &  Unilateral   \\
 &  &  &  &  &  &  \\
Observations                                  	&       3,919   &       3,919   &       3,919   &         930   &         930   &         930   \\
 &  &  &  &  &  &  \\
Adj. $R^2$                            		&       0.376   &       0.376   &       0.419   &       0.365   &       0.365   &       0.396   \\
 &  &  &  &  &  &  \\
$MSE$ \textit{(sample)}                    	&      13.520   &      13.518   &      12.515   &      13.345   &      13.340   &      12.609   \\
$MSE$ \textit{(2-fold $\times$ valid)}          &      13.593   &      13.600   &      12.675   &      13.409   &      13.428   &      12.876   \\
$MSE$ \textit{(future $|$ present)}             &      14.174   &      14.172   &      13.286   &      10.978   &      10.988   &      10.652   \\
$MSE$ \textit{(present $|$ future)}             &      13.176   &      13.196   &      12.234   &      20.951   &      20.934   &      19.948   \\
 &  &  &  &  &  &  \\
$\%CorrMaxCP$ \textit{(sample)}             	&      80.684   &      80.786   &      82.138   &      81.398   &      81.398   &      82.258   \\
$\%CorrMaxCP$ \textit{(2-fold $\times$ valid)}  &      80.888   &      80.888   &      82.164   &      81.828   &      82.043   &      82.688   \\
$\%CorrMaxCP$ \textit{(future $|$ present)}     &      79.612   &      79.669   &      81.496   &      86.630   &      86.630   &      86.351   \\
$\%CorrMaxCP$ \textit{(present $|$ future)}     &      81.411   &      81.458   &      82.426   &      72.504   &      73.205   &      73.905   \\
 &  &  &  &  &  &  \\
$Avg\,\mathbb{P}(OK\,Pred)$ \textit{(sample)}         		&      72.828   &      72.830   &      74.869   &      73.434   &      73.439   &      74.820   \\
$Avg\,\mathbb{P}(OK\,Pred)$ \textit{(2-fold $\times$ valid)}    &      72.814   &      72.808   &      74.800   &      73.362   &      73.428   &      74.584   \\
$Avg\,\mathbb{P}(OK\,Pred)$ \textit{(future $|$ present)}  	&      70.942   &      70.900   &      72.855   &      76.135   &      76.127   &      76.265   \\
$Avg\,\mathbb{P}(OK\,Pred)$ \textit{(present $|$ future)}  	&      74.356   &      74.313   &      76.490   &      68.593   &      68.451   &      69.762   \\
&  &  &  &  &  &  \\\hline\hline
\multicolumn{7}{p{14.5cm}}{\scriptsize\textit{Notes.} The table reports average marginal effects of binary logits. The response variable is an indicator of whether an offer was accepted. Standard errors are clustered at the participant level. *, **, and *** indicate significance at the 90\%, 95\%, and 99\% confidence level, respectively. The table also reports the mean-squared error (\textit{MSE}) of the predicted probability, percentage of choices in which the predicted probability of the alternative chosen in the data is the greatest among all alternatives ($\%CorrMaxCP$), and the average probability of correctly predicting the data ($Avg\,\mathbb{P}(OK\,Pred)$). Each is computed in the estimation sample and out of the sample using random two-fold cross-validation, predicting the final five rounds with the first five rounds, and the first five rounds using the final five.}
\end{tabular}
\end{table}

In our main treatments, larger monetary gains are associated with higher acceptance probabilities, even for proposers who are ranked similarly, echoing the importance of cardinal payoffs in our markets. However, when receivers cannot make proposals themselves, as in our unilateral treatments, they place a higher weight on avoiding losses. History matters for receivers: they are more likely to accept offers from proposers with whom they have already matched, but less likely to accept offers from those whom they have already rejected. In our main treatments, receivers are less likely to accept offers as they receive more of them, a pattern that reverse in our unilateral treatments. Similar to proposers, receivers seem to be drawn to proposers who are stable partners, accepting their offers at higher rates.\footnote{Adding participant fixed effects to these regressions does not alter the results and decreases their predictive power. This suggests that there is no significant unobserved heterogeneity across participants acting as receivers; see the Online Appendix for details.}

\section{Discussion and Conclusions}\label{conclusion}

This paper presents an experimental investigation of decentralized matching markets. There are three main findings. First, in line with the cooperative solution, decentralized markets often culminate in stable matchings. They do so quickly, in terms of both time and market activity. Second, the median stable matching has very strong drawing power and is frequently selected. Furthermore, cardinal incentives impact the distribution over non-median selected matchings. Roughly speaking, the side of the market that has ``more to lose'' from forgoing their favorite stable matching, is more likely to implement it. Last, in terms of dynamics, by and large, participants form successive blocking pairs. However, participants are more sophisticated than suggested by some of the na\"ive dynamics the theoretical literature has focused on, with proposers strategically targeting receivers who value them highly and receivers taking into account past market activity. As a consequence, cycles of blocking pairs are relatively rare in our data.

Our findings are important from a market design perspective. On the one hand, they reinforce the drive to create institutions that implement a stable matching. Indeed, the results suggest that decentralized processes that precede, or follow, a centralized matching protocol would push outcomes toward stability (and, if not generated by the centralized mechanism in place, may produce inefficiencies having to do with the changes yielding an ultimate stable matching). On the other hand, our findings suggest some important comparisons that underlie the decision to use a centralized clearinghouse to begin with. Many implemented centralized matching markets use a variation of the DA algorithm that, absent incentive compatibility concerns, generate extremal stable matchings, favored by one side of the market. Our decentralized processes did not culminate in an extremal matching often. Therefore, the use of DA should be accompanied with a justification for the choice of stable matching to implement, or at least an appeal to incentive compatibility.

Our results also have implications for the theory of dynamic stabilization in matching markets. Existing models that generate stable matchings through the sequential formation of blocking pairs (such as \citeip{roth1990random}, or \citeip{ackermann2011uncoordinated}), or which prescribe an order for the offers made by agents (such as the DA algorithm), do not explain basic features of our data. There is therefore room for further theoretical work that provides foundational guidance on the selection of stable matchings in decentralized markets and takes into account cardinal, not only ordinal, assessments of partners.

Our experiments are designed as a benchmark for decentralized interactions in an idealized setting, allowing a clean examination of the cooperative theoretical predictions. Nonetheless, in applications, a variety of frictions might be at play: incomplete information, offer and time costs, disutility from rejection, and so on. We hope our study opens the door for further investigations of decentralized markets incorporating such frictions. 
 
\clearpage
\bibliographystyle{ecta}
\bibliography{matching_experiment}

\clearpage

\begin{center}
    {\LARGE \textbf{Online Appendix}}
    
    \bigskip
    
    {\Large \textbf{[Not Intended For Publication]}}
\end{center}

\setcounter{table}{0}
\renewcommand{\thetable}{A\arabic{table}}

\setcounter{figure}{0}
\renewcommand{\thefigure}{A\arabic{figure}}

\bigskip

{\small


\begin{definition*}[Match-level cycle]
    Let $S=(f_t,c_t)_{t=1}^T$ denote the ordered sequence of all matches formed (offers accepted) in a market, experimental or simulated. That is, the $t$-th pair formed in the market is between food $f_t\in F$ and color $c_t\in C$, and $T$ is the total number of matches formed. Two pairs, $(f,c)$ and $(f',c')$, are \emph{connected} if $f=f'$, or $c=c'$, and not both. A \emph{match-level cycle} is a subsequence of $S$, $(f_{t_k},c_{t_k})_{k=1}^K$, such that (i) the initial and final pairs coincide, i.e., $(f_{t_1},c_{t_1})=(f_{t_K},c_{t_K})$; (ii) the initial pair is not reached in the interim, i.e., $(f_{t_k},c_{t_k})\neq(f_{t_1},c_{t_1})$ for every $k=2,3,\dots,K-1$; (iii) contiguous pairs along the subsequence are connected, i.e., $(f_{t_k},c_{t_k})$ and $(f_{t_{k+1}},c_{t_{k+1}})$ are connected for every $k=1,2,\dots,K-1$; and (iv) it is of maximal length, i.e., there does not exist a longer subsequence of $S$, $(f_{t'_k},c_{t'_k})_{k=1}^{K'}$ with $K'>K$, that satisfies (i)--(iii), and starts at the same point in $S$, i.e., with $t'_1=t_1$. The \emph{length} of a match-level cycle $(f_{t_k},c_{t_k})_{k=1}^K$ is $K$.
\end{definition*}

\paragraph{Discussion and examples of match-level cycles.} Under our definition, more than one match-level cycle can stem from the same pair. Moreover, match-level cycles can overlap or be contained in each other. Below we include three examples to illustrate these features of match-level cycles and discuss the motivation underlying the definition.

\begin{enumerate}

    \item \textbf{Example of two match-level cycles linking the same repeated pair.} This example is the same one given in the main text (see footnote~\ref*{foot_match_level_cycle}). Suppose we observe the following sequence of matches in $S$: $(f,c)$, $(f,c')$, $(f',c)$, and $(f,c)$. This means we observe that $f$ and $c$ break their tentative pairing to match with $c'$ and $f'$, respectively, and then break these two matches to pair with one another again. There are two match-level cycles of length 3 in this sequence, given by: $(f,c)\to(f,c')\to(f,c)$, and $(f,c)\to(f',c)\to(f,c)$. Our definition captures the plausible perceptions of agents on ensuing paths that revert back to the starting point. Indeed, with the observed sequence, there are two possible paths doing so.
    
    \item \textbf{Example of two overlapping match-level cycles.} Suppose we observe the following sequence of matches in $S$: $(f,c)$, $(f,c')$, $(f,c)$, and $(f,c')$. That is, $f$ matches with $c$, then this match breaks, and $f$ matches with $c'$, after which $f$ matches again with $c$, followed by $f$ matching with $c'$ yet again. In this sequence, there are two match-level cycles of length 3: one is $(f,c)\to(f,c')\to(f,c)$, the other is $(f,c')\to(f,c)\to(f,c')$. This captures the fact that agent $f$ switches back to her original partner from two separate starting points. 
    
    \item \textbf{Example of a match-level cycle contained in another match-level cycle.} Suppose we observe the following sequence of matches in $S$: $(f,c)$, $(f,c')$, $(f',c)$, $(f',c')$, $(f',c)$, and $(f,c)$. In this sequence, there are three match-level cycles, two of length 5, and one of length 3. The two match-level cycles of length 5 link $(f,c)$ back to itself, and are: $(f,c)\to(f,c')\to(f',c')\to(f',c)\to(f,c)$, and $(f,c)\to(f',c)\to(f',c')\to(f',c)\to(f,c)$. The match-level cycle of length 3 is contained in the second of these two cycles: $(f',c)\to(f',c')\to(f',c)$. The first two match-level cycles of length 5 are indeed of maximal length since the subsequences $(f,c)\to(f,c')\to(f,c)$, and $(f,c)\to(f',c)\to(f,c)$, also link $(f,c)$ to itself but do not satisfy the maximality restriction. Our definition captures the idea that agents involved in the pair $(f,c)$ and those involved in the pair $(f',c)$ perceive cyclical behavior as plausible, and do so via different, albeit overlapping, paths.

\end{enumerate}


\paragraph{Deferred Acceptance with Two-Sided Random Proposers (2RDA).} In the 2RDA, an agent proposes to an agent in each round. If the offer is accepted, both agents are matched tentatively. As in the standard DA, agents go down their rank-ordered lists as they propose to other agents. However, at every round, the proposer may be chosen from either side of the market, and only agents who are not matched to their most preferred agent among the ones they have not proposed to (i.e., agents who are ``active'') may be chosen as proposers. In contrast to standard DA, this algorithm may result in unstable matchings. Nonetheless, by construction, all offers are ``Gale-Shapley,'' ``downward,'' and do not ``skip someone'' (as defined in the main text). As our implementation of this procedure is not standard, we now provide our algorithm's details. To write down the algorithm, it is convenient to keep track of two ranks for every agent, one for their next offer (if chosen as proposers), and one for their current match (if any).

{\singlespacing

\begin{algorithm*}[2RDA]
Set the match-rank of every agent at $\infty$, and their offer-rank at 1. An agent is \textit{active} if their match-rank is strictly greater than their offer-rank. If there are no active agents, stop; else, proceed in steps:

\begin{itemize}
    \item Choose an active agent at random. Say, food $f$ is chosen (analogously if a color is chosen).
    \item Food $f$ proposes to the color $c$ that they rank at their current offer-rank (if $f$'s offer-rank equals 1, they propose to their top-choice, if it's 2, to their second choice, and so on).
    \item Color $c$ accepts the offer if they find fruit $f$ acceptable and rank them above their match-rank (i.e., the rank of food $f$ in the list of color $c$ is strictly less than their match-rank).
    \item If the offer is accepted, match $f$ and $c$, and
    \begin{itemize}
        \item set $f$'s match-rank to their current offer-rank ($c$'s rank in their rank-ordered list).
        \item set $c$'s match-rank to the rank corresponding to $f$ in their rank-ordered list.
        \item set the match-rank of their previous partners (if any) to $\infty$.
    \end{itemize}
    \item Increase food $f$'s offer-rank by one.
\end{itemize}
\end{algorithm*}

}

\noindent \autoref{tab:simul:2RDA} reports simulation results from a variation of 2RDA that captures agents' cardinal incentives to make offers. Specifically, we run simulations in which an active agent $a$ is chosen to be a proposer with probability proportional to $g_a$ or $\exp(\lambda g_a)$, where $g_a$ is the gain $a$ would obtain if their next offer were accepted, and $\lambda >0$ is a fixed parameter. The resulting simulations generate a higher frequency of stable matchings, but they fail to replicate other aspects of the data, in particular the volume of offers.


\paragraph{Deferred Acceptance with Compensation Chains (DACC).} 

This algorithm, inspired by \cite{dworczak21}, is similar to 2RDA with one critical adjustment. At every round, a proposer is chosen uniformly at random from either side of the market. Say agent $i$ is chosen as proposer. Agent $i$ then makes a proposal to agent $j$, where $j$ is $i$'s most preferred agent among those that $i$ has not proposed to or who have not been matched to $i$ and broken the match. Agent $j$ accepts the proposal if they prefer agent $i$ to their current match. If agent $j$ accepts the proposal, $i$ and $j$ are tentatively matched and any existing match they are involved in is severed. An agent is \textit{deceived} if their current match, who originally proposed to them, breaks the match to match with another agent (e.g., if $j$ accepts $i$'s proposal while being matched to $i'$, whom $j$ proposed to previously, then $i'$ is deceived). Whenever an agent is deceived, they are ``compensated'' and are chosen as proposers in the next round. Compensating an agent can cause further agents to be deceived, triggering a ``compensation chain.'' \cite{dworczak21} proved that, unlike 2RDA, this algorithm converges with probability one to a stable matching, and that every stable matching can be reached by a sequence of proposers.

\autoref{tab:simul:DACC} reports simulation results from a variation of DACC that captures agents' cardinal incentives to make offers. Specifically, we run simulations in which an agent $a$ is chosen to be a proposer with probability proportional to $g_a$ or $\exp(\lambda g_a)$, where $g_a$ is the gain $a$ would obtain if their next offer were accepted, and $\lambda >0$ is a fixed parameter. The resulting simulations tend to increase the speed of convergence to stability of the algorithm, which is already faster than our experimental markets in the benchmark version of DACC. Furthermore, simulation results are further away from median stable outcomes.


\paragraph{Random Paths to Stability (RPS).} This model assumes that blocking pairs are formed at random. Starting from some matching at time $t$, say $\mu_{t}$, the set of all blocking pairs is tabulated, and one is formed uniformly at random. That is, the corresponding color and food in that blocking pair get matched and their partners in $\mu_{t}$ (if they exist) are unmatched. The resulting matching is $\mu_{t+1}$, and the process continues iteratively. \cite{roth1990random} proved that these dynamics converge to a stable matching with probability one.\footnote{\cite{rudov2022fragile} shows that, in fact, this prediction cannot be refined further: under mild conditions, any unstable matching can reach any stable matching through these dynamics.}

As discussed in the main text (see \autoref{table:mktdynandsimulations}), in markets with five stable matchings, RPS does a poor job at predicting the offer volume and the distribution of ultimate stable matchings that we see in our experimental data. In order to give na\"ive dynamics such as RPS a chance at explaining our data, we consider versions of RPS in which the probability that a blocking pair forms depends on the welfare gain for the agents participating. In particular, we consider a version in which the probability that a blocking pair forms is proportional to the sum of payoff gains of the blocking partners. We also consider a version in which that probability is logistic. Namely, the probability that any blocking pair $(f,c)$ forms is proportional to $\exp (\lambda g_{f,c})$, where $g_{f,c}$ is the sum of $f$ and $c$'s payoffs from matching and $\lambda$ is a sensitivity parameter. 

\autoref{tab:simul:RPS} reports results from simulations of the original uniform RPS, as well as its two variants, including alternative values of the sensitivity parameter $\lambda$ of the logistic variant. The logistic model seems to fit the data of markets with multiple stable matchings best, perhaps due to the additional degree of freedom its sensitivity parameter affords.


\paragraph{Random Best Response (RBR).} This dynamic model, due
to \cite{ackermann2011uncoordinated}, is an alternative to RPS in which, instead of randomly choosing blocking pairs, a random agent is selected at each stage. That agent's most preferred blocking pair is then formed, if one exists. Specifically, given a matching $\mu_t$, we tabulate the set of agents that have at least one blocking partner, and choose one uniformly at random. The next matching, $\mu_{t+1}$, is obtained by matching the chosen agent with their most preferred blocking partner. RBR converges with probability one to a stable matching, just as RPS. 

We also consider versions of RBR in which cardinal payoff information is allowed to determine the probability with which a blocking pair is chosen. Similar to RPS, albeit to a lesser degreee, the standard uniform version of RBR generates a higher offer volume observed in our data for markets with five stable matchings. We consider two variants analogous to those we consider for RPS. Let $g_a$ denote the net gain agent $a$ would obtain if matched to their most preferred blocking partner. In the two variants of RBR, we choose each agent $a$ who is part of a blocking pair with probability proportional to $g_a$, or $\exp(\lambda g_a$), where $\lambda$ is a sensitivity parameter. \autoref{tab:simul:RBR} reports simulation results for these different variants of RBR, allowing for an array of $\lambda$ values. 

\begin{table}[ht!]
\centering\scriptsize
\caption{Simulations---Two-Sided Random Deferred Acceptance (2RDA)
\label{tab:simul:2RDA}\centering}
\begin{tabular}{lccccccccc}
 &  &  &  &  &  &  &  &  &   \\\hline\hline
 & & & & \multicolumn{6}{c}{\scriptsize\textit{Exponential ($\lambda$)}} \\
 & 
{\scriptsize\textit{Experiment}} & 
{\scriptsize\textit{Uniform}} & 
{\scriptsize\textit{Proportional}} & 
0.005 & 
0.0075 & 
0.01 & 
0.05 & 
0.1 & 
0.5 \\\hline
 &  &  &  &  &  &  &  &  &   \\
\multicolumn{3}{l}{\textit{Unique stable matching}} &  &  &  &  &  &  &   \\
 &  &  &  &  &  &  &  &  &  \\
\hspace{0.2cm} \# \textit{offers} & 44.8 & 36.2 & 36.3 & 36.2 & 36.4 & 36.4 & 36.5 & 36.5 & 36.5 \\
\hspace{0.2cm} \# \textit{matches} & 15.9 & 17.2 & 16.5 & 16.4 & 16.1 & 15.9 & 15.0 & 14.8 & 14.7 \\
\hspace{0.2cm} \% \textit{accepted offers} & 41.3 & 51.6 & 48.8 & 48.6 & 47.5 & 46.8 & 44.3 & 43.6 & 43.4 \\
\hspace{0.2cm} \# \textit{accepted $\mid$ to BP} & 15.2 & 17.2 & 16.5 & 16.4 & 16.1 & 15.9 & 15.0 & 14.8 & 14.7 \\
\hspace{0.2cm} \% \textit{repeated matches} & 14.0 & 4.7 & 1.7 & 1.5 & 0.9 & 0.6 & 0.1 & 0.0 & 0.0 \\
\hspace{0.2cm} \% \textit{repeated matchings} & 6.0 & 0.0 & 0.0 & 0.0 & 0.0 & 0.0 & 0.0 & 0.0 & 0.0 \\
\hspace{0.2cm} \# \textit{match-level cycles} & 2.5 & 1.4 & 0.5 & 0.4 & 0.3 & 0.2 & 0.0 & 0.0 & 0.0 \\
\hspace{0.2cm} \textit{avg.\ match-level cycle length} & 3.3 & 3.3 & 3.3 & 3.3 & 3.2 & 3.2 & 3.2 & 3.1 & 3.0 \\
\hspace{0.2cm} \% \textit{final matching is stable} & 90.0 & 89.0 & 91.8 & 93.2 & 94.3 & 95.1 & 98.6 & 99.6 & 100.0 \\
\hspace{0.2cm} \% \textit{final pairs are stable} & 94.7 & 98.4 & 98.7 & 98.9 & 99.0 & 99.0 & 99.6 & 99.9 & 100.0 \\
 &  &  &  &  &  &  &  &  &  \\
\multicolumn{3}{l}{\textit{Two embedded 4-by-4 markets}} &  &  &  &  &  &  & \\
 &  &  &  &  &  &  &  &  & \\
\hspace{0.2cm} \# \textit{offers} & 39.7 & 25.4 & 23.4 & 23.6 & 23.1 & 22.8 & 21.8 & 21.7 & 21.7 \\
\hspace{0.2cm} \# \textit{matches} & 15.9 & 15.9 & 13.1 & 13.3 & 12.6 & 12.3 & 11.6 & 11.6 & 11.6 \\
\hspace{0.2cm} \% \textit{accepted offers} & 46.5 & 62.3 & 55.7 & 56.1 & 54.6 & 53.8 & 53.0 & 53.3 & 53.3 \\
\hspace{0.2cm} \# \textit{accepted $\mid$ to BP} & 15.1 & 15.9 & 13.1 & 13.3 & 12.6 & 12.3 & 11.6 & 11.6 & 11.6 \\
\hspace{0.2cm} \% \textit{repeated matches} & 19.3 & 9.3 & 3.1 & 3.5 & 2.2 & 1.5 & 0.3 & 0.2 & 0.2 \\
\hspace{0.2cm} \% \textit{repeated matchings} & 8.0 & 0.1 & 0.1 & 0.1 & 0.1 & 0.1 & 0.0 & 0.0 & 0.0 \\
\hspace{0.2cm} \# \textit{match-level cycles} & 4.2 & 2.6 & 0.8 & 0.9 & 0.6 & 0.4 & 0.1 & 0.1 & 0.1 \\
\hspace{0.2cm} \textit{avg.\ match-level cycle length} & 3.2 & 3.3 & 3.2 & 3.2 & 3.2 & 3.2 & 3.1 & 3.1 & 3.1 \\
\hspace{0.2cm} \% \textit{final matching is stable} & 94.3 & 83.3 & 93.6 & 92.9 & 95.3 & 96.6 & 99.3 & 99.6 & 99.7 \\
\hspace{0.2cm} \% \textit{final pairs are stable} & 99.6 & 96.5 & 98.7 & 98.5 & 99.0 & 99.3 & 99.9 & 99.9 & 99.9 \\
 &  &  &  &  &  &  &  &  &   \\
\multicolumn{3}{l}{\textit{5 stable matchings \& 3 stable partners}} &  &  &  &  &  &  &   \\
 &  &  &  &  &  &  &  &  &   \\
\hspace{0.2cm} \# \textit{offers} & 59.2 & 54.2 & 51.7 & 51.7 & 50.0 & 48.7 & 43.3 & 42.2 & 42.1 \\
\hspace{0.2cm} \# \textit{matches} & 24.6 & 30.7 & 25.5 & 25.3 & 23.3 & 21.9 & 15.4 & 13.8 & 13.1 \\
\hspace{0.2cm} \% \textit{accepted offers} & 40.4 & 56.3 & 48.7 & 48.2 & 45.7 & 43.9 & 34.6 & 31.9 & 30.5 \\
\hspace{0.2cm} \# \textit{accepted $\mid$ to BP} & 22.6 & 30.7 & 25.5 & 25.3 & 23.3 & 21.9 & 15.4 & 13.8 & 13.1 \\
\hspace{0.2cm} \% \textit{repeated matches} & 27.2 & 12.9 & 8.1 & 8.2 & 6.5 & 5.2 & 1.5 & 1.0 & 0.9 \\
\hspace{0.2cm} \% \textit{repeated matchings} & 6.2 & 0.1 & 0.1 & 0.1 & 0.1 & 0.0 & 0.0 & 0.0 & 0.0 \\
\hspace{0.2cm} \# \textit{match-level cycles} & 10.4 & 6.9 & 3.9 & 4.0 & 3.0 & 2.4 & 0.7 & 0.4 & 0.4 \\
\hspace{0.2cm} \textit{avg.\ match-level cycle length} & 3.5 & 3.9 & 3.7 & 3.7 & 3.6 & 3.6 & 3.7 & 3.6 & 3.6 \\
\hspace{0.2cm} \% \textit{final matching is stable} & 75.0 & 46.9 & 55.1 & 56.6 & 63.1 & 68.5 & 88.4 & 91.6 & 90.5 \\
\hspace{0.2cm} \% \textit{final pairs are stable} & 96.9 & 93.8 & 95.2 & 95.3 & 96.1 & 96.7 & 98.9 & 99.1 & 98.7 \\
 &  &  &  &  &  &  &  &  &   \\
\multicolumn{3}{l}{\textit{\hspace{0.3cm}Market-level outcomes (matchings)}} &  &  &  &  &  &  &   \\
 &  &  &  &  &  &  &  &  &   \\
\hspace{0.5cm} \% \textit{median $\mid$ stable} & 80.0 & 64.1 & 43.2 & 42.7 & 36.7 & 33.5 & 13.9 & 7.2 & 4.4  \\
\hspace{0.5cm} \% \textit{non-extremal $\mid$ stable} & 100.0 & 89.1 & 77.2 & 75.5 & 69.9 & 65.7 & 29.9 & 17.5 & 11.9  \\
\hspace{0.5cm} \% \textit{food-optimal $\mid$ stable} & 0.0 & 5.5 & 7.3 & 7.3 & 8.5 & 9.4 & 23.6 & 25.3 & 25.0  \\
\hspace{0.5cm} \% \textit{color-optimal $\mid$ stable} & 0.0 & 5.3 & 15.5 & 17.2 & 21.6 & 24.9 & 46.5 & 57.1 & 63.1  \\
 &  &  &  &  &  &  &  &  &   \\
\multicolumn{3}{l}{\textit{\hspace{0.3cm}Individual-level outcomes (matches)}} &  &  &  &  &  &  &   \\
 &  &  &  &  &  &  &  &  &   \\
\hspace{0.5cm} \% \textit{median $\mid$ stable} & 76.8 & 52.9 & 48.0 & 47.7 & 45.3 & 43.3 & 21.0 & 12.7 & 9.7  \\
\hspace{0.5cm} \% \textit{fruit-optimal $\mid$ stable} & 2.9 & 22.0 & 19.8 & 19.2 & 17.9 & 17.1 & 27.8 & 28.3 & 27.6  \\
\hspace{0.5cm} \% \textit{color-optimal $\mid$ stable} & 20.4 & 24.7 & 32.0 & 32.8 & 36.6 & 39.5 & 51.2 & 59.0 & 62.6  \\
 &  &  &  &  &  &  &  &  &   \\\hline\hline
\multicolumn{10}{p{16cm}}{\scriptsize\textit{Notes.} The table reports averages across 10,000 simulations of every market in our main treatments using the Two-Sided Random Deferred Acceptance (2RDA) algorithm. The first column reports the experimental averages for reference. Each column from the second reports the results using a distinct distribution to choose the proposer on each round. \textit{Uniform} corresponds to uniformly at random; \textit{Proportional} to probability proportional to $g_a$, and \textit{Exponential} to probability proportional to $\exp(\lambda g_a)$, where $g_a$ denotes the net gain of active agent $a$ if their next proposal were accepted.}
\end{tabular}
\end{table}

\clearpage

\begin{table}[ht!]
\centering\scriptsize
\caption{Simulations---Deferred Acceptance with Compensation Chains (DACC)
\label{tab:simul:DACC}\centering}
\begin{tabular}{lcccccccc}
 &  &  &  &  &  &  &  &  \\\hline\hline
& & & & \multicolumn{5}{c}{\scriptsize\textit{Exponential ($\lambda$)}} \\
 & 
{\scriptsize\textit{Experiment}} & 
{\scriptsize\textit{Uniform}} & 
{\scriptsize\textit{Proportional}} & 
0.005 & 
0.0075 & 
0.01 & 
0.05 & 
0.075 \\\hline
 &  &  &  &  &  &  &  &  \\
\multicolumn{3}{l}{\textit{Unique stable matching}} &  &  &  &  &  &  \\
 &  &  &  &  &  &  &  &  \\
\hspace{0.2cm} \# \textit{offers} & 44.8 & 35.3 & 35.5 & 35.6 & 35.9 & 36.1 & 36.9 & 37.2 \\
\hspace{0.2cm} \# \textit{matches} & 15.9 & 17.3 & 16.6 & 16.5 & 16.2 & 15.9 & 15.0 & 14.9 \\
\hspace{0.2cm} \% \textit{accepted offers} & 41.3 & 53.5 & 50.2 & 49.7 & 48.3 & 47.4 & 43.8 & 43.0 \\
\hspace{0.2cm} \# \textit{accepted $\mid$ to BP} & 15.2 & 17.3 & 16.6 & 16.5 & 16.2 & 15.9 & 15.0 & 14.9 \\
\hspace{0.2cm} \% \textit{repeated matches} & 14.0 & 5.1 & 2.0 & 1.8 & 1.2 & 0.8 & 0.1 & 0.1 \\
\hspace{0.2cm} \% \textit{repeated matchings} & 6.0 & 0.0 & 0.0 & 0.0 & 0.0 & 0.0 & 0.0 & 0.0 \\
\hspace{0.2cm} \# \textit{match-level cycles} & 2.5 & 1.6 & 0.7 & 0.6 & 0.4 & 0.3 & 0.0 & 0.0 \\
\hspace{0.2cm} \textit{avg.\ match-level cycle length} & 3.3 & 3.3 & 3.2 & 3.2 & 3.2 & 3.2 & 3.0 & 3.0 \\
\hspace{0.2cm} \% \textit{final matching is stable} & 90.0 & 100.0 & 100.0 & 100.0 & 100.0 & 100.0 & 100.0 & 100.0 \\
\hspace{0.2cm} \% \textit{final pairs are stable} & 94.7 & 100.0 & 100.0 & 100.0 & 100.0 & 100.0 & 100.0 & 100.0 \\
 &  &  &  &  &  &  &  &  \\
\multicolumn{3}{l}{\textit{Two embedded 4-by-4 markets}} &  &  &  &  &  &  \\
 &  &  &  &  &  &  &  &  \\
\hspace{0.2cm} \# \textit{offers} & 39.7 & 23.8 & 22.7 & 22.8 & 22.6 & 22.4 & 21.7 & 21.7 \\
\hspace{0.2cm} \# \textit{matches} & 15.9 & 15.1 & 12.8 & 12.9 & 12.4 & 12.1 & 11.5 & 11.5 \\
\hspace{0.2cm} \% \textit{accepted offers} & 46.5 & 63.3 & 56.1 & 56.6 & 55.0 & 54.1 & 53.2 & 53.3 \\
\hspace{0.2cm} \# \textit{accepted $\mid$ to BP} & 15.1 & 15.1 & 12.8 & 12.9 & 12.4 & 12.1 & 11.5 & 11.5 \\
\hspace{0.2cm} \% \textit{repeated matches} & 19.3 & 7.9 & 2.5 & 2.8 & 1.7 & 1.2 & 0.3 & 0.2 \\
\hspace{0.2cm} \% \textit{repeated matchings} & 8.0 & 0.0 & 0.0 & 0.0 & 0.0 & 0.0 & 0.0 & 0.0 \\
\hspace{0.2cm} \# \textit{match-level cycles} & 4.2 & 1.9 & 0.5 & 0.6 & 0.4 & 0.2 & 0.1 & 0.0 \\
\hspace{0.2cm} \textit{avg.\ match-level cycle length} & 3.2 & 3.2 & 3.1 & 3.1 & 3.1 & 3.1 & 3.0 & 3.0 \\
\hspace{0.2cm} \% \textit{final matching is stable} & 94.3 & 100.0 & 100.0 & 100.0 & 100.0 & 100.0 & 100.0 & 100.0 \\
\hspace{0.2cm} \% \textit{final pairs are stable} & 99.6 & 100.0 & 100.0 & 100.0 & 100.0 & 100.0 & 100.0 & 100.0 \\
 &  &  &  &  &  &  &  &  \\
\multicolumn{3}{l}{\textit{5 stable matchings \& 3 stable partners}} &  &  &  &  &  &  \\
 &  &  &  &  &  &  &  &  \\
\hspace{0.2cm} \# \textit{offers} & 59.2 & 48.4 & 46.9 & 47.0 & 46.3 & 45.6 & 42.5 & 41.9 \\
\hspace{0.2cm} \# \textit{matches} & 24.6 & 27.9 & 23.3 & 23.2 & 21.6 & 20.5 & 15.3 & 14.2 \\
\hspace{0.2cm} \% \textit{accepted offers} & 40.4 & 57.3 & 49.0 & 48.7 & 45.9 & 44.1 & 35.1 & 33.2 \\
\hspace{0.2cm} \# \textit{accepted $\mid$ to BP} & 22.6 & 27.9 & 23.3 & 23.2 & 21.6 & 20.5 & 15.3 & 14.2 \\
\hspace{0.2cm} \% \textit{repeated matches} & 27.2 & 12.1 & 8.5 & 8.7 & 7.4 & 6.3 & 2.3 & 1.8 \\
\hspace{0.2cm} \% \textit{repeated matchings} & 6.2 & 0.0 & 0.0 & 0.0 & 0.0 & 0.0 & 0.0 & 0.0 \\
\hspace{0.2cm} \# \textit{match-level cycles} & 10.4 & 5.7 & 3.7 & 3.9 & 3.2 & 2.7 & 1.0 & 0.8 \\
\hspace{0.2cm} \textit{avg.\ match-level cycle length} & 3.5 & 3.5 & 3.3 & 3.3 & 3.2 & 3.2 & 3.2 & 3.2 \\
\hspace{0.2cm} \% \textit{final matching is stable} & 75.0 & 100.0 & 100.0 & 100.0 & 100.0 & 100.0 & 100.0 & 100.0 \\
\hspace{0.2cm} \% \textit{final pairs are stable} & 96.9 & 100.0 & 100.0 & 100.0 & 100.0 & 100.0 & 100.0 & 100.0 \\
 &  &  &  &  &  &  &  &  \\
\multicolumn{3}{l}{\textit{\hspace{0.3cm}Market-level outcomes (matchings)}} &  &  &  &  &  &  \\
 &  &  &  &  &  &  &  &  \\
\hspace{0.5cm} \% \textit{median $\mid$ stable} & 80.0 & 50.4 & 38.5 & 39.0 & 36.1 & 34.3 & 17.3 & 11.6 \\
\hspace{0.5cm} \% \textit{non-extremal $\mid$ stable} & 100.0 & 75.1 & 68.3 & 68.1 & 64.8 & 61.6 & 32.2 & 24.1 \\
\hspace{0.5cm} \% \textit{food-optimal $\mid$ stable} & 0.0 & 19.4 & 17.9 & 17.1 & 16.3 & 16.2 & 26.2 & 27.2 \\
\hspace{0.5cm} \% \textit{color-optimal $\mid$ stable} & 0.0 & 5.4 & 13.8 & 14.8 & 19.0 & 22.2 & 41.6 & 48.6 \\
 &  &  &  &  &  &  &  &  \\
\multicolumn{3}{l}{\textit{\hspace{0.3cm}Individual-level outcomes (matches)}} &  &  &  &  &  &  \\
 &  &  &  &  &  &  &  &  \\
\hspace{0.5cm} \% \textit{median $\mid$ stable} & 76.8 & 62.8 & 53.4 & 53.5 & 50.5 & 48.0 & 24.7 & 17.8 \\
\hspace{0.5cm} \% \textit{fruit-optimal $\mid$ stable} & 2.9 & 19.4 & 17.9 & 17.1 & 16.3 & 16.2 & 26.2 & 27.2 \\
\hspace{0.5cm} \% \textit{color-optimal $\mid$ stable} & 20.4 & 17.8 & 28.7 & 29.4 & 33.3 & 35.9 & 49.1 & 54.9 \\
 &  &  &  &  &  &  &  &  \\\hline\hline
\multicolumn{9}{p{15.5cm}}{\scriptsize\textit{Notes.} The table reports averages across 10,000 simulations of every market in our main treatments using the Deferred Acceptance with Compensation Chains (DACC) algorithm of \cite{dworczak21}. The first column reports the experimental averages for reference. Each column from the second reports the results using a distinct distribution to choose the proposer on each round. \textit{Uniform} corresponds to uniformly at random; \textit{Proportional} to probability proportional to $g_a$, and \textit{Exponential} to probability proportional to $\exp(\lambda g_a)$, where $g_a$ denotes the net gain of active agent $a$ if their next proposal were accepted.}
\end{tabular}
\end{table}

\clearpage

\begin{table}[ht!]
\centering\scriptsize
\caption{Simulations---Random Paths to Stability (RPS)
\label{tab:simul:RPS}\centering}
\begin{tabular}{lccccccccc}
 &  &  &  &  &  &  &  &  &  \\\hline\hline
& & & & \multicolumn{6}{c}{\scriptsize\textit{Exponential ($\lambda$)}} \\ 
 & 
{\scriptsize\textit{Experiment}} & 
{\scriptsize\textit{Uniform}} & 
{\scriptsize\textit{Proportional}} & 
0.005 & 
0.01 & 
0.0175 & 
0.02 & 
0.03 & 
0.05 \\\hline
 &  &  &  &  &  &  &  &  &  \\
\multicolumn{3}{l}{\textit{Unique stable matching}} &  &  &  &  &  &  &  \\
 &  &  &  &  &  &  &  &  &  \\
\hspace{0.2cm} \# \textit{offers} & 44.8 & 36.4 & 27.0 & 20.2 & 15.7 & 13.1 & 12.6 & 11.5 & 10.8 \\
\hspace{0.2cm} \# \textit{matches} & 15.9 & 36.4 & 27.0 & 20.2 & 15.7 & 13.1 & 12.6 & 11.5 & 10.8 \\
\hspace{0.2cm} \% \textit{accepted offers} & 41.3 & 100.0 & 100.0 & 100.0 & 100.0 & 100.0 & 100.0 & 100.0 & 100.0 \\
\hspace{0.2cm} \# \textit{accepted $\mid$ to BP} & 15.2 & 36.4 & 27.0 & 20.2 & 15.7 & 13.1 & 12.6 & 11.5 & 10.8 \\
\hspace{0.2cm} \% \textit{repeated matches} & 14.0 & 20.2 & 12.4 & 6.9 & 3.0 & 1.4 & 1.2 & 0.8 & 0.4 \\
\hspace{0.2cm} \% \textit{repeated matchings} & 6.0 & 0.6 & 0.2 & 0.1 & 0.0 & 0.0 & 0.0 & 0.0 & 0.0 \\
\hspace{0.2cm} \# \textit{match-level cycles} & 2.5 & 13.5 & 6.0 & 2.5 & 0.9 & 0.3 & 0.3 & 0.2 & 0.1 \\
\hspace{0.2cm} \textit{avg.\ match-level cycle length} & 3.3 & 4.4 & 3.9 & 3.6 & 3.6 & 3.7 & 3.7 & 3.7 & 4.0 \\
\hspace{0.2cm} \% \textit{final matching is stable} & 90.0 & 100.0 & 100.0 & 100.0 & 100.0 & 100.0 & 100.0 & 100.0 & 100.0 \\
\hspace{0.2cm} \% \textit{final pairs are stable} & 94.7 & 100.0 & 100.0 & 100.0 & 100.0 & 100.0 & 100.0 & 100.0 & 100.0 \\
 &  &  &  &  &  &  &  &  &  \\
\multicolumn{3}{l}{\textit{Two embedded 4-by-4 markets}} &  &  &  &  &  &  &  \\
 &  &  &  &  &  &  &  &  &  \\
\hspace{0.2cm} \# \textit{offers} & 39.7 & 28.1 & 19.8 & 15.6 & 12.9 & 11.4 & 11.1 & 10.1 & 8.9 \\
\hspace{0.2cm} \# \textit{matches} & 15.9 & 28.1 & 19.8 & 15.6 & 12.9 & 11.4 & 11.1 & 10.1 & 8.9 \\
\hspace{0.2cm} \% \textit{accepted offers} & 46.5 & 100.0 & 100.0 & 100.0 & 100.0 & 100.0 & 100.0 & 100.0 & 100.0 \\
\hspace{0.2cm} \# \textit{accepted $\mid$ to BP} & 15.1 & 28.1 & 19.8 & 15.6 & 12.9 & 11.4 & 11.1 & 10.1 & 8.9 \\
\hspace{0.2cm} \% \textit{repeated matches} & 19.3 & 17.1 & 9.6 & 5.1 & 2.0 & 1.0 & 0.8 & 0.4 & 0.1 \\
\hspace{0.2cm} \% \textit{repeated matchings} & 8.0 & 0.6 & 0.2 & 0.1 & 0.0 & 0.0 & 0.0 & 0.0 & 0.0 \\
\hspace{0.2cm} \# \textit{match-level cycles} & 4.2 & 8.3 & 3.3 & 1.5 & 0.5 & 0.2 & 0.2 & 0.1 & 0.0 \\
\hspace{0.2cm} \textit{avg.\ match-level cycle length} & 3.2 & 3.9 & 3.5 & 3.4 & 3.5 & 3.7 & 3.7 & 3.8 & 3.8 \\
\hspace{0.2cm} \% \textit{final matching is stable} & 94.3 & 100.0 & 100.0 & 100.0 & 100.0 & 100.0 & 100.0 & 100.0 & 100.0 \\
\hspace{0.2cm} \% \textit{final pairs are stable} & 99.6 & 100.0 & 100.0 & 100.0 & 100.0 & 100.0 & 100.0 & 100.0 & 100.0 \\
 &  &  &  &  &  &  &  &  &  \\
\multicolumn{3}{l}{\textit{5 stable matchings \& 3 stable partners}} &  &  &  &  &  &  &  \\
 &  &  &  &  &  &  &  &  &  \\
\hspace{0.2cm} \# \textit{offers} & 59.2 & 682.0 & 162.5 & 69.5 & 45.8 & 37.7 & 35.9 & 30.7 & 25.1 \\
\hspace{0.2cm} \# \textit{matches} & 24.6 & 682.0 & 162.5 & 69.5 & 45.8 & 37.7 & 35.9 & 30.7 & 25.1 \\
\hspace{0.2cm} \% \textit{accepted offers} & 40.4 & 100.0 & 100.0 & 100.0 & 100.0 & 100.0 & 100.0 & 100.0 & 100.0 \\
\hspace{0.2cm} \# \textit{accepted $\mid$ to BP} & 22.6 & 682.0 & 162.5 & 69.5 & 45.8 & 37.7 & 35.9 & 30.7 & 25.1 \\
\hspace{0.2cm} \% \textit{repeated matches} & 27.2 & 79.6 & 55.3 & 34.7 & 23.0 & 17.6 & 16.5 & 13.2 & 9.7 \\
\hspace{0.2cm} \% \textit{repeated matchings} & 6.2 & 1.3 & 1.6 & 1.1 & 0.9 & 0.8 & 0.8 & 0.6 & 0.3 \\
\hspace{0.2cm} \# \textit{match-level cycles} & 10.4 & 8.23e21 & 1.10e14 & 4714.4 & 33.4 & 19.3 & 17.2 & 11.5 & 6.1 \\
\hspace{0.2cm} \textit{avg.\ match-level cycle length} & 3.5 & 96.3 & 19.9 & 7.4 & 6.1 & 6.0 & 6.0 & 5.8 & 5.5 \\
\hspace{0.2cm} \% \textit{final matching is stable} & 75.0 & 100.0 & 100.0 & 100.0 & 100.0 & 100.0 & 100.0 & 100.0 & 100.0 \\
\hspace{0.2cm} \% \textit{final pairs are stable} & 96.9 & 100.0 & 100.0 & 100.0 & 100.0 & 100.0 & 100.0 & 100.0 & 100.0 \\
 &  &  &  &  &  &  &  &  &  \\
\multicolumn{3}{l}{\textit{\hspace{0.3cm}Market-level outcomes (matchings)}} &  &  &  &  &  &  &  \\
 &  &  &  &  &  &  &  &  &  \\
\hspace{0.5cm} \% \textit{median $\mid$ stable} & 80.0 & 39.4 & 40.7 & 48.4 & 56.2 & 59.3 & 59.5 & 59.1 & 58.6 \\
\hspace{0.5cm} \% \textit{non-extremal $\mid$ stable} & 100.0 & 77.5 & 80.6 & 82.5 & 84.2 & 84.0 & 83.4 & 81.7 & 78.8 \\
\hspace{0.5cm} \% \textit{food-optimal $\mid$ stable} & 0.0 & 10.8 & 5.0 & 2.7 & 1.1 & 0.6 & 0.5 & 0.5 & 0.4 \\
\hspace{0.5cm} \% \textit{color-optimal $\mid$ stable} & 0.0 & 11.6 & 14.4 & 14.8 & 14.8 & 15.4 & 16.1 & 17.8 & 20.8 \\
 &  &  &  &  &  &  &  &  &  \\
\multicolumn{3}{l}{\textit{\hspace{0.3cm}Individual-level outcomes (matches)}} &  &  &  &  &  &  &  \\
 &  &  &  &  &  &  &  &  &  \\
\hspace{0.5cm} \% \textit{median $\mid$ stable} & 76.8 & 58.5 & 60.6 & 65.4 & 70.2 & 71.6 & 71.5 & 70.4 & 68.7 \\
\hspace{0.5cm} \% \textit{fruit-optimal $\mid$ stable} & 2.9 & 10.8 & 5.0 & 2.7 & 1.1 & 0.6 & 0.5 & 0.5 & 0.4 \\
\hspace{0.5cm} \% \textit{color-optimal $\mid$ stable} & 20.4 & 30.7 & 34.3 & 31.9 & 28.8 & 27.8 & 28.0 & 29.1 & 30.9 \\
 &  &  &  &  &  &  &  &  &  \\\hline\hline
\multicolumn{10}{p{16.5cm}}{\scriptsize\textit{Notes.} The table reports averages across 10,000 simulations of every market in our main treatments using the Random Paths to Stability (RPS) algorithm of \cite{roth1990random}. The first column reports the experimental average for reference. Each column from the second reports the results using a distinct distribution to choose a blocking pair on each round. \textit{Uniform} corresponds to uniformly at random; \textit{Proportional} to probability proportional to $g_{f,c}$, and \textit{Exponential} to probability proportional to $\exp(\lambda g_{f,c})$, where $g_{f,c}$ refers to the total net gain of blocking pair $(f,c)$.}
\end{tabular}
\end{table}

\clearpage

\begin{table}[ht!]
\centering\scriptsize
\caption{Simulations---Random Best Response (RBR)
\label{tab:simul:RBR}\centering}
\begin{tabular}{lccccccccc}
 &  &  &  &  &  &  &  &  &  \\\hline\hline
& & & & \multicolumn{6}{c}{\scriptsize\textit{Exponential ($\lambda$)}} \\ 
 & 
{\scriptsize\textit{Experiment}} & 
{\scriptsize\textit{Uniform}} & 
{\scriptsize\textit{Proportional}} & 
0.001 & 
0.0025 & 
0.0035 & 
0.00425 & 
0.005 & 
0.0075 \\\hline
 &  &  &  &  &  &  &  &  &  \\
\multicolumn{3}{l}{\textit{Unique stable matching}} &  &  &  &  &  &  &  \\
 &  &  &  &  &  &  &  &  &  \\
\hspace{0.2cm} \# \textit{offers} & 44.8 & 19.8 & 17.5 & 19.1 & 18.4 & 18.0 & 17.8 & 17.5 & 16.9 \\
\hspace{0.2cm} \# \textit{matches} & 15.9 & 19.8 & 17.5 & 19.1 & 18.4 & 18.0 & 17.8 & 17.5 & 16.9 \\
\hspace{0.2cm} \% \textit{accepted offers} & 41.3 & 100.0 & 100.0 & 100.0 & 100.0 & 100.0 & 100.0 & 100.0 & 100.0 \\
\hspace{0.2cm} \# \textit{accepted $\mid$ to BP} & 15.2 & 19.8 & 17.5 & 19.1 & 18.4 & 18.0 & 17.8 & 17.5 & 16.9 \\
\hspace{0.2cm} \% \textit{repeated matches} & 14.0 & 10.9 & 4.0 & 8.9 & 6.6 & 5.4 & 4.7 & 4.0 & 2.6 \\
\hspace{0.2cm} \% \textit{repeated matchings} & 6.0 & 0.3 & 0.0 & 0.2 & 0.1 & 0.0 & 0.0 & 0.0 & 0.0 \\
\hspace{0.2cm} \# \textit{match-level cycles} & 2.5 & 3.8 & 1.3 & 3.0 & 2.1 & 1.7 & 1.5 & 1.3 & 0.8 \\
\hspace{0.2cm} \textit{avg.\ match-level cycle length} & 3.3 & 3.4 & 3.3 & 3.4 & 3.3 & 3.3 & 3.3 & 3.3 & 3.3 \\
\hspace{0.2cm} \% \textit{final matching is stable} & 90.0 & 100.0 & 100.0 & 100.0 & 100.0 & 100.0 & 100.0 & 100.0 & 100.0 \\
\hspace{0.2cm} \% \textit{final pairs are stable} & 94.7 & 100.0 & 100.0 & 100.0 & 100.0 & 100.0 & 100.0 & 100.0 & 100.0 \\
 &  &  &  &  &  &  &  &  &  \\
\multicolumn{3}{l}{\textit{Two embedded 4-by-4 markets}} &  &  &  &  &  &  &  \\
 &  &  &  &  &  &  &  &  &  \\
\hspace{0.2cm} \# \textit{offers} & 39.7 & 15.3 & 12.4 & 14.5 & 13.5 & 13.1 & 12.8 & 12.6 & 12.2 \\
\hspace{0.2cm} \# \textit{matches} & 15.9 & 15.3 & 12.4 & 14.5 & 13.5 & 13.1 & 12.8 & 12.6 & 12.2 \\
\hspace{0.2cm} \% \textit{accepted offers} & 46.5 & 100.0 & 100.0 & 100.0 & 100.0 & 100.0 & 100.0 & 100.0 & 100.0 \\
\hspace{0.2cm} \# \textit{accepted $\mid$ to BP} & 15.1 & 15.3 & 12.4 & 14.5 & 13.5 & 13.1 & 12.8 & 12.6 & 12.2 \\
\hspace{0.2cm} \% \textit{repeated matches} & 19.3 & 9.6 & 2.0 & 7.4 & 5.0 & 3.8 & 3.1 & 2.5 & 1.3 \\
\hspace{0.2cm} \% \textit{repeated matchings} & 8.0 & 0.2 & 0.0 & 0.1 & 0.0 & 0.0 & 0.0 & 0.0 & 0.0 \\
\hspace{0.2cm} \# \textit{match-level cycles} & 4.2 & 2.6 & 0.4 & 1.9 & 1.2 & 0.9 & 0.7 & 0.6 & 0.3 \\
\hspace{0.2cm} \textit{avg.\ match-level cycle length} & 3.2 & 3.2 & 3.1 & 3.2 & 3.2 & 3.1 & 3.1 & 3.1 & 3.1 \\
\hspace{0.2cm} \% \textit{final matching is stable} & 94.3 & 100.0 & 100.0 & 100.0 & 100.0 & 100.0 & 100.0 & 100.0 & 100.0 \\
\hspace{0.2cm} \% \textit{final pairs are stable} & 99.6 & 100.0 & 100.0 & 100.0 & 100.0 & 100.0 & 100.0 & 100.0 & 100.0 \\
 &  &  &  &  &  &  &  &  &  \\
\multicolumn{3}{l}{\textit{5 stable matchings \& 3 stable partners}} &  &  &  &  &  &  &  \\
 &  &  &  &  &  &  &  &  &  \\
\hspace{0.2cm} \# \textit{offers} & 59.2 & 145.0 & 32.9 & 89.7 & 54.2 & 43.3 & 38.0 & 34.3 & 27.6 \\
\hspace{0.2cm} \# \textit{matches} & 24.6 & 145.0 & 32.9 & 89.7 & 54.2 & 43.3 & 38.0 & 34.3 & 27.6 \\
\hspace{0.2cm} \% \textit{accepted offers} & 40.4 & 100.0 & 100.0 & 100.0 & 100.0 & 100.0 & 100.0 & 100.0 & 100.0 \\
\hspace{0.2cm} \# \textit{accepted $\mid$ to BP} & 22.6 & 145.0 & 32.9 & 89.7 & 54.2 & 43.3 & 38.0 & 34.3 & 27.6 \\
\hspace{0.2cm} \% \textit{repeated matches} & 27.2 & 58.7 & 19.3 & 47.6 & 34.1 & 27.7 & 24.0 & 20.9 & 14.5 \\
\hspace{0.2cm} \% \textit{repeated matchings} & 6.2 & 1.0 & 0.5 & 0.8 & 0.7 & 0.6 & 0.5 & 0.5 & 0.4 \\
\hspace{0.2cm} \# \textit{match-level cycles} & 10.4 & 2.61e07 & 14.9 & 695.3 & 52.9 & 29.5 & 21.0 & 16.3 & 8.9 \\
\hspace{0.2cm} \textit{avg.\ match-level cycle length} & 3.5 & 12.2 & 4.0 & 7.6 & 5.1 & 4.5 & 4.2 & 4.1 & 3.8 \\
\hspace{0.2cm} \% \textit{final matching is stable} & 75.0 & 100.0 & 100.0 & 100.0 & 100.0 & 100.0 & 100.0 & 100.0 & 100.0 \\
\hspace{0.2cm} \% \textit{final pairs are stable} & 96.9 & 100.0 & 100.0 & 100.0 & 100.0 & 100.0 & 100.0 & 100.0 & 100.0 \\
 &  &  &  &  &  &  &  &  &  \\
\multicolumn{3}{l}{\textit{\hspace{0.3cm}Market-level outcomes (matchings)}} &  &  &  &  &  &  &  \\
 &  &  &  &  &  &  &  &  &  \\
\hspace{0.5cm} \% \textit{median $\mid$ stable} & 80.0 & 46.2 & 56.1 & 50.7 & 54.2 & 55.0 & 55.2 & 55.0 & 54.0 \\
\hspace{0.5cm} \% \textit{non-extremal $\mid$ stable} & 100.0 & 79.1 & 83.5 & 82.3 & 84.1 & 84.0 & 83.6 & 82.9 & 80.1 \\
\hspace{0.5cm} \% \textit{food-optimal $\mid$ stable} & 0.0 & 8.1 & 3.3 & 5.8 & 4.2 & 3.9 & 3.7 & 3.8 & 4.2 \\
\hspace{0.5cm} \% \textit{color-optimal $\mid$ stable} & 0.0 & 12.9 & 13.2 & 11.9 & 11.6 & 12.1 & 12.7 & 13.4 & 15.7 \\
 &  &  &  &  &  &  &  &  &  \\
\multicolumn{3}{l}{\textit{\hspace{0.3cm}Individual-level outcomes (matches)}} &  &  &  &  &  &  &  \\
 &  &  &  &  &  &  &  &  &  \\
\hspace{0.5cm} \% \textit{median $\mid$ stable} & 76.8 & 62.6 & 69.8 & 66.5 & 69.2 & 69.5 & 69.4 & 68.9 & 67.1 \\
\hspace{0.5cm} \% \textit{fruit-optimal $\mid$ stable} & 2.9 & 8.1 & 3.3 & 5.8 & 4.2 & 3.9 & 3.7 & 3.8 & 4.2 \\
\hspace{0.5cm} \% \textit{color-optimal $\mid$ stable} & 20.4 & 29.3 & 26.9 & 27.7 & 26.6 & 26.6 & 26.9 & 27.3 & 28.8 \\
 &  &  &  &  &  &  &  &  &  \\\hline\hline
\multicolumn{10}{p{17cm}}{\scriptsize\textit{Notes.} The table reports averages across 10,000 simulations of every market in our main treatments using the the Random Best Response (RBR) algorithm  of \cite{ackermann2011uncoordinated}. The first column reports the experimental averages for reference. Each column from the second reports the results using a distinct distribution to choose an agent among the ones with at least one blocking partner on each round. \textit{Uniform} corresponds to uniformly at random; \textit{Proportional} to probability proportional to $g_{a}$, and \textit{Exponential} to probability proportional to $\exp(\lambda g_{a})$, where $g_{a}$ refers to $a$'s maximum net gain among all their blocking partners.}
\end{tabular}
\end{table}

\clearpage


{\small

\paragraph{Predictive measures.} Let $y_{ij}=1\{\text{alternative $j$ is chosen in choice $i$}\}$, where $i=1,\dots,N$, with $N$ equal to the number of choices in the sample, and $j=1,\dots,J$, with $J$ equal to the number of alternatives in every choice. Denote the choices in the sample as $S=\{1,\dots,N\}$. We specify a parametric model for $P(y_{ij}=1\mid x_{ij},\beta)$, where $x_{ij}$ is a vector of covariates associated to alternative $j$ in choice $i$, and $\beta$ is an unknown parameter vector. To evaluate the fit of the model, we define a \textit{training dataset} $D\subset S$ and a \textit{test dataset} $T\subset S$. Let $N_T=|T|$, where $N_T\leq N$. Given the training data $D$, we estimate the parameter vector $\hat{\beta}(D)$ and compute the predicted choice probabilities $\hat{y}_{ij}(D)=P(y_{ij}=1\mid x_{ij},\hat{\beta}(D))$. We use the following measures to evaluate the fit of the estimated model on the test data: (i) the mean-squared error of the predicted choice probabilities (\textit{MSE}), (ii) the percentage of choices in which the predicted probability of the alternative chosen in the data is the greatest among all alternatives ($\%CorrMaxCP$), and (iii) the average probability of correctly predicting the data ($Avg\,\mathbb{P}(OK\,Pred)$), respectively given by:
\begin{align*}
    MSE\,(T\mid D) &= \frac{1}{N_T\times J}\sum_{i\in T}\sum_{j=1}^{J}(y_{ij}-\hat{y}_{ij}(D))^2\\
    \%CorrMaxCP\,(T\mid D) &= \frac{1}{N_T}\sum_{i\in T}\sum_{j=1}^{J}1\{j=\text{arg max}_j \hat{y}_{ij}(D)\}\cdot 1\{y_{ij}=1\}\\
    Avg\,\mathbb{P}(OK\,Pred)\,(T\mid D) &= \frac{1}{N_T}\sum_{i\in T}\sum_{j=1}^{J}(\hat{y}_{ij}(D))^{y_{ij}}
\end{align*}
We evaluate the in-sample fit of the model with the following: 
\begin{align*}
    MSE\,(sample) &= MSE\,(S\mid S)\\
    \%CorrMaxCP\,(sample) &= \%CorrMaxCP\,(S\mid S)\\
    Avg\,\mathbb{P}(OK\,Pred)\,(sample) &= Avg\,\mathbb{P}(OK\,Pred)\,(S\mid S)
\end{align*}
To evaluate the fit of the model out of the sample, we compute the same measures in three distinct ways. First, we use two-fold cross validation, which consists in partitioning the choices in the sample into two sets of equal size uniformly at random, denoted by $I_1$ and $I_2$, then using both as training and test datasets and averaging the resulting measures. That is,
\begin{align*}
    MSE\,(\text{\textit{2-fold $\times$ valid}}) &= [MSE\,(I_1\mid I_2)+MSE\,(I_2\mid I_1)]/2\\
    \%CorrMaxCP\,(\text{\textit{2-fold $\times$ valid}}) &= [\%CorrMaxCP\,(I_1\mid I_2)+\%CorrMaxCP\,(I_2\mid I_1)]/2\\
    Avg\,\mathbb{P}(OK\,Pred)\,(\text{\textit{2-fold $\times$ valid}}) &= [Avg\,\mathbb{P}(OK\,Pred)\,(I_1\mid I_2)+ Avg\,\mathbb{P}(OK\,Pred)\,(I_2\mid I_1)]/2
\end{align*}
Second, we split the sample into the choices made within the first five rounds of each session, $S_1$, and those made during the final five rounds, $S_2$. We evaluate the fit of the model in the final rounds, using the data of the first rounds:
\begin{align*}
    MSE\,(\text{\textit{future $|$ present}}) &= MSE\,(S_2\mid S_1)\\
    \%CorrMaxCP\,(\text{\textit{future $|$ present}}) &= \%CorrMaxCP\,(S_2\mid S_1)\\
    Avg\,\mathbb{P}(OK\,Pred)\,(\text{\textit{future $|$ present}}) &= Avg\,\mathbb{P}(OK\,Pred)\,(S_2\mid S_1)
\end{align*}
Finally, we also report the fit of the model in the first rounds using the final rounds as training set:
\begin{align*}
    MSE\,(\text{\textit{present $|$ future}}) &= MSE\,(S_1\mid S_2)\\
    \%CorrMaxCP\,(\text{\textit{present $|$ future}}) &= \%CorrMaxCP\,(S_1\mid S_2)\\
    Avg\,\mathbb{P}(OK\,Pred)\,(\text{\textit{present $|$ future}}) &= Avg\,\mathbb{P}(OK\,Pred)\,(S_1\mid S_2)
\end{align*}

}

\begin{table}
\centering\scriptsize
\caption{Proposal conditional logit estimations: ordinal vs.\ cardinal payoffs
\label{final_table_7a}}
\begin{tabular}{lcccccc}
 &  &  &  &  &  & \\\hline\hline
  &  &  &  &  &  & \\
 & (1) & (2) & (3) & (4) & (5) & (6)\\\hline
 &  &  &  &  &  & \\
\textit{Proposer's PA $> 0$}                    &       0.248***&       0.228***&       0.261***&       0.253***&       0.229***&       0.184***\\
                                   		&     (0.036)   &     (0.043)   &     (0.053)   &     (0.075)   &     (0.053)   &     (0.063)   \\
\textit{Receiver's rank (in prop's list)}       &      -0.074***&      -0.075***&      -0.080***&      -0.087***&      -0.038** &      -0.042** \\
                                   		&     (0.001)   &     (0.002)   &     (0.005)   &     (0.004)   &     (0.018)   &     (0.018)   \\
\textit{$\max\{\text{Prop's PA},0\}$}           &      -0.008   &      -0.015   &               &       0.056***&       0.066***\\
                                   		&               &     (0.014)   &     (0.015)   &               &     (0.009)   &     (0.013)   \\
\textit{$\min\{\text{Prop's PA},0\}$}           &               &               &      -0.038   &               &               &       0.104   \\
                                   		&               &               &     (0.040)   &               &               &     (0.125)   \\
  &  &  &  &  &  & \\\hline
 &  &  &  &  &  & \\
Treatment                          		&        Main   &        Main   &        Main   &  Unilateral   &  Unilateral   &  Unilateral   \\
  &  &  &  &  &  & \\
Observations                                  	&      29,760   &      29,760   &      29,760   &       7,246   &       7,246   &       7,246   \\
  &  &  &  &  &  & \\
Adj. $R^2$                            	&       0.316   &       0.316   &       0.317   &       0.323   &       0.329   &       0.330   \\
  &  &  &  &  &  & \\
$MSE$ \textit{(sample)}                    	&       8.628   &       8.627   &       8.648   &       8.840   &       8.809   &       8.803   \\
$MSE$ \textit{(2-fold $\times$ valid)}          &       8.630   &       8.632   &       8.658   &       8.845   &       8.818   &       8.823   \\
$MSE$ \textit{(future $|$ present)}             &       8.209   &       8.211   &       8.251   &       7.928   &       7.846   &       7.842   \\
$MSE$ \textit{(present $|$ future)}             &       9.242   &       9.312   &       9.366   &      10.508   &      10.234   &      10.203   \\
  &  &  &  &  &  & \\
$\%CorrMaxCP$ \textit{(sample)}             	&      45.380   &      45.380   &      45.380   &      40.538   &      40.538   &      40.538   \\
$\%CorrMaxCP$ \textit{(2-fold $\times$ valid)}  &      45.380   &      45.380   &      45.380   &      40.538   &      40.538   &      40.538   \\
$\%CorrMaxCP$ \textit{(future $|$ present)}     &      48.800   &      48.800   &      48.800   &      50.696   &      50.696   &      50.696   \\
$\%CorrMaxCP$ \textit{(present $|$ future)}     &      42.620   &      42.343   &      42.113   &      34.151   &      34.151   &      34.151   \\
  &  &  &  &  &  & \\
$Avg\,\mathbb{P}(OK\,Pred)$ \textit{(sample)}         		&      34.591   &      34.586   &      34.542   &      32.233   &      32.629   &      32.651   \\
$Avg\,\mathbb{P}(OK\,Pred)$ \textit{(2-fold $\times$ valid)}    &      34.571   &      34.571   &      34.516   &      32.223   &      32.607   &      32.684   \\
$Avg\,\mathbb{P}(OK\,Pred)$ \textit{(future $|$ present)}  	&      33.656   &      33.659   &      33.414   &      32.841   &      33.686   &      33.808   \\
$Avg\,\mathbb{P}(OK\,Pred)$ \textit{(present $|$ future)}  	&      35.995   &      35.916   &      35.951   &      32.122   &      31.590   &      30.754   \\
 &  &  &  &  &  & \\\hline\hline
\multicolumn{7}{p{14.5cm}}{\scriptsize\textit{Notes.} The table reports average marginal effects of conditional logits. The response variable indicates the receiver of every offer in the data. Standard errors are clustered at participant level. *, **, and *** stand for 90\%, 95\%, and 99\% confidence levels, respectively. The table also reports the mean-squared error (\textit{MSE}) of the predicted choice probability, percentage of choices in which the predicted probability of the alternative chosen in the data is the greatest among all alternatives ($\%CorrMaxCP$), and the average probability of correctly predicting the data ($Avg\,\mathbb{P}(OK\,Pred)$). Each is computed in the estimation sample and out of the sample using: random two-fold cross-validation, predicting the final five rounds with the first five rounds, and the first five rounds using the final five. See the Appendix for more details.}
\end{tabular}
\end{table}

\clearpage

\begin{table}
\vspace{-1cm}
\centering\scriptsize
\caption{Proposal conditional logits estimations with proposer's cardinal payoffs
\label{final_table_7b}}
\begin{tabular}{lcccccc}
 &  &  &  &  &  &  \\\hline\hline
 &  &  &  &  &  &  \\
 & (1) & (2) & (3) & (4) & (5) & (6) \\\hline
 &  &  &  &  &  &  \\
\textit{Proposer's PA $> 0$}                    &       0.162***&       0.163***&       0.168***&       0.060*  &       0.065*  &       0.082***\\
                                   		&     (0.023)   &     (0.019)   &     (0.017)   &     (0.034)   &     (0.034)   &     (0.031)   \\
\textit{$\max\{\text{Prop's PA},0\}$}           &       0.063***&       0.059***&       0.042***&       0.059***&       0.057***&       0.042***\\
                                   		&     (0.004)   &     (0.004)   &     (0.003)   &     (0.009)   &     (0.011)   &     (0.008)   \\
\textit{$\min\{\text{Prop's PA},0\}$}           &       0.017   &       0.002   &      -0.005   &       0.034   &       0.032   &       0.022   \\
                                   		&     (0.016)   &     (0.012)   &     (0.012)   &     (0.040)   &     (0.034)   &     (0.036)   \\
\textit{Receiver is matched}                    &       0.035***&       0.004   &       0.010   &       0.022** &      -0.000   &       0.003   \\
                                   		&     (0.013)   &     (0.010)   &     (0.011)   &     (0.009)   &     (0.009)   &     (0.011)   \\
\textit{Are blocking pair (BP)}                 &       0.053***&       0.063***&       0.064***&       0.055***&       0.049***&       0.053***\\
                                   		&     (0.011)   &     (0.011)   &     (0.012)   &     (0.011)   &     (0.013)   &     (0.015)   \\
\textit{$\max\{\text{Rec's PA},0\}$}            &       0.014***&       0.007***&       0.011***&       0.011***&       0.006** &       0.010***\\
                                   		&     (0.003)   &     (0.002)   &     (0.003)   &     (0.003)   &     (0.003)   &     (0.003)   \\
\textit{$\min\{\text{Rec's PA},0\}$}            &       0.038***&       0.030***&       0.024***&       0.018** &       0.010   &       0.008   \\
                                   		&     (0.005)   &     (0.005)   &     (0.005)   &     (0.009)   &     (0.008)   &     (0.009)   \\
\textit{Matched previously}                     &               &       0.024***&       0.035***&               &       0.001   &       0.015***\\
                                   		&               &     (0.006)   &     (0.007)   &               &     (0.008)   &     (0.006)   \\
\textit{\# previous offers (prop.\ to rec.)}    &               &       0.010***&       0.013***&               &      -0.009   &       0.004   \\
                                   		&               &     (0.002)   &     (0.003)   &               &     (0.010)   &     (0.004)   \\
\textit{Are stable partners (SP)}               &               &       0.059***&       0.061***&               &       0.024***&       0.035***\\
                                   		&               &     (0.004)   &     (0.006)   &               &     (0.006)   &     (0.009)   \\
\textit{Offer is downward}                      &               &               &       0.041***&               &               &      -0.001   \\
                                   		&               &               &     (0.015)   &               &               &     (0.008)   \\
\textit{Offer is Gale-Shapley}                  &               &               &       0.050***&               &               &       0.040***\\
                                   		&               &               &     (0.018)   &               &               &     (0.009)   \\
\textit{Offer skips someone}                    &               &               &      -0.042*  &               &               &      -0.020   \\
                                   		&               &               &     (0.023)   &               &               &     (0.013)   \\
  &  &  &  &  &  &  \\\hline
  &  &  &  &  &  &  \\
Treatment                          		&        Main   &        Main   &        Main   &  Unilateral   &  Unilateral   &  Unilateral   \\
 &  &  &  &  &  &  \\
Observations                                  	&      29,760   &      29,760   &      29,760   &       7,246   &       7,246   &       7,246   \\
 &  &  &  &  &  &  \\
Adj. $R^2$                            		&       0.369   &       0.407   &       0.438   &       0.449   &       0.464   &       0.504   \\
 &  &  &  &  &  &  \\
$MSE$ \textit{(sample)}                    	&       7.918   &       7.570   &       7.228   &       7.051   &       6.887   &       6.337   \\
$MSE$ \textit{(2-fold $\times$ valid)}          &       7.938   &       7.603   &       7.257   &       7.088   &       6.969   &       6.408   \\
$MSE$ \textit{(future $|$ present)}             &       7.651   &       7.344   &       6.849   &       5.721   &       5.646   &       4.814   \\
$MSE$ \textit{(present $|$ future)}             &       8.212   &       7.989   &       7.768   &       8.471   &       8.745   &       8.074   \\
 &  &  &  &  &  &  \\
$\%CorrMaxCP$ \textit{(sample)}             	&      55.564   &      56.202   &      58.933   &      61.075   &      59.892   &      63.011   \\
$\%CorrMaxCP$ \textit{(2-fold $\times$ valid)}  &      55.360   &      56.023   &      58.882   &      60.860   &      58.602   &      62.688   \\
$\%CorrMaxCP$ \textit{(future $|$ present)}     &      57.371   &      57.714   &      62.171   &      73.538   &      68.524   &      76.323   \\
$\%CorrMaxCP$ \textit{(present $|$ future)}     &      53.506   &      52.906   &      54.797   &      49.037   &      50.263   &      53.765   \\
 &  &  &  &  &  &  \\
$Avg\,\mathbb{P}(OK\,Pred)$ \textit{(sample)}         		&      39.543   &      42.742   &      45.714   &      44.270   &      45.503   &      50.302   \\
$Avg\,\mathbb{P}(OK\,Pred)$ \textit{(2-fold $\times$ valid)}    &      39.466   &      42.654   &      45.646   &      44.184   &      45.249   &      50.054   \\
$Avg\,\mathbb{P}(OK\,Pred)$ \textit{(future $|$ present)}  	&      40.533   &      43.475   &      46.887   &      50.625   &      52.281   &      58.977   \\
$Avg\,\mathbb{P}(OK\,Pred)$ \textit{(present $|$ future)}  	&      39.386   &      41.865   &      44.375   &      40.090   &      40.024   &      43.191   \\
 &  &  &  &  &  &  \\\hline\hline
\multicolumn{7}{p{14.5cm}}{\scriptsize\textit{Notes.} The table reports average marginal effects of conditional logits. The response variable indicates the receiver of every offer in the data. Standard errors are clustered at participant level. *, **, and *** stand for 90\%, 95\%, and 99\% confidence levels, respectively. The table also reports the mean-squared error (\textit{MSE}) of the predicted choice probability, percentage of choices in which the predicted probability of the alternative chosen in the data is the greatest among all alternatives ($\%CorrMaxCP$), and the average probability of correctly predicting the data ($Avg\,\mathbb{P}(OK\,Pred)$). Each is computed in the estimation sample and out of the sample using: random two-fold cross-validation, predicting the final five rounds with the first five rounds, and the first five rounds using the final five. See the Appendix for more details.}
\end{tabular}
\end{table}

\clearpage

\begin{table}
\centering\scriptsize
\caption{Acceptance binary logit estimations with fixed effects
\label{final_table_8_FEs}}
\begin{tabular}{lcccccc}
 &  &  &  &  &  &  \\\hline\hline
  &  &  &  &  &  &  \\
 & (1) & (2) & (3) & (4) & (5) & (6) \\\hline
 &  &  &  &  &  &  \\
\textit{Receiver's PA $> 0$}       		&       0.522***&       0.522***&       0.472***&       0.505***&       0.508***&       0.556***\\
                                   		&     (0.034)   &     (0.034)   &     (0.028)   &     (0.103)   &     (0.104)   &     (0.115)   \\
\textit{Receiver is matched}       		&       0.167***&       0.166***&       0.149***&       0.091*  &       0.093*  &       0.002   \\
                                   		&     (0.034)   &     (0.034)   &     (0.030)   &     (0.051)   &     (0.052)   &     (0.058)   \\
\textit{$\max\{\text{Rec's PA},0\}$}		&       0.029***&       0.029***&       0.028***&       0.037** &       0.037** &       0.017   \\
                                   		&     (0.009)   &     (0.009)   &     (0.008)   &     (0.018)   &     (0.018)   &     (0.017)   \\
\textit{$\min\{\text{Rec's PA},0\}$}    	&       0.007   &       0.008   &      -0.011   &       0.452** &       0.449** &       0.397** \\
                                   		&     (0.026)   &     (0.026)   &     (0.017)   &     (0.201)   &     (0.202)   &     (0.165)   \\
\textit{Proposer's rank (in rec's list)}	&      -0.042***&      -0.042***&      -0.020***&      -0.063***&      -0.064***&      -0.062***\\
                                   		&     (0.004)   &     (0.004)   &     (0.004)   &     (0.008)   &     (0.008)   &     (0.009)   \\
\textit{Proposer's PA $> 0$}            	&               &      -0.011   &      -0.053** &               &       0.115   &       0.143   \\
                                   		&               &     (0.029)   &     (0.026)   &               &     (0.144)   &     (0.112)   \\
\textit{Matched previously}             	&               &               &       0.073***&               &               &       0.137** \\
                                   		&               &               &     (0.018)   &               &               &     (0.069)   \\
\textit{\# previous offers (prop.\ to rec.)}	&               &               &      -0.034***&               &               &      -0.057***\\
                                   		&               &               &     (0.010)   &               &               &     (0.018)   \\
\textit{\# previous offers (total to rec.)}     &       &        &      -0.005   &               &               &       0.030***\\
                                   		&               &               &     (0.004)   &               &               &     (0.010)   \\
\textit{Are stable partners (SP)}               &               &               &       0.132***&               &               &       0.048   \\
                                   		&               &               &     (0.013)   &               &               &     (0.030)   \\
 &  &  &  &  &  &  \\\hline
 &  &  &  &  &  &  \\
Treatment                          		&        Main   &        Main   &        Main   &  Unilateral   &  Unilateral   &  Unilateral   \\
 &  &  &  &  &  &  \\
Observations                                  	&       3,919   &       3,919   &       3,919   &         877   &         877   &         877   \\
Participant fixed effects                    	&           1   &           1   &           1   &           1   &           1   &           1   \\
 &  &  &  &  &  &  \\
Adj. $R^2$                            		&       0.455   &       0.455   &       0.484   &       0.495   &       0.496   &       0.515   \\
 &  &  &  &  &  &  \\
$MSE$ \textit{(sample)}                    	&      11.560   &      11.558   &      10.919   &      10.673   &      10.682   &      10.195   \\
$MSE$ \textit{(2-fold $\times$ valid)}          &      13.846   &      13.855   &      13.411   &      14.465   &      14.451   &      14.961   \\
$MSE$ \textit{(future $|$ present)}             &      15.147   &      15.133   &      14.073   &      12.684   &      12.621   &      13.141   \\
$MSE$ \textit{(present $|$ future)}             &      14.607   &      14.606   &      13.700   &      15.936   &      16.112   &      18.741   \\
 &  &  &  &  &  &  \\
$\%CorrMaxCP$ \textit{(sample)}             	&      83.669   &      83.695   &      84.307   &      86.129   &      85.914   &      86.022   \\
$\%CorrMaxCP$ \textit{(2-fold $\times$ valid)}  &      79.663   &      79.714   &      80.454   &      79.032   &      79.247   &      78.065   \\
$\%CorrMaxCP$ \textit{(future $|$ present)}     &      78.070   &      77.955   &      79.098   &      80.780   &      81.337   &      82.173   \\
$\%CorrMaxCP$ \textit{(present $|$ future)}     &      77.814   &      77.814   &      79.336   &      72.154   &      71.278   &      69.177   \\
 &  &  &  &  &  &  \\
$Avg\,\mathbb{P}(OK\,Pred)$ \textit{(sample)}         		&      76.775   &      76.778   &      78.049   &      78.557   &      78.561   &      79.495   \\
$Avg\,\mathbb{P}(OK\,Pred)$ \textit{(2-fold $\times$ valid)}    &      75.265   &      75.260   &      76.410   &      75.393   &      75.559   &      75.703   \\
$Avg\,\mathbb{P}(OK\,Pred)$ \textit{(future $|$ present)}  	&      72.916   &      72.937   &      74.339   &      78.208   &      78.349   &      78.227   \\
$Avg\,\mathbb{P}(OK\,Pred)$ \textit{(present $|$ future)}  	&      75.702   &      75.702   &      77.238   &      74.558   &      74.258   &      72.607   \\
&  &  &  &  &  &  \\\hline\hline
\multicolumn{7}{p{14.5cm}}{\scriptsize\textit{Notes.} The table reports average marginal effects of binary logits with participant fixed effects. The response variable is an indicator of whether an offer was accepted. Standard errors are clustered at participant level. *, **, and *** stand for significantly different to zero at a 90\%, 95\%, and 99\% confidence level, respectively. The table also reports the mean-squared error (\textit{MSE}) of the predicted probability, percentage of choices in which the predicted probability of the alternative chosen in the data is the greatest among all alternatives ($\%CorrMaxCP$), and the average probability of correctly predicting the data ($Avg\,\mathbb{P}(OK\,Pred)$). Each is computed in the estimation sample and out of the sample using: random two-fold cross-validation, predicting the final five rounds with the first five rounds, and the first five rounds using the final five. See the Appendix for more details.}
\end{tabular}
\end{table}

\clearpage

\begin{figure}[h]
	\centering
	\includegraphics[width=0.49\textwidth]{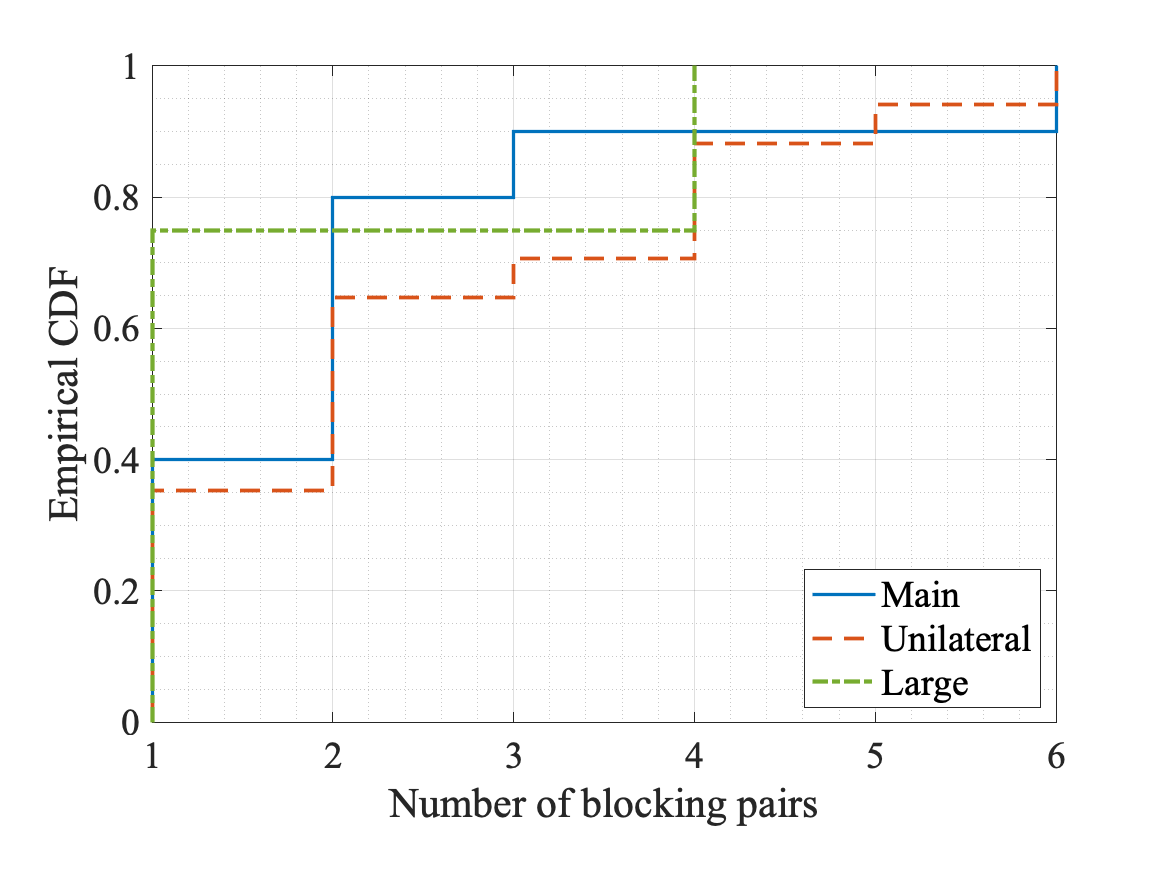}
	\includegraphics[width=0.49\textwidth]{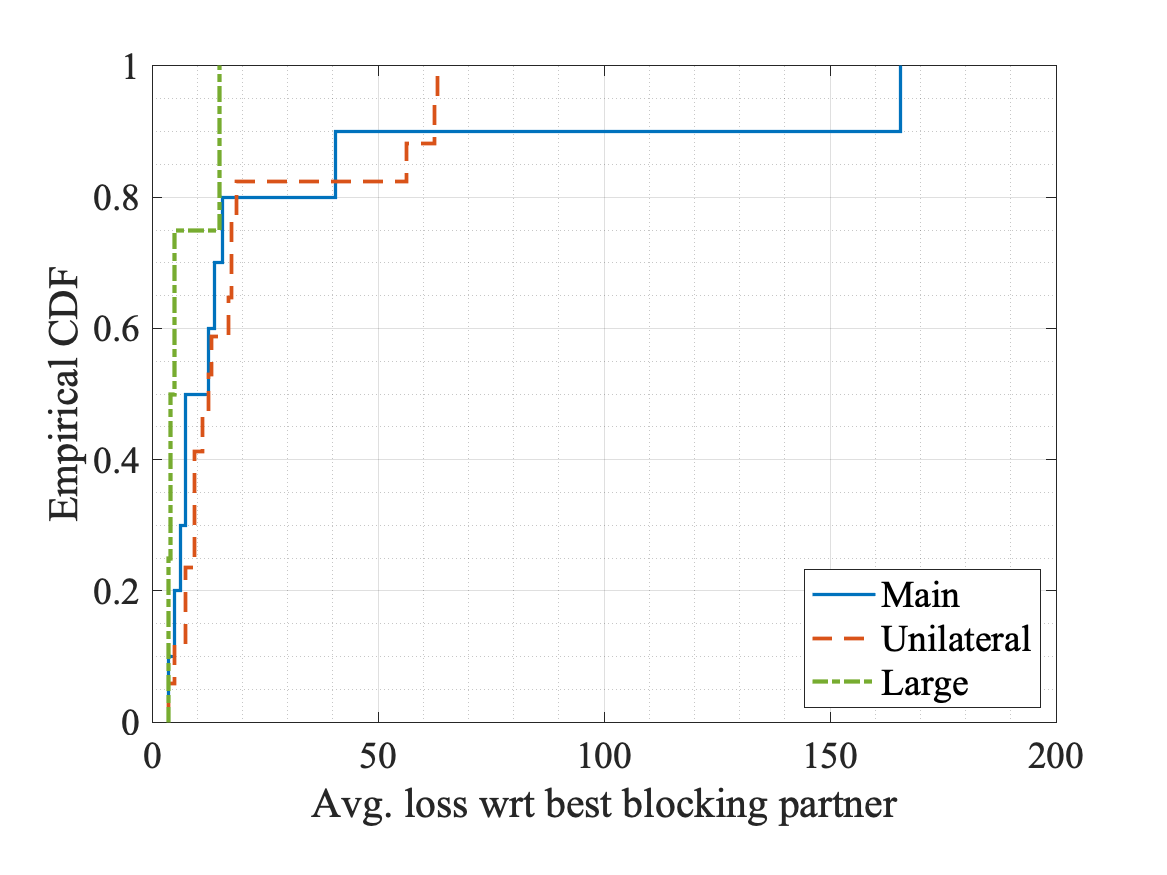}
	
	\vspace{0.5cm}
	
	\includegraphics[width=0.49\textwidth]{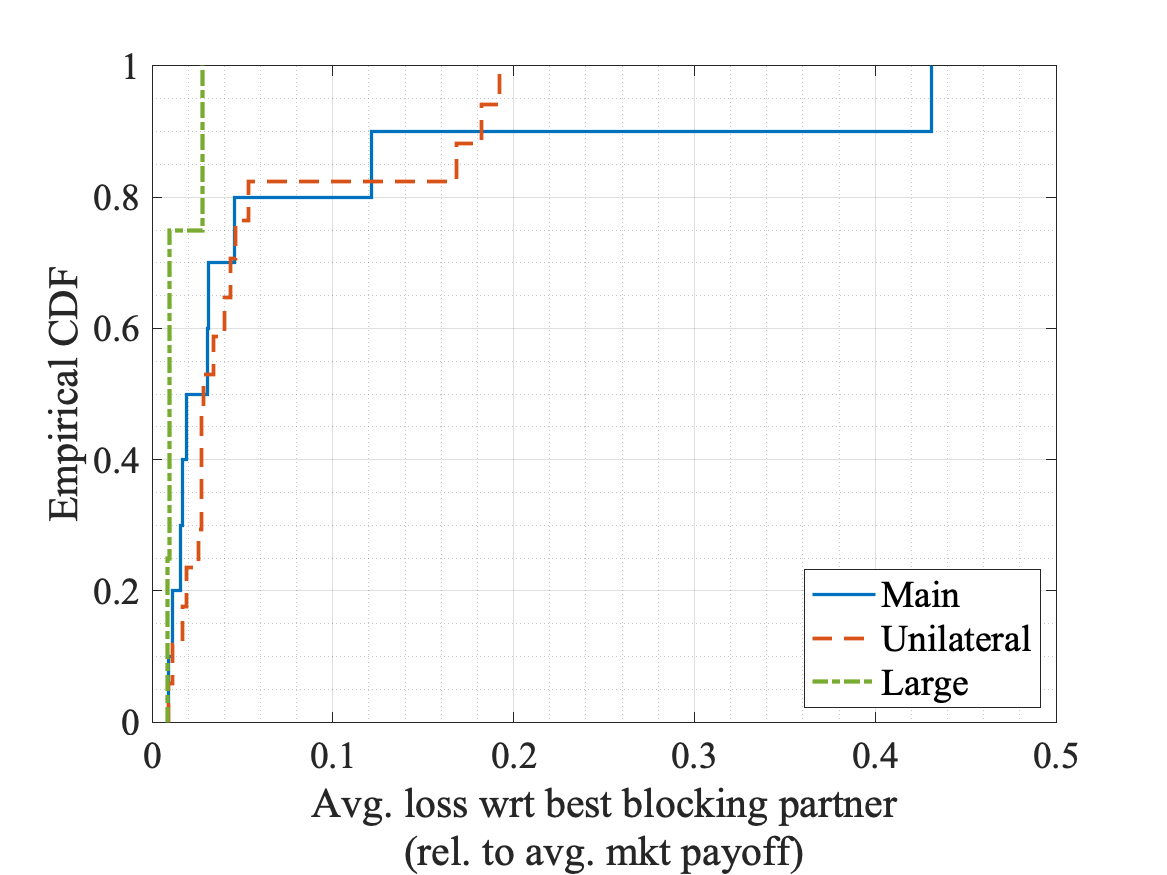}
	\includegraphics[width=0.49\textwidth]{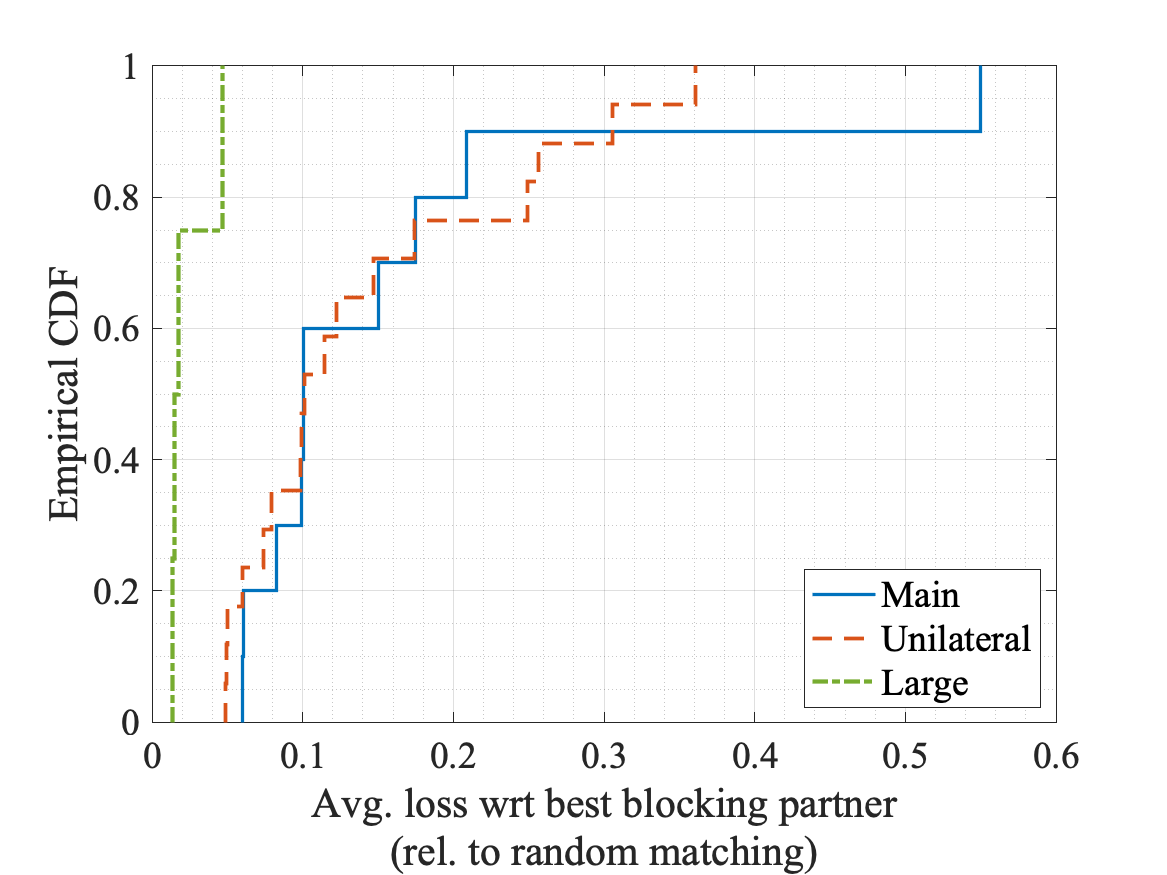}
	\caption{Distance to stability in unstable markets (alternative measures) 
    \label{fig_final_A1}}
\end{figure}

\clearpage

\begin{figure}[ht!]
	\centering
	
    \includegraphics[width=0.49\textwidth]{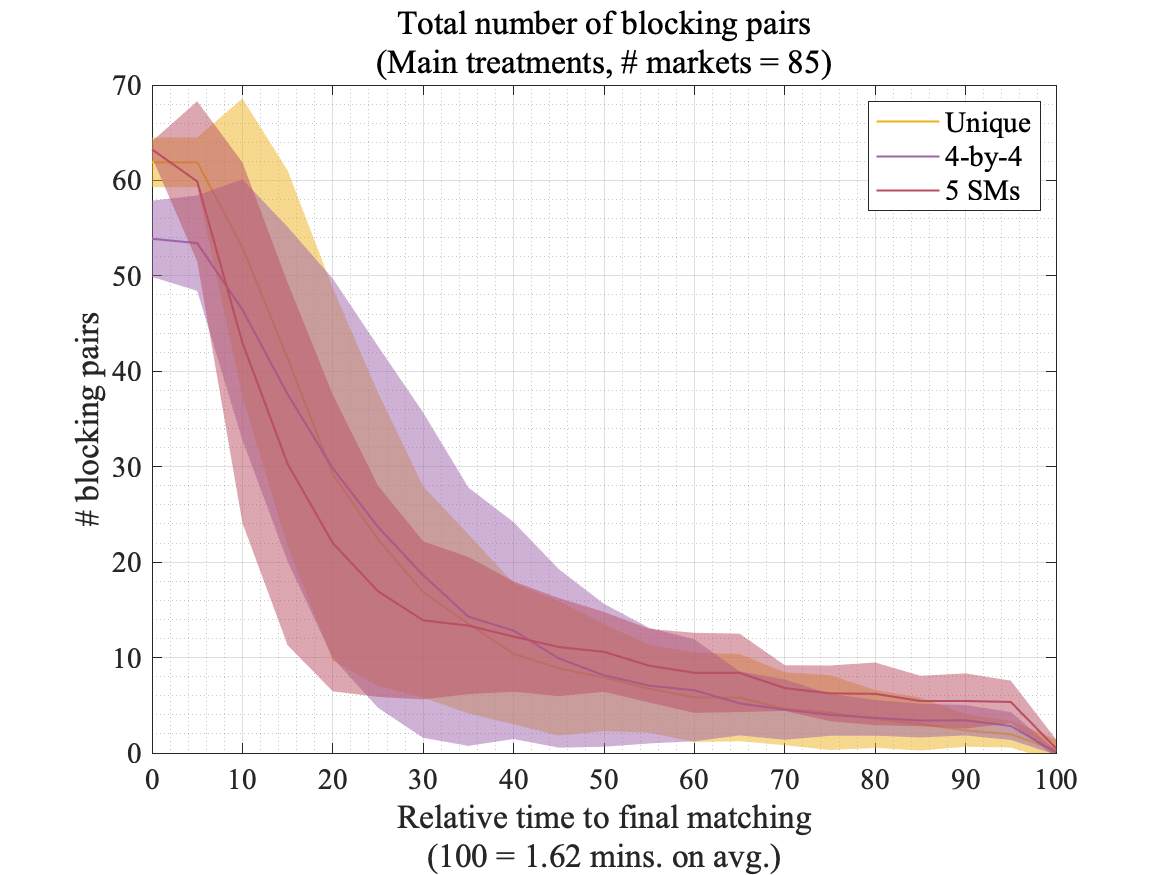}
    \includegraphics[width=0.49\textwidth]{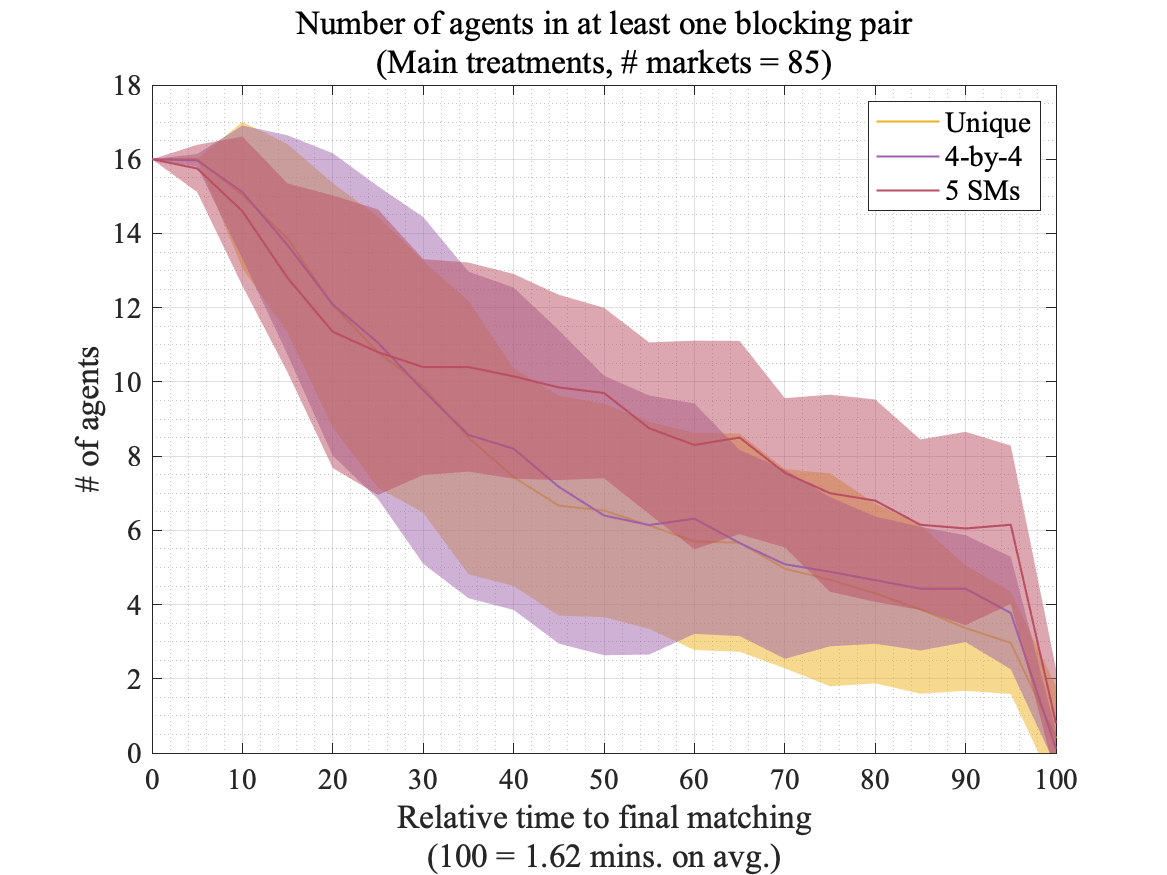}
    
    \vspace{0.75cm}

    \includegraphics[width=0.49\textwidth]{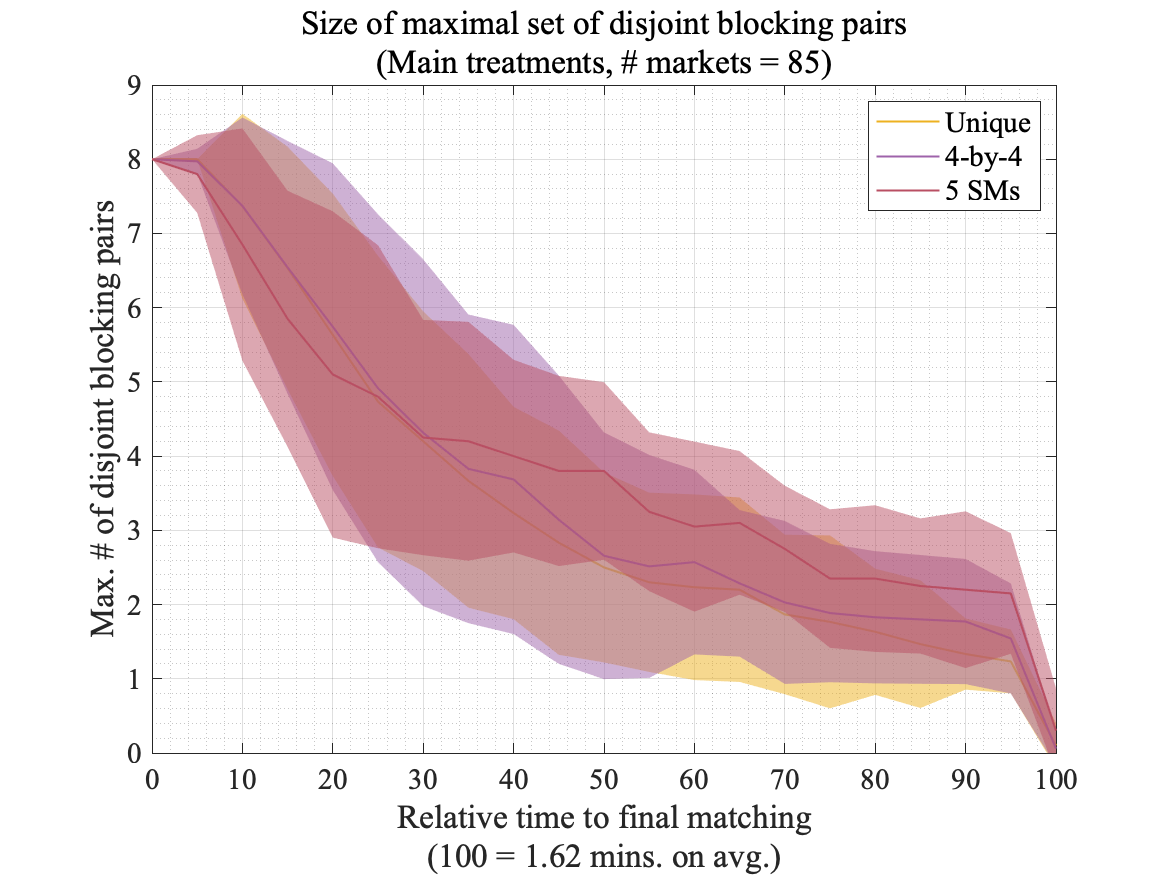}
    \includegraphics[width=0.49\textwidth]{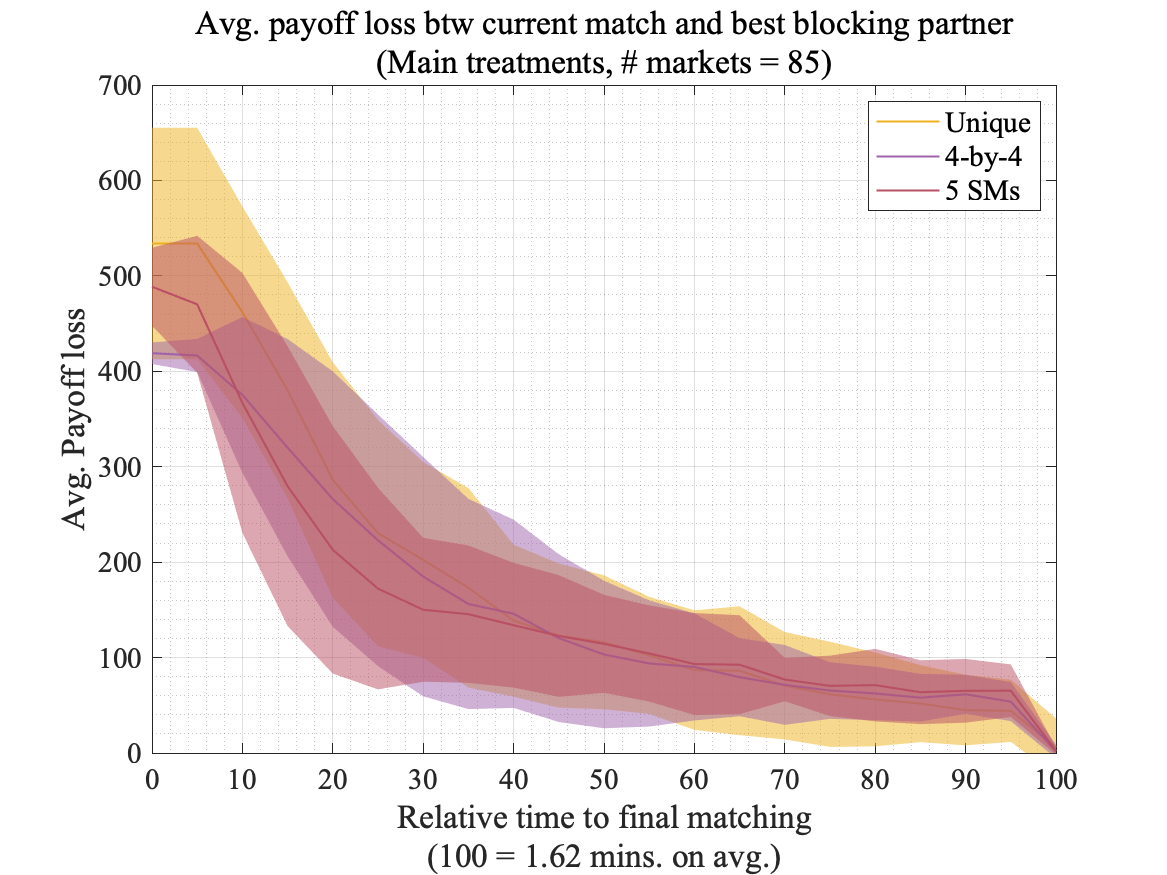}

    \vspace{0.25cm}
    
    \caption{Distance to stable matching over time (by each main treatment) 
    \label{Dist2SMoverTime_more}}
\end{figure}

\clearpage

\renewcommand{\wdth}{0.35\textwidth}
\begin{landscape}
\begin{figure}[h!]
    \centering
    \subfloat[Main: unique]
    {\includegraphics[width=\wdth]{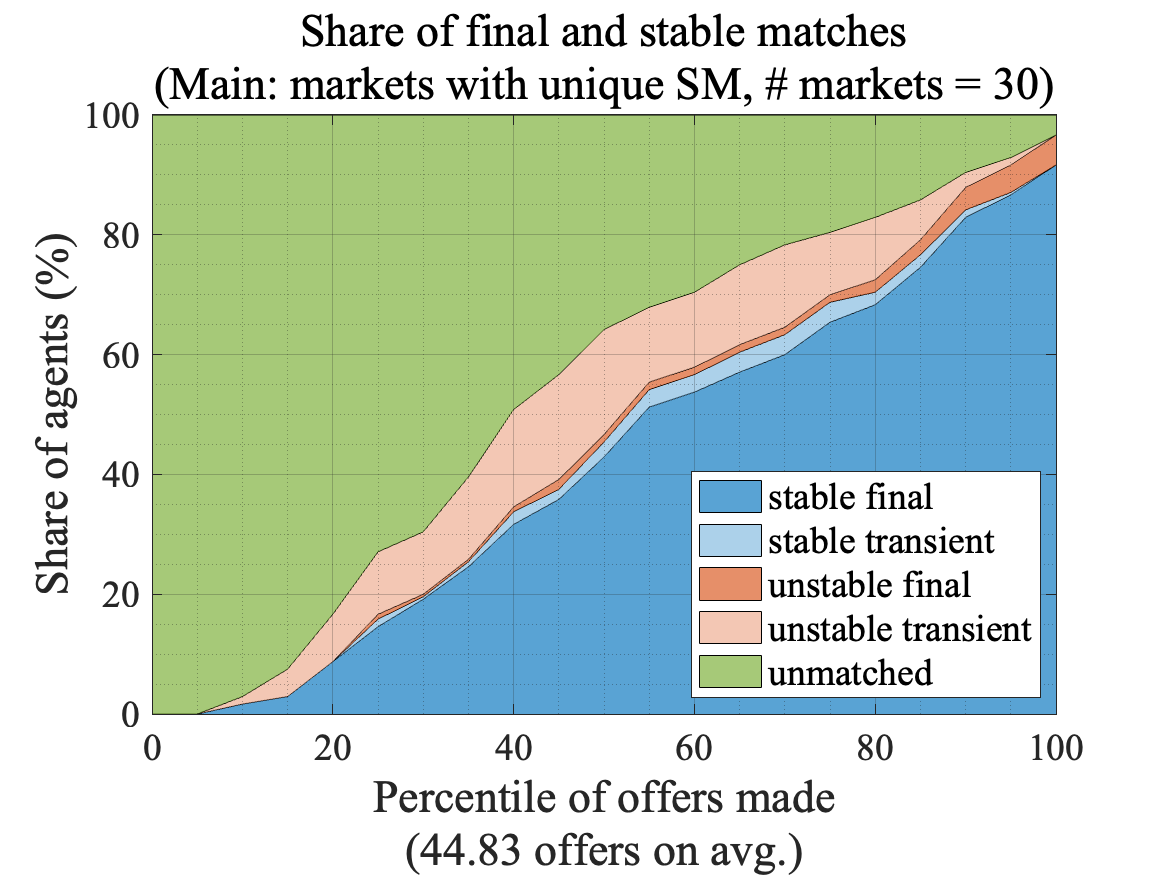}}
    \subfloat[Main: four-by-four]
    {\includegraphics[width=\wdth]{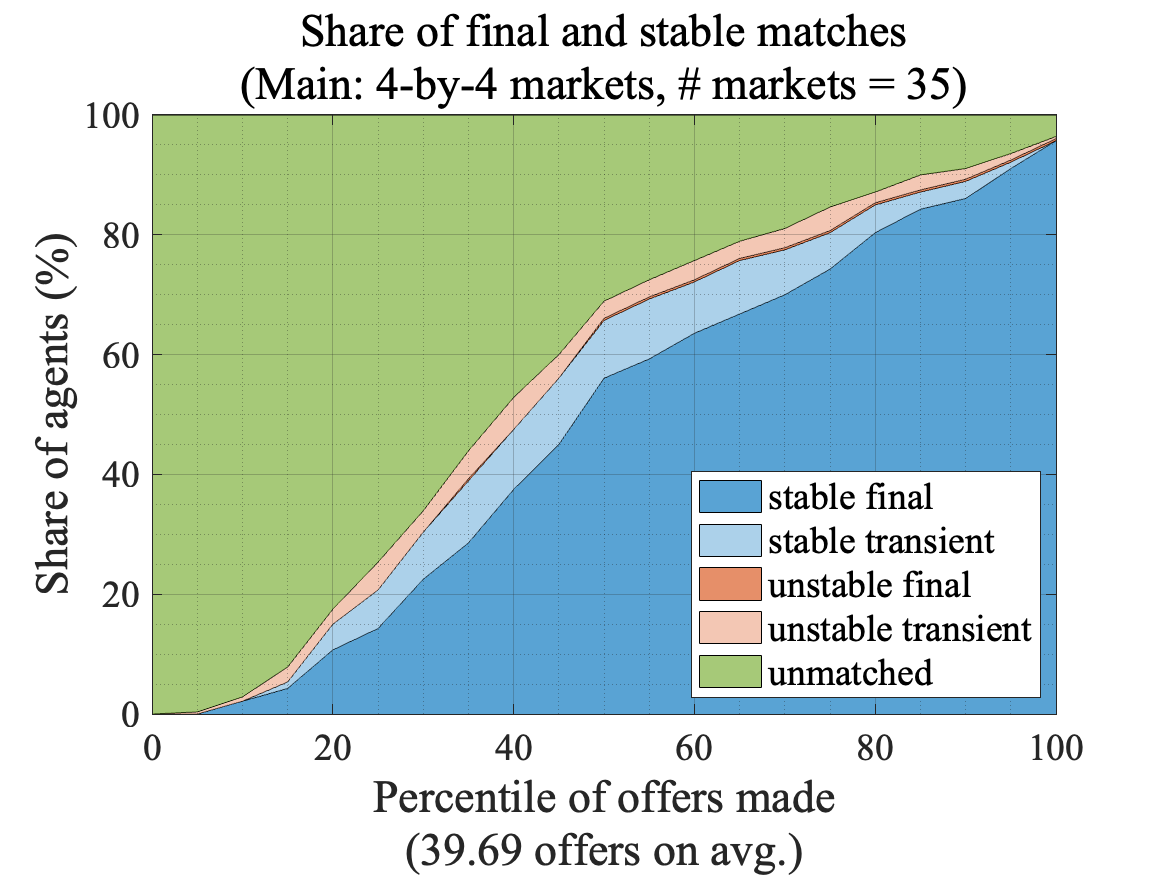}}  
    \subfloat[Main: 5 SMs]{\includegraphics[width=\wdth]{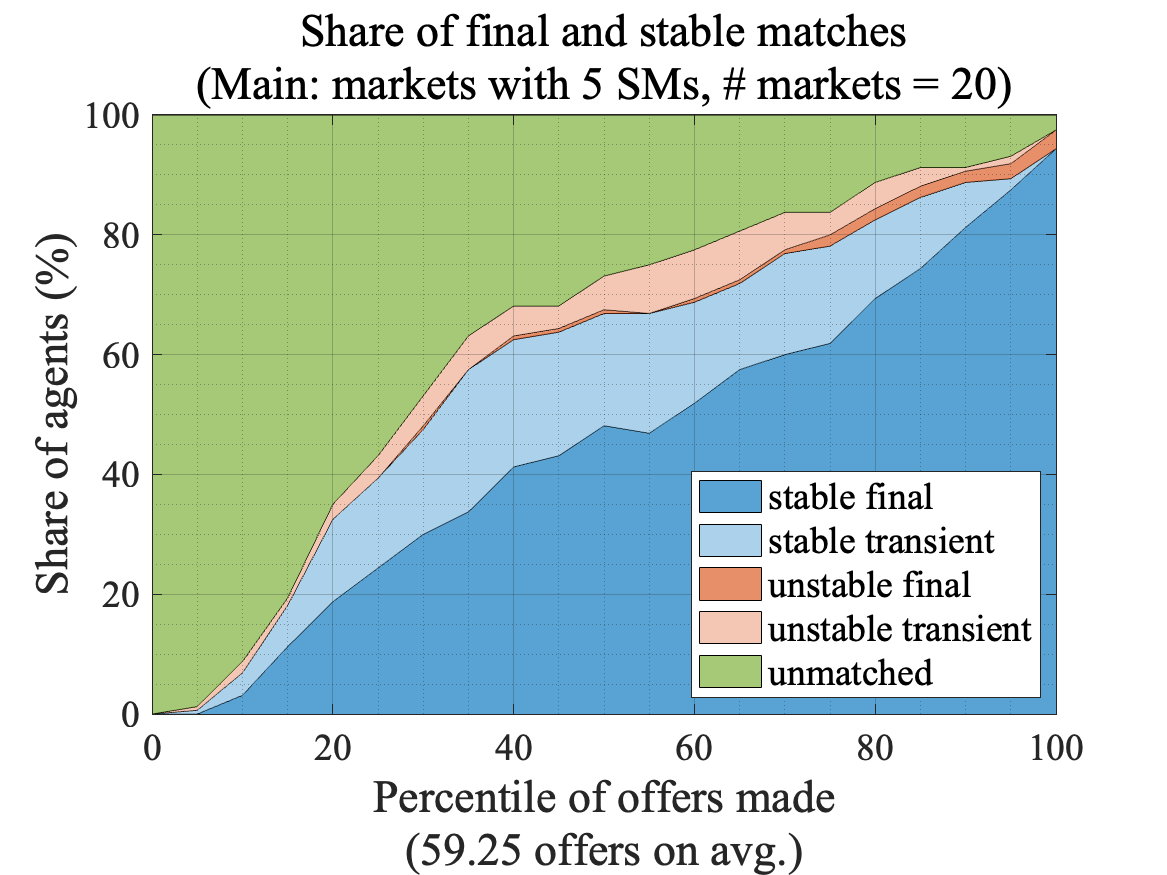}}  
    \subfloat[Main: 5 SMs]{\includegraphics[width=\wdth]{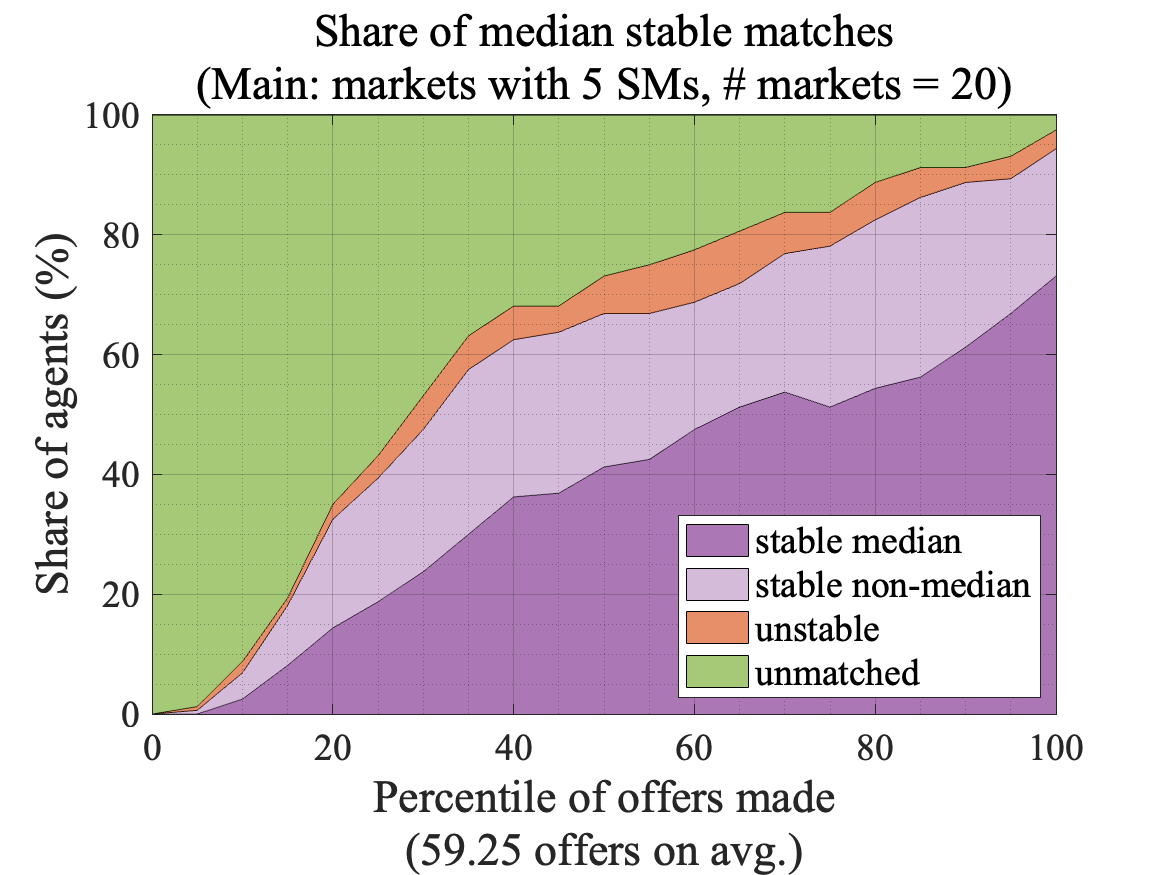}}

    \subfloat[Unilateral: unique]
    {\includegraphics[width=\wdth]{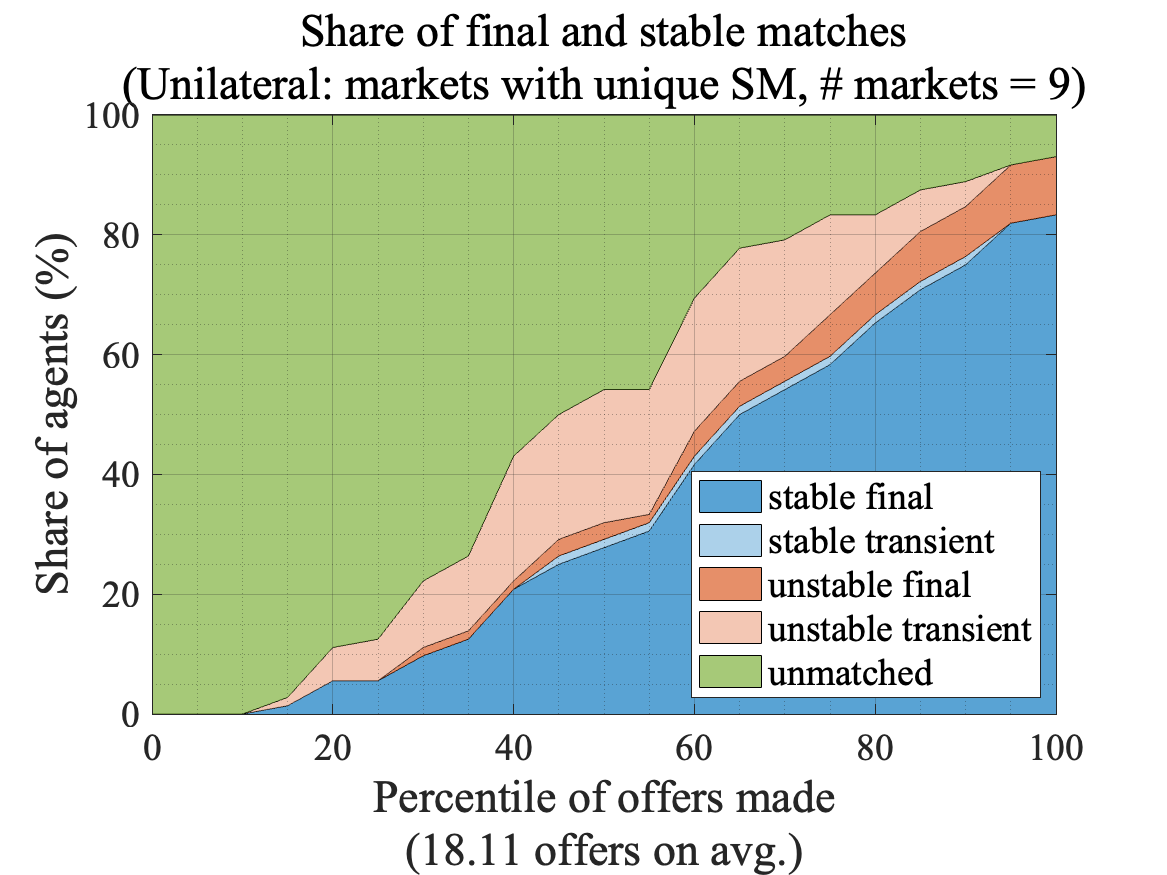}}
    \subfloat[Unilateral: four-by-four]
    {\includegraphics[width=\wdth]{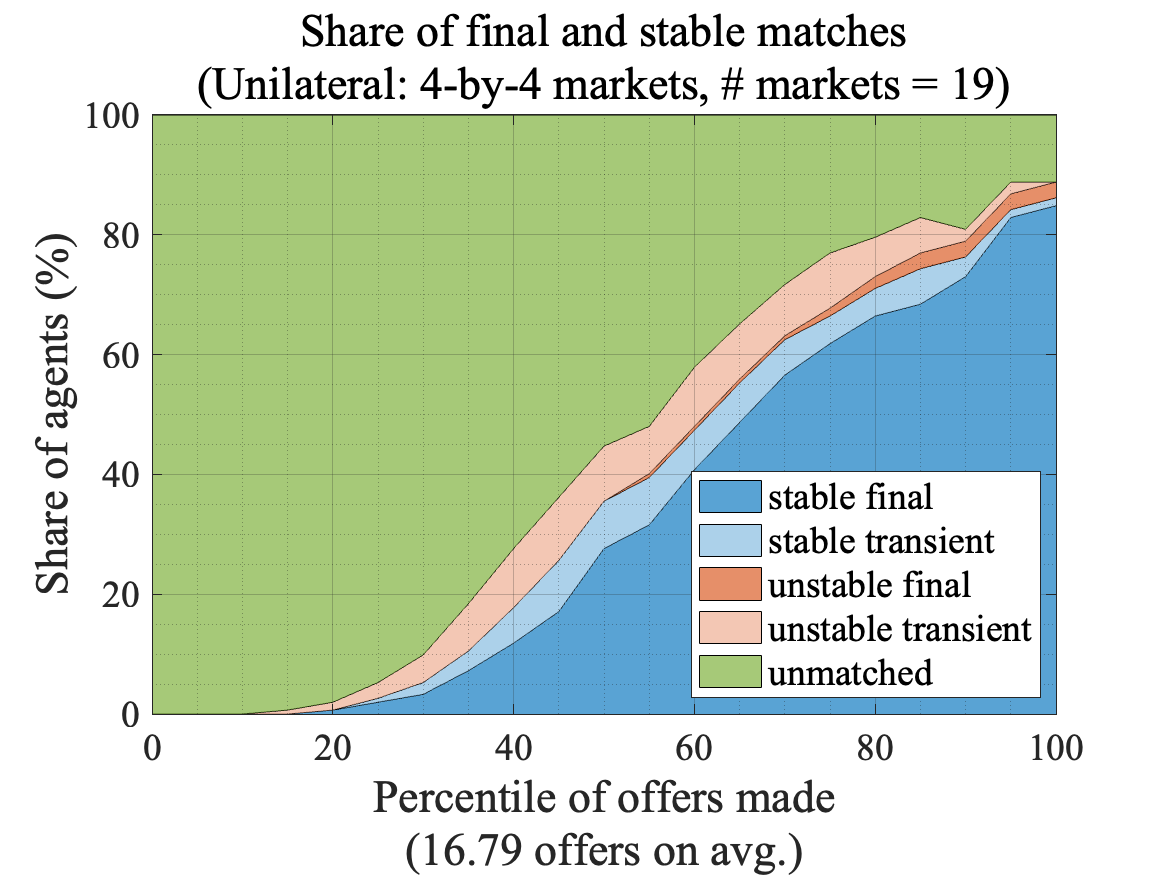}}  
    \subfloat[Unilateral: 5 SMs]{\includegraphics[width=\wdth]{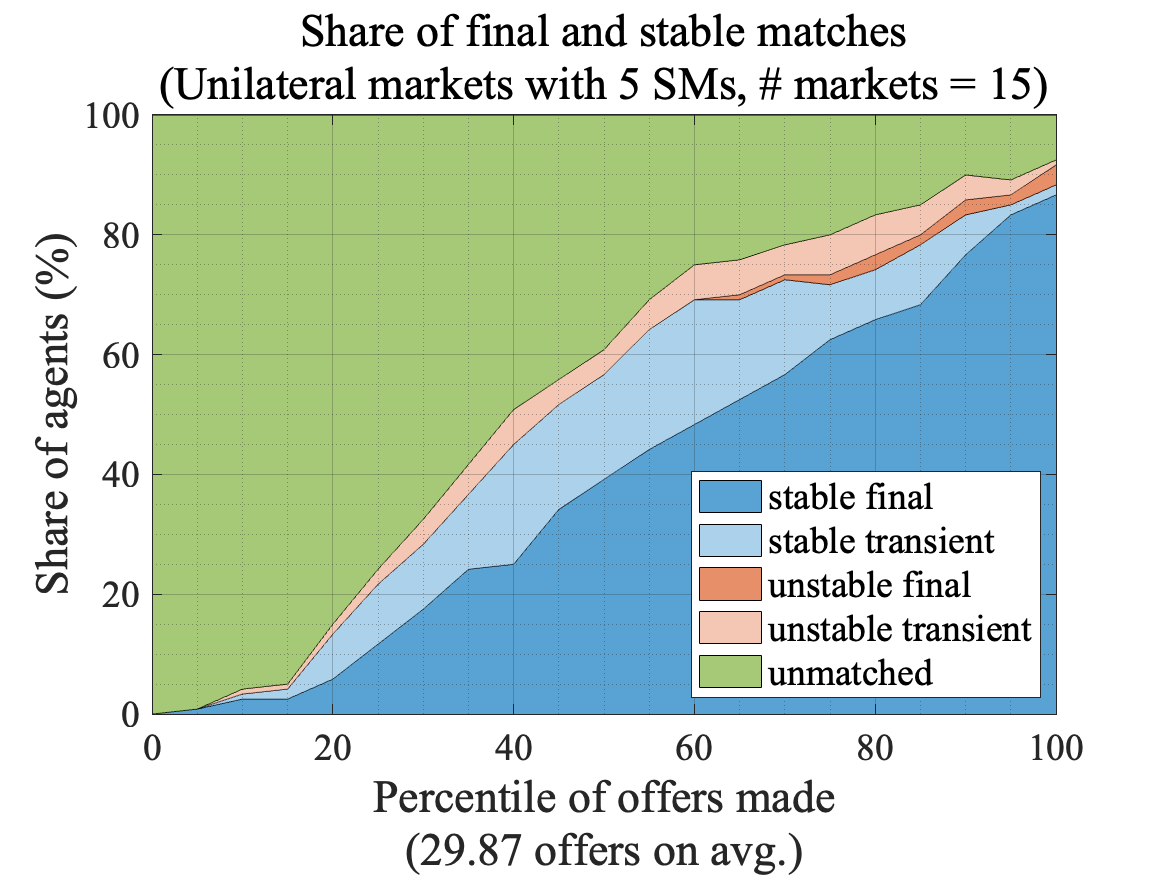}}  
    \subfloat[Unilateral: 5 SMs]{\includegraphics[width=\wdth]{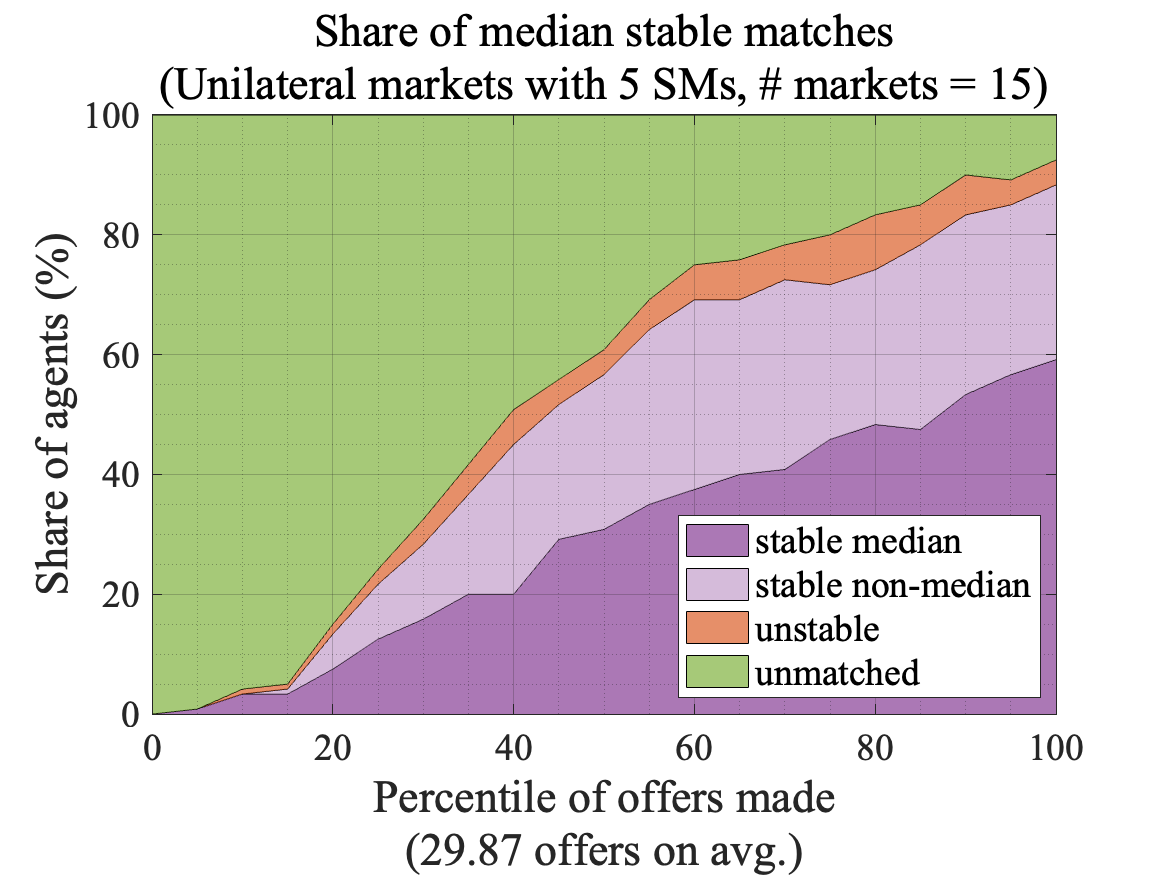}}

    \subfloat[Large: unique]
    {\includegraphics[width=\wdth]{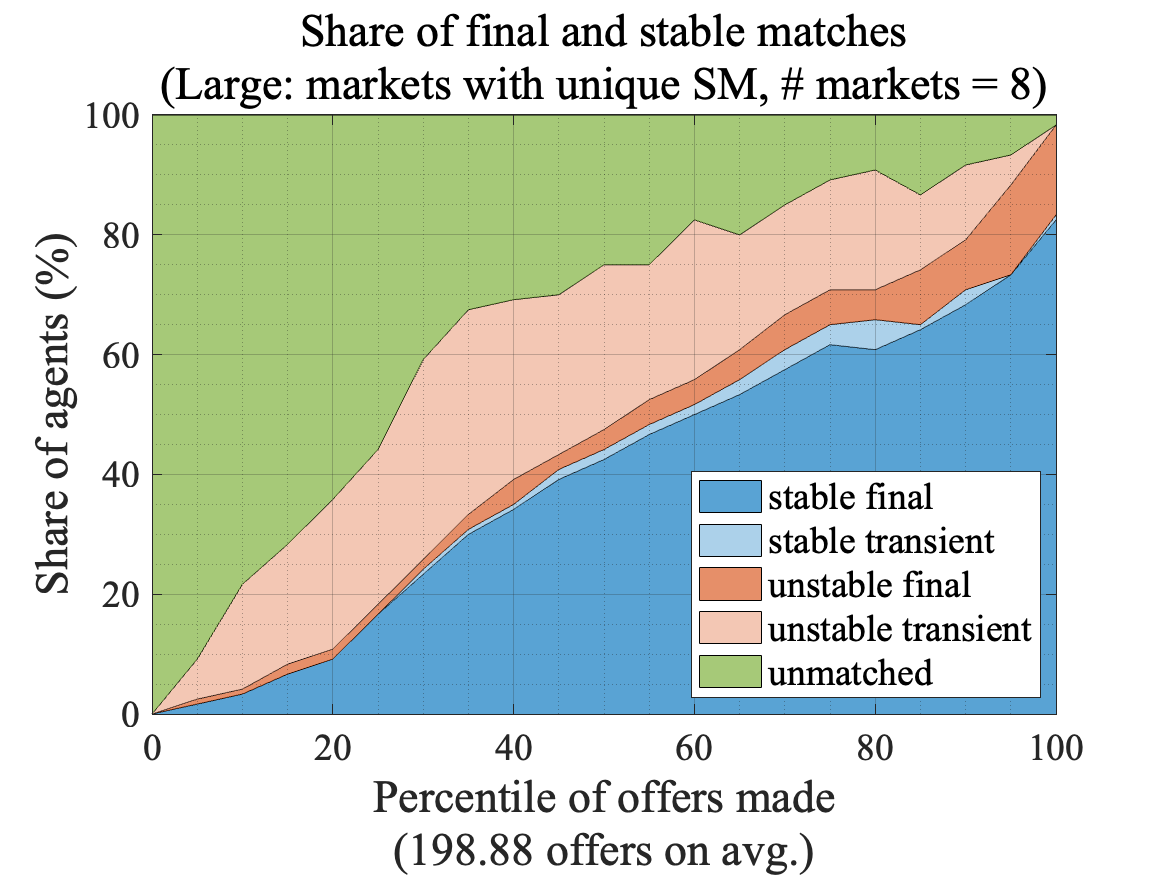}}
    \subfloat[Large: 3 SPs]{\includegraphics[width=\wdth]{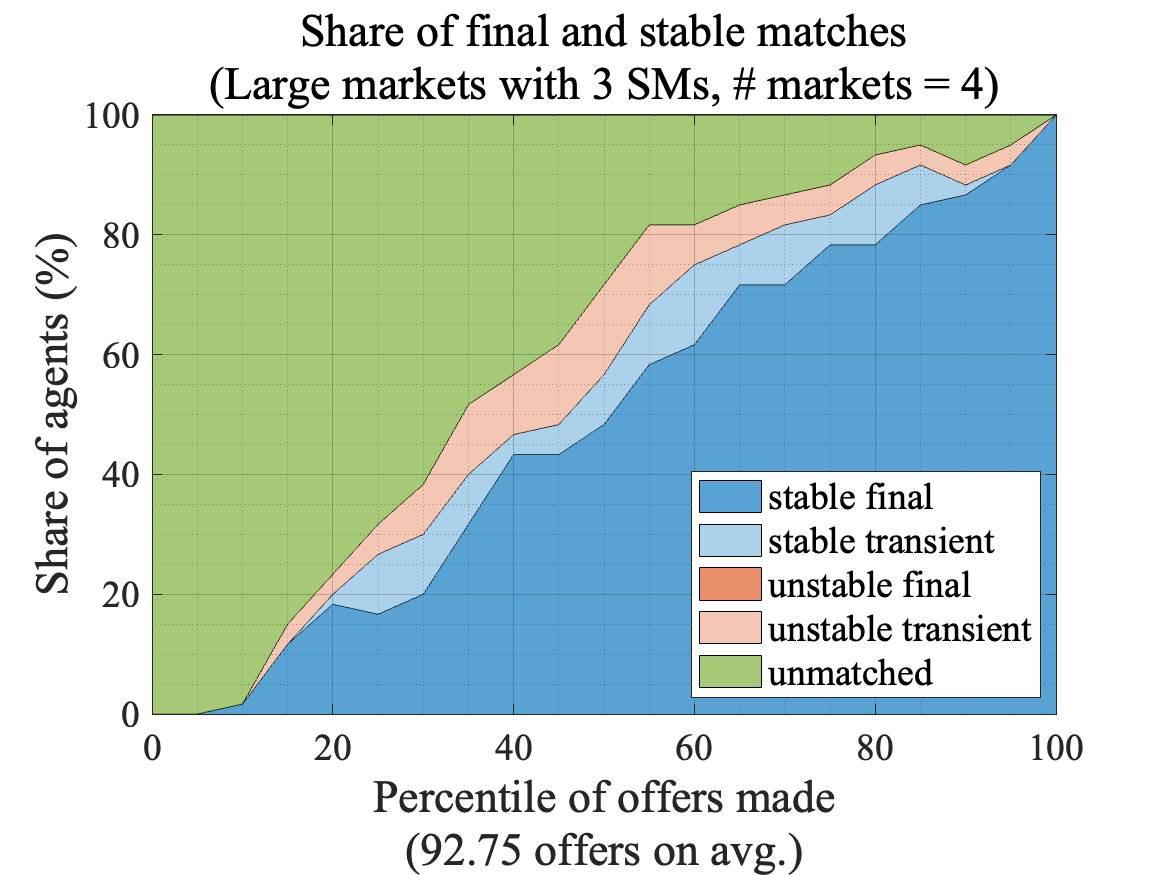}}  
    \subfloat[Large: 3 SPs]{\includegraphics[width=\wdth]{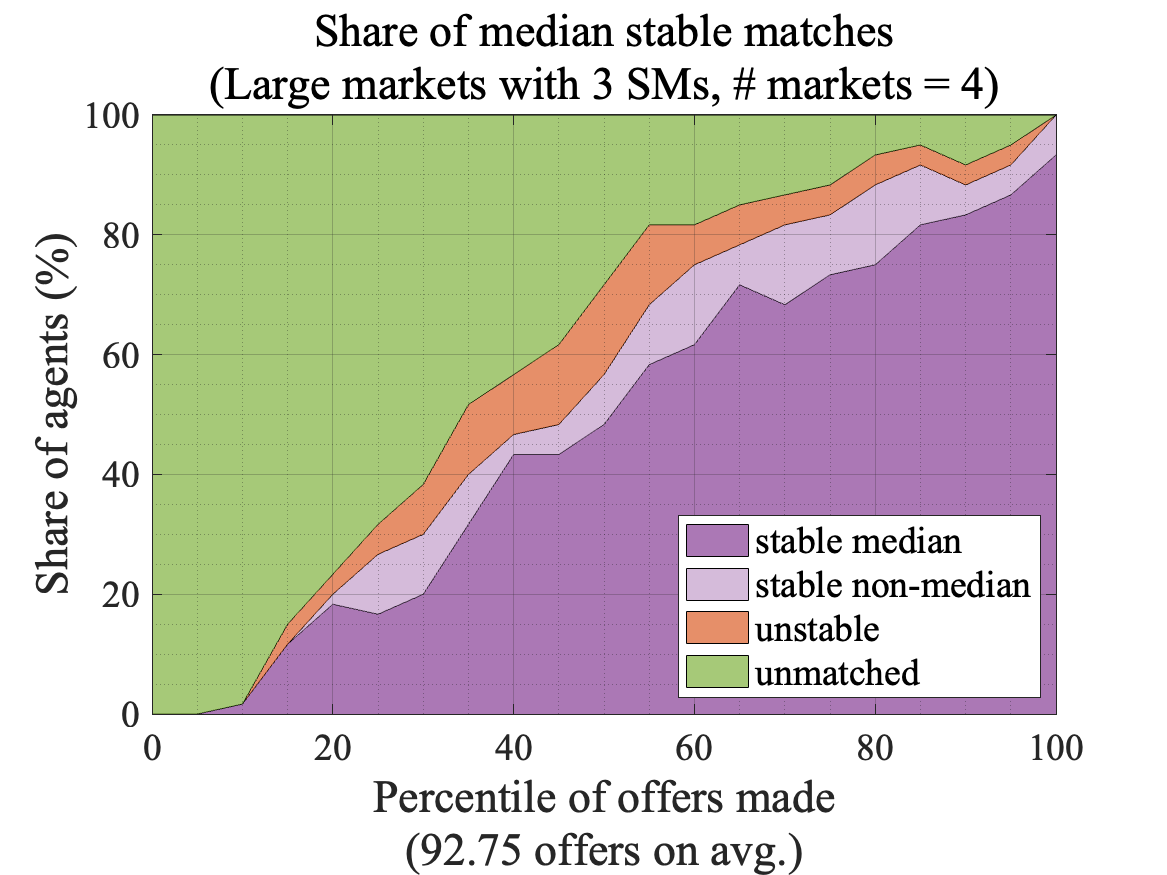}}
    
    \caption{Progression of final, stable, and median stable matches as offers are made
    \label{fig:progression_over_time_all_app}}
\end{figure}
\end{landscape}

\clearpage

\begin{figure}[htbp!]
    \centering
    \subfloat[Main]{\includegraphics[width=0.5\textwidth]{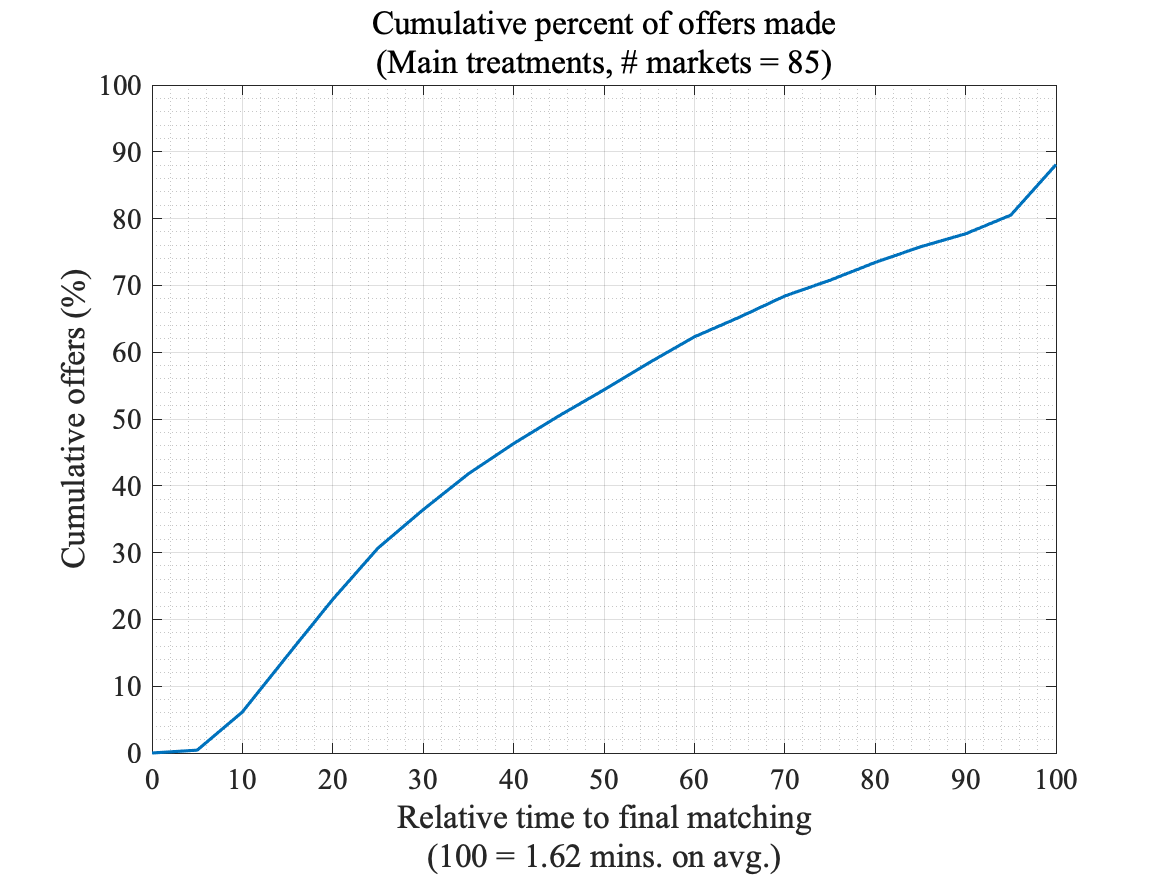}}  
    \subfloat[Unilateral]{\includegraphics[width=0.5\textwidth]{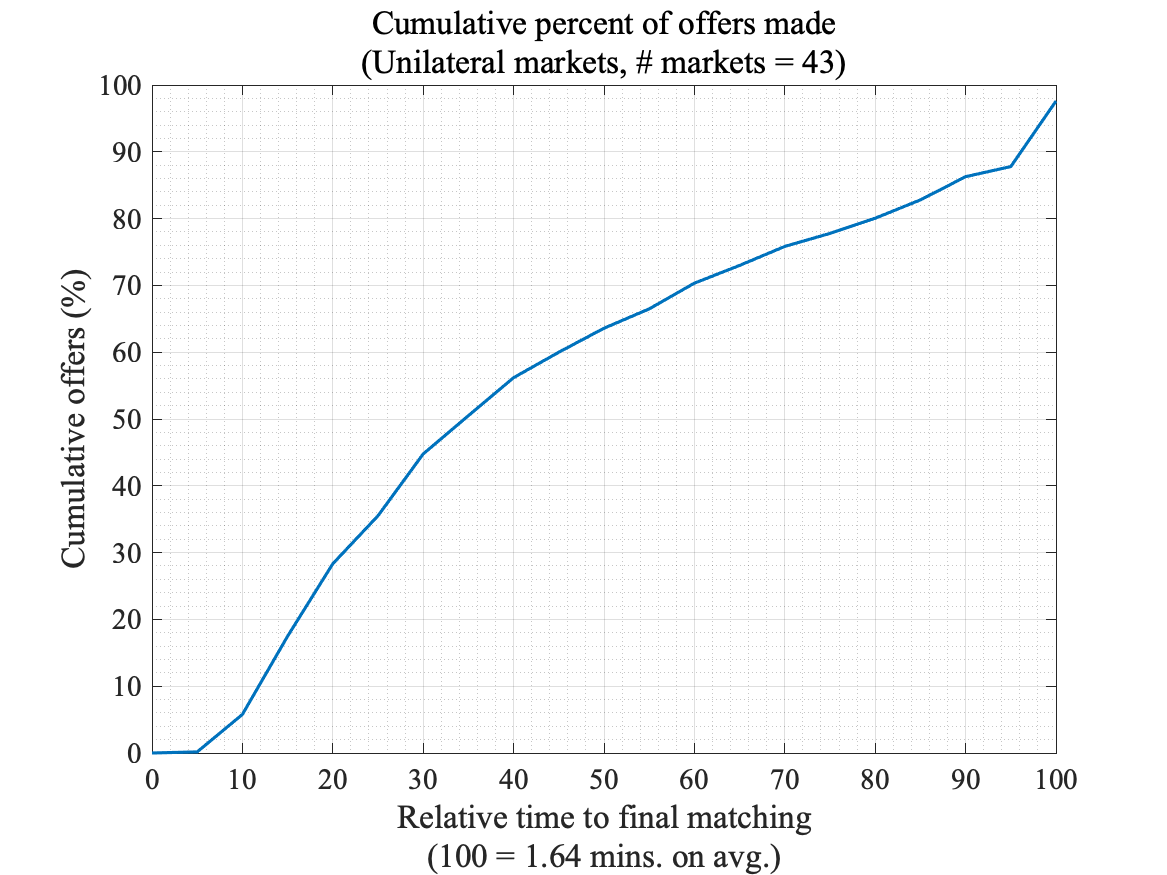}}  

    \subfloat[Large]{\includegraphics[width=0.5\textwidth]{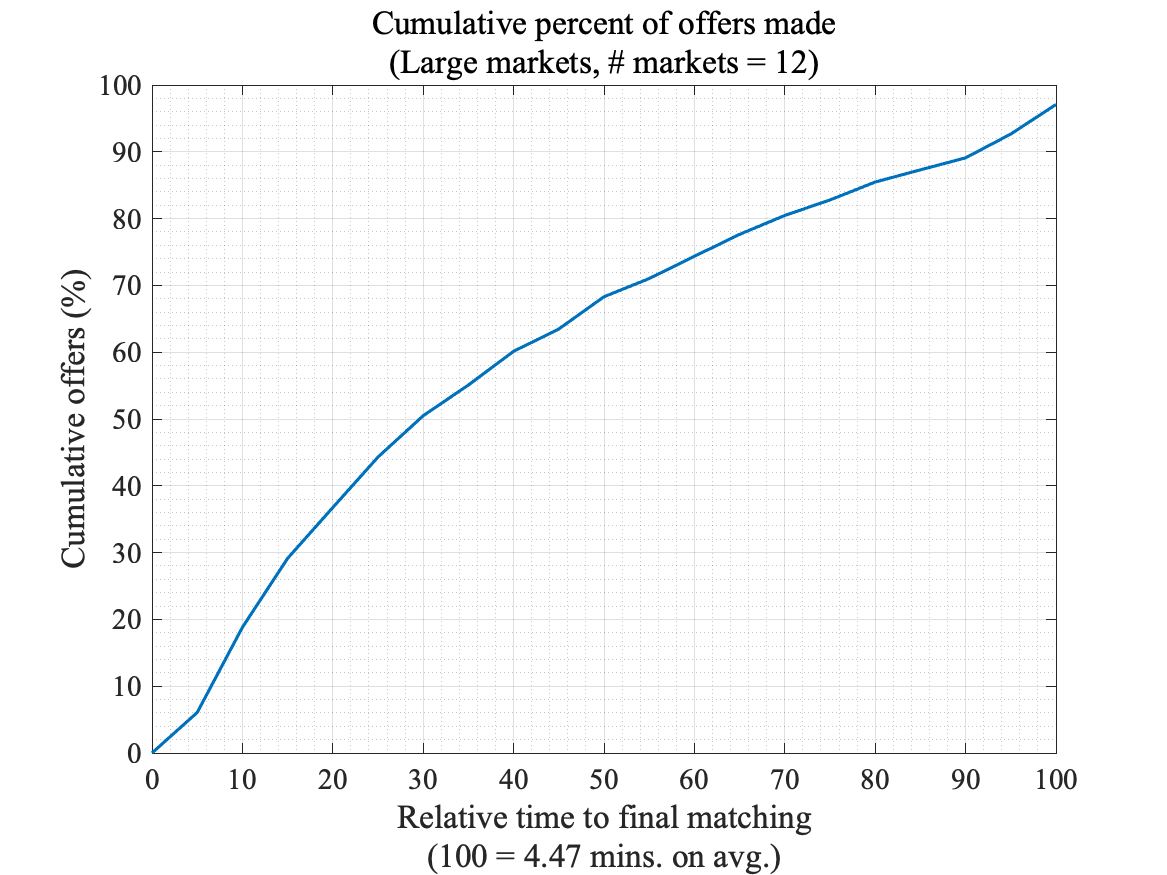}}  
    
    \caption{Percentage of offers made over time\label{percent_offers_time}}
\end{figure}

\clearpage

\renewcommand{\cwidth}{\hspace{0.1cm}\,}
\renewcommand{\spone}{\hspace{0.25cm}}
\renewcommand{\sptwo}{\hspace{0.25cm}}
\begin{table}[h!]
\centering\footnotesize
\caption{Proposing and accepting behavior across time in unilateral markets\label{tab:dynamics:time:unilateral}\centering}
\begin{tabular}{lcccc}
 &  &  &  &  \\\hline\hline
 &  &  &  &  \\
 & 
\cwidth \textit{1st/3} \cwidth& 
\cwidth \textit{2nd/3} \cwidth& 
\cwidth \textit{3rd/3} \cwidth& 
\textit{All} \\\hline
 &  &  &  &  \\
\spone\% \textit{offers} & 53.4 & 24.5 & 22.0 & 100.0 \\
\spone\% \textit{accepted offers} & (68.8) & (68.8) & (63.4) & (68.1) \\
 &  &  &  &  \\
\spone\textit{offers per minute (avg.)} & 25.7 & 6.2 & 3.2 & 14.5 \\
\spone\textit{matches per minute (avg.)} & 15.6 & 4.0 & 2.2 & 10.0 \\
 &  &  &  &  \\
\spone\% \textit{matches formed are repeated} & 1.4 & 9.6 & 25.3 & 8.6 \\
\spone\% \textit{matches formed break} & 14.2 & 8.3 & 2.7 & 10.5 \\
\spone\% \textit{break $\mid$ final} & 9.2 & 3.8 & 0.6 & 5.0 \\
\spone\% \textit{break $\mid$ stable} & 11.1 & 4.8 & 2.6 & 8.0 \\
&  &  &  &  \\ 
\multicolumn{5}{l}{\textit{Characteristics of offers}} \\
\multicolumn{5}{l}{\textit{(\% accepted)}} \\
 &  &  &  &  \\ 
\spone\% \textit{offers to blocking pairs} & 93.2 & 82.3 & 70.5 & 85.4 \\
 & (73.4) & (83.7) & (92.1) & (78.7) \\
\spone\% \textit{offers only-proposer beneficial} & 6.2 & 17.3 & 28.0 & 13.7 \\
 & (6.0) & (0.0) & (5.2) & (4.0) \\
\spone\% \textit{offers are repeated} & 4.1 & 27.4 & 49.6 & 19.8 \\
 & (66.0) & (49.9) & (51.8) & (51.7) \\
\spone\% \textit{offers to previous match} & 1.3 & 10.1 & 29.8 & 10.1 \\
 & (68.8) & (46.7) & (58.2) & (56.1) \\
 &  &  &  &  \\ 
\multicolumn{5}{l}{\textit{In markets with three stable partners:}} \\
 &  &  &  &  \\
\sptwo\% \textit{offers to best stable partner} & 35.7 & 28.9 & 25.5 & 30.6 \\
 & (55.8) & (40.5) & (28.1) & (47.8) \\
\sptwo\% \textit{offers to median stable partner} & 33.1 & 34.0 & 26.8 & 32.9 \\
 & (81.9) & (68.8) & (70.9) & (79.6) \\
\sptwo\% \textit{offers to worst stable partner} & 6.3 & 12.5 & 9.3 & 9.0 \\
 & (93.8) & (83.3) & (87.5) & (85.6) \\
 &  &  &  &  \\ \hline\hline
\multicolumn{5}{p{12cm}}{\scriptsize\textit{Notes.} The table reports averages across unilateral markets (``thirds'' are relative to the time of the last offer) of: 
avg.\ number of offers made and accepted per minute;
\% of repeated matches;
\% of matches that break;
\% of offers made and accepted (\% accepted shown in parentheses): 
to blocking partners, 
that are only beneficial to the proposer,
repeated,
and
to a previous match.
For markets that have five stable matchings and three stable partners, the table also reports the average \% of offers made and accepted (\% accepted shown in parentheses) to the proposer's best, median, and worst stable partner.
}
\end{tabular}
\end{table}

\clearpage

\afterpage{
\begin{landscape}
\begin{table}[htbp!]
\centering\footnotesize
\caption{\centering Summary statistics of dynamics in main treatment with 5 stable matchings and 3 stable partners, by cardinal payoffs and final matching\label{tab:dynam_5sms}}
\begin{tabular}{lcccc|ccc}\hline\hline
 & 
20--20 & 
20--20$_{+100}$ & 
20--70 & 
70--70 & 
\textit{\makecell[b]{ Stable \\ median }} & 
\textit{\makecell[b]{ Stable \\ non-median }} & 
\textit{Unstable} \\\hline
 &  &  &  &  &  &  &  \\
\# \textit{Mkts.} & 5 & 5 & 5 & 5 & 12 & 3 & 5 \\
 & (25.0\%) & (25.0\%) & (25.0\%) & (25.0\%) & (60.0\%) & (15.0\%) & (25.0\%) \\
 &  &  &  &  &  &  &  \\
\# \textit{offers} & 69.0 & 67.6 & 49.4 & 51.0 & 56.2 & 58.3 & 67.0 \\
\# \textit{offers per agent (avg.)} & 4.3 & 4.2 & 3.1 & 3.2 & 3.5 & 3.6 & 4.2 \\
\# \textit{matches} & 30.2 & 31.2 & 18.0 & 19.0 & 22.7 & 22.3 & 30.6 \\
\# \textit{matches per agent (avg.)} & 3.8 & 3.9 & 2.2 & 2.4 & 2.8 & 2.8 & 3.8 \\
 &  &  &  &  &  &  &  \\
\textit{time to final matching (mins.)} & 2.85 & 2.77 & 1.56 & 1.72 & 2.07 & 2.21 & 2.60 \\
\textit{time to last proposal (mins.)} & 2.98 & 3.60 & 2.55 & 2.44 & 2.56 & 3.31 & 3.45 \\
 &  &  &  &  &  &  &  \\
\% \textit{accepted offers} & 42.7 & 45.4 & 35.4 & 38.0 & 39.5 & 37.6 & 44.1 \\
 &  &  &  &  &  &  &  \\
\% \textit{offers are repeated} & 36.1 & 39.0 & 35.3 & 36.9 & 37.0 & 36.7 & 36.5 \\
\% \textit{accepted $\mid$ repeated} & 31.0 & 35.4 & 22.6 & 20.3 & 26.2 & 22.8 & 32.8 \\
\% \textit{matches are repeated} & 27.5 & 34.6 & 20.2 & 26.7 & 27.2 & 18.7 & 32.4 \\
 &  &  &  &  &  &  &  \\
\% \textit{offers to blocking pairs} & 72.6 & 56.8 & 63.8 & 60.6 & 63.2 & 65.3 & 62.8 \\
\% \textit{accepted $\mid$ to BP} & 57.8 & 68.1 & 53.6 & 58.5 & 59.2 & 55.2 & 62.8 \\
\% \textit{offers only-proposer beneficial} & 26.8 & 32.9 & 36.2 & 38.7 & 36.0 & 34.7 & 27.3 \\
\% \textit{accepted $\mid$ only-prop beneficial} & 2.3 & 5.5 & 3.3 & 5.3 & 3.9 & 4.0 & 4.6 \\
 &  &  &  &  &  &  &  \\
\% \textit{proposer is active} & 90.8 & 83.9 & 84.0 & 80.1 & 84.3 & 84.3 & 85.9 \\
\% \textit{offer is downward} & 59.3 & 58.9 & 62.8 & 63.4 & 61.7 & 61.4 & 59.7 \\
\% \textit{offer is Gale-Shapley} & 30.0 & 34.3 & 31.0 & 31.9 & 30.8 & 35.4 & 32.2 \\
\% \textit{offer skips someone} & 41.9 & 40.1 & 42.9 & 39.2 & 41.1 & 33.6 & 45.4 \\
 &  &  &  &  &  &  &  \\
\% \textit{broken match $\mid$ final} & 12.7 & 14.9 & 7.9 & 7.4 & 12.3 & 5.2 & 10.3 \\
\% \textit{final match $\mid$ broken} & 36.2 & 25.0 & 39.3 & 25.7 & 37.2 & 25.0 & 18.0 \\
\% \textit{broken match $\mid$ stable} & 13.0 & 21.9 & 9.3 & 14.1 & 15.3 & 5.0 & 18.6 \\
\% \textit{stable match $\mid$ broken} & 65.8 & 70.3 & 71.4 & 66.7 & 75.0 & 41.7 & 71.1 \\
 &  &  &  &  &  &  &  \\\hline\hline
\multicolumn{8}{p{17.5cm}}{\scriptsize\textit{Notes.} The table reports averages across markets in our main treatment with five stable matchings and three stable partners, split by (a) cardinal payoffs (first four columns), and (b) whether the market reached a median stable matching, a non-median stable matching, or an unstable matching (last three columns).}
\end{tabular}
\end{table}
\end{landscape}
}

\clearpage

\renewcommand{\wdth}{0.5\textwidth}
\afterpage{
\begin{landscape}
\begin{figure}
    \centering
    \includegraphics[width=0.6\textwidth]{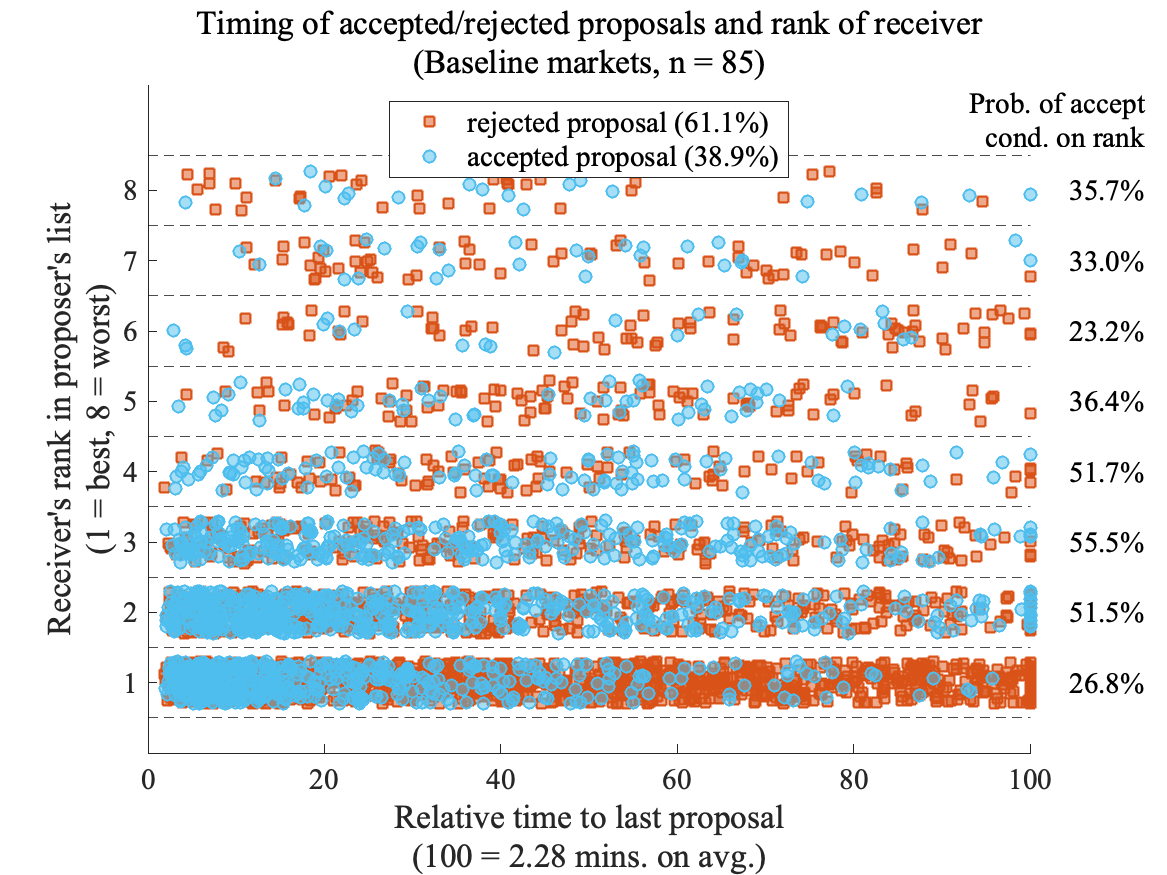}
    \hspace{0.75cm}
    \includegraphics[width=0.55\textwidth]{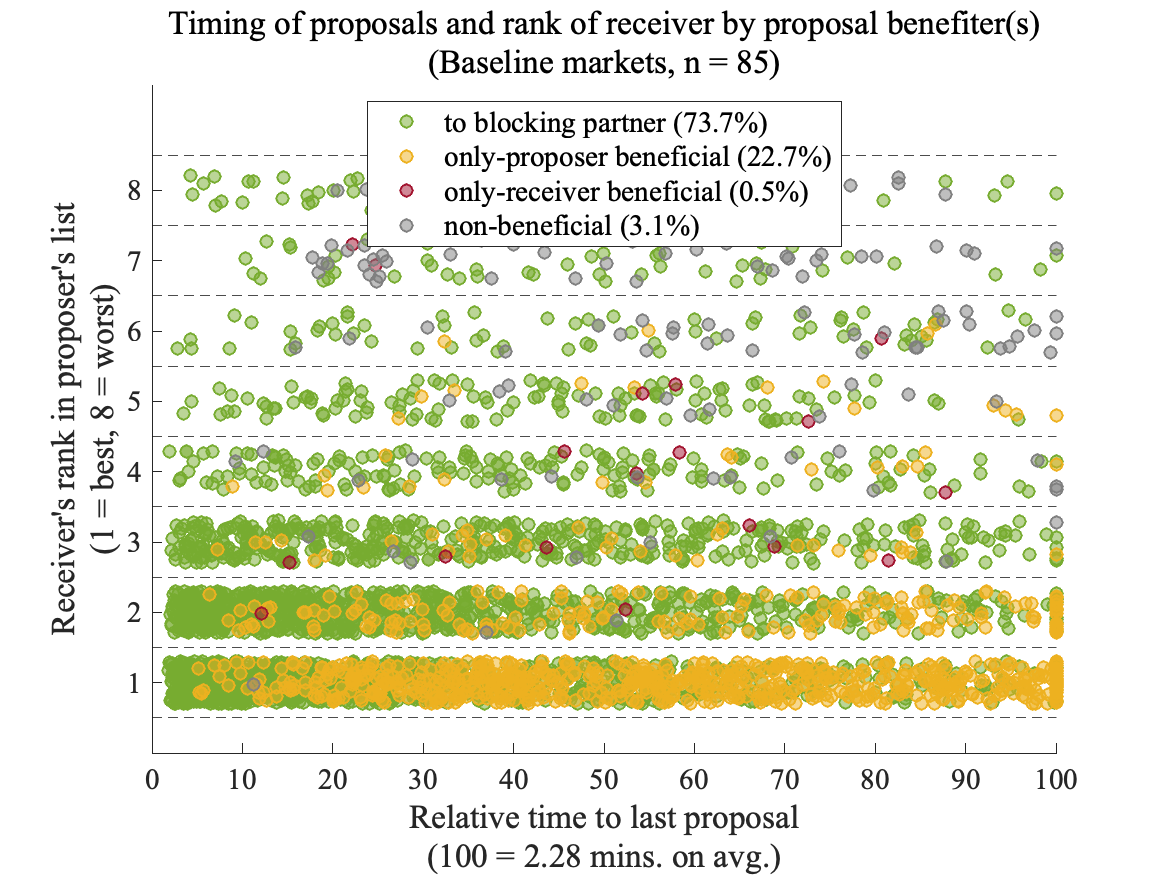}
    
    \vspace{0.8cm}
    
    \includegraphics[width=0.6\textwidth]{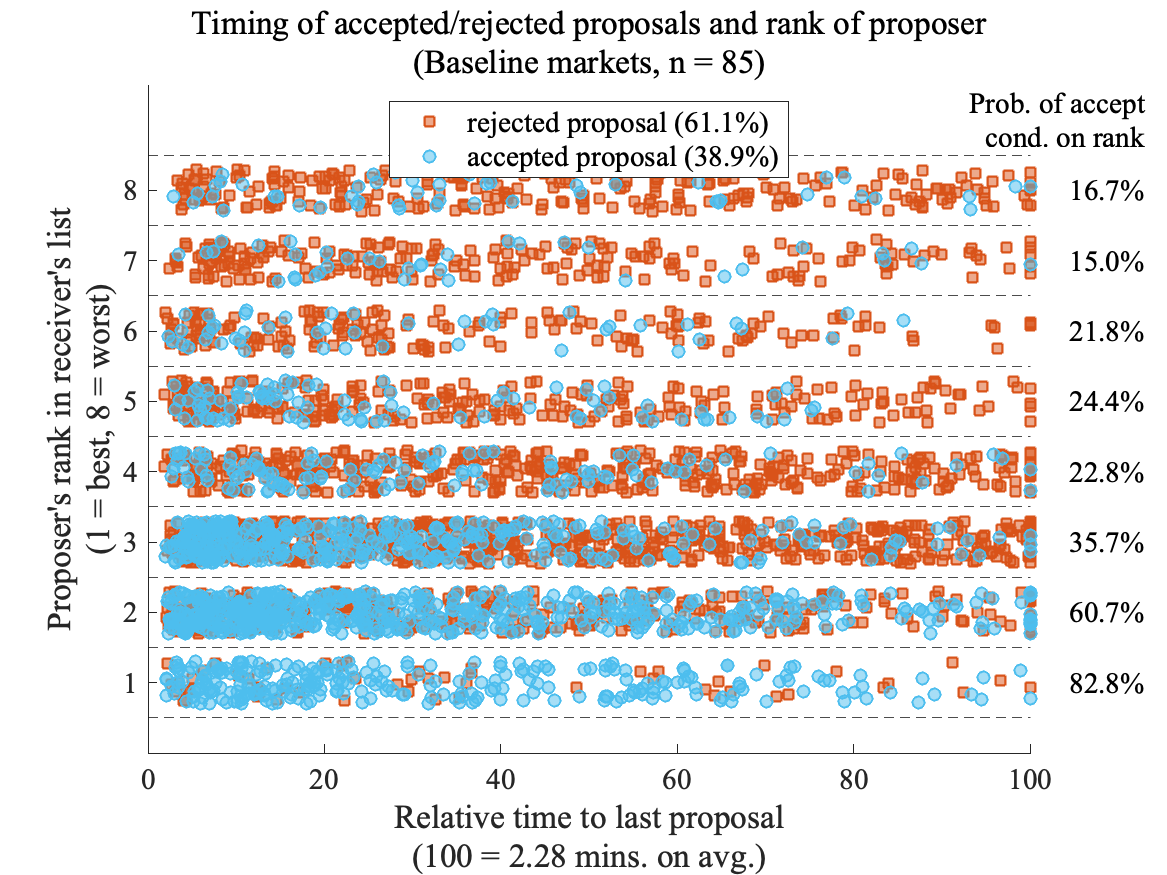}
    \hspace{0.75cm}
    \includegraphics[width=0.56\textwidth]{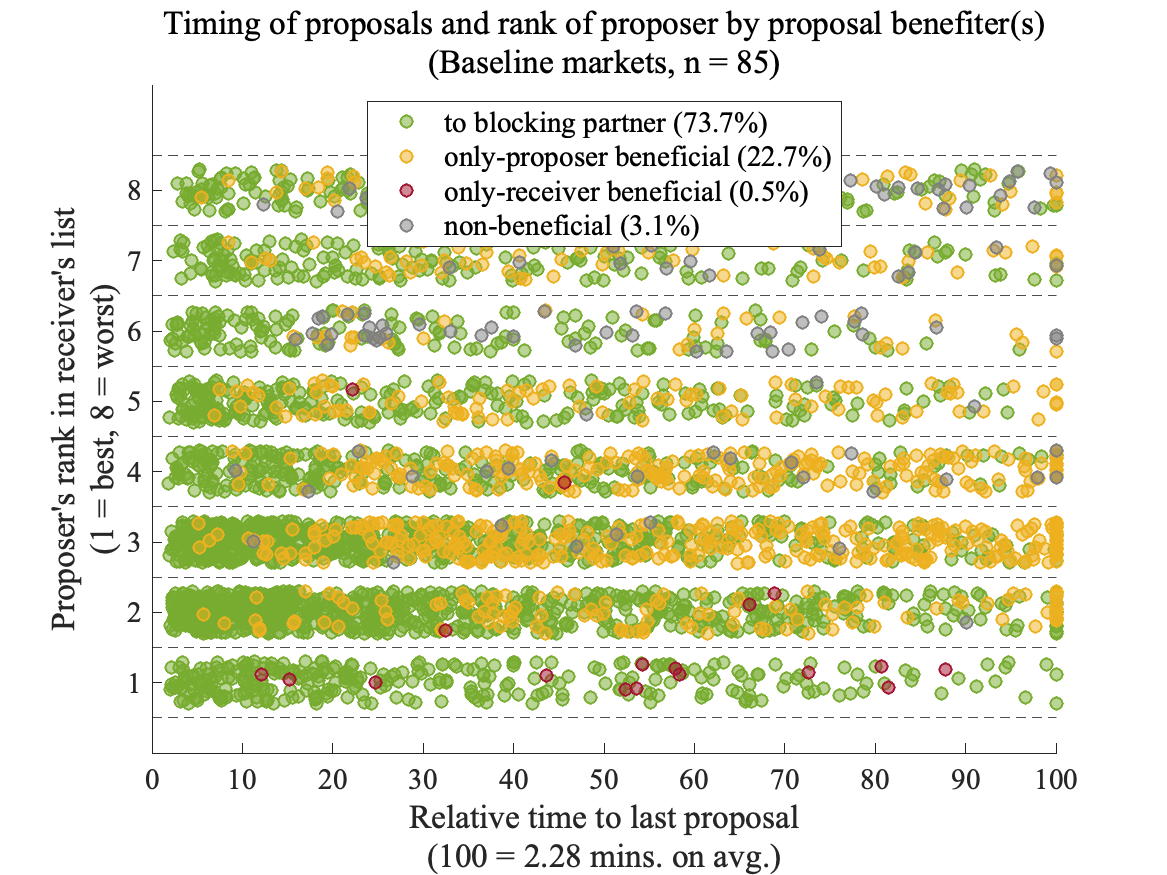}

    \vspace{0.25cm}
    
    \caption{Timing of offers, and rank of receiver (top two panels) and of proposer (bottom two panels) 
    \label{fig4}}
\end{figure}
\end{landscape}
}

\clearpage

\begin{figure}[t]
    \centering
    \vspace{-0.5cm}
    \subfloat[Avg.\ across all baseline]{\includegraphics[width=0.5\textwidth]{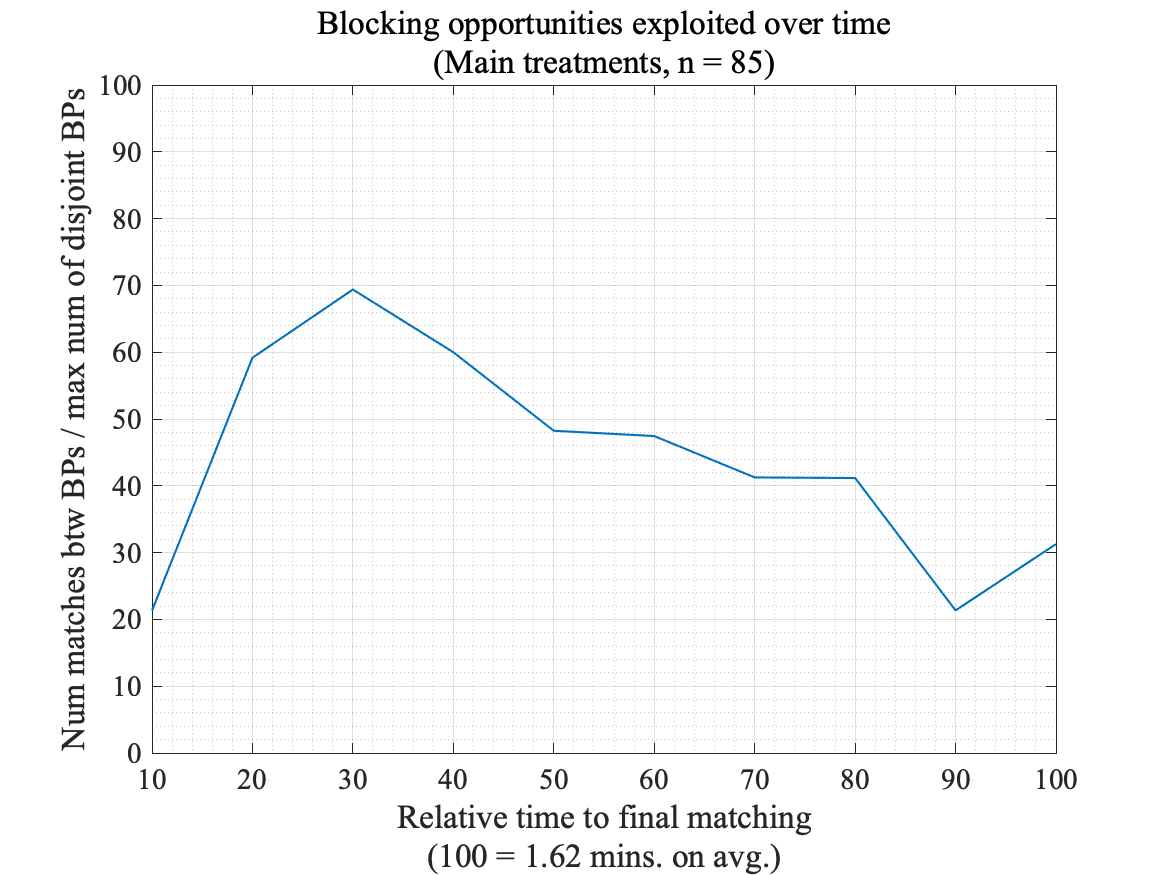}}  
    \subfloat[Avg.\ across all baseline w/ 25p and 75p]{\includegraphics[width=0.5\textwidth]{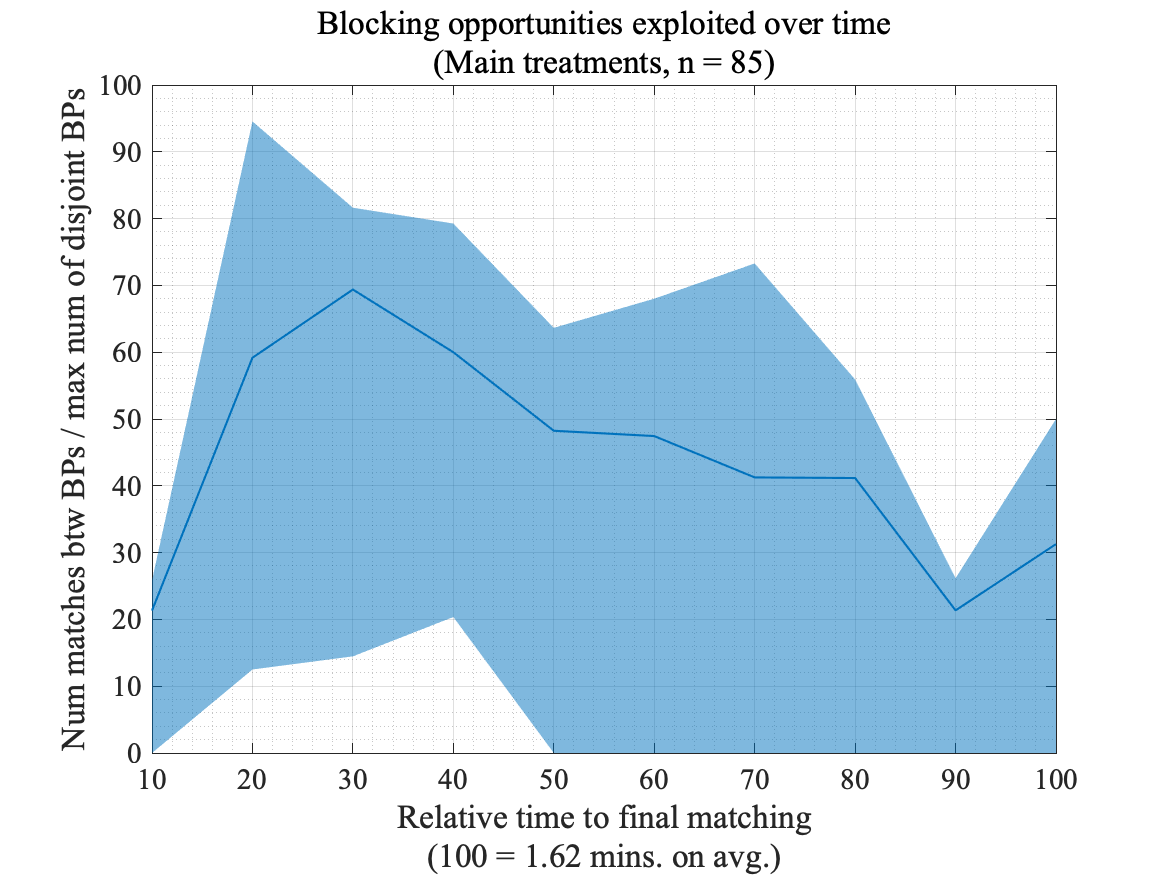}}  
    \caption{Blocking opportunities exploited over time\label{blocking_over_time}}
    \caption*{\scriptsize\raggedright\textit{Notes.} We split every market in ten segments of equal duration. Within each segment, we compute the ratio between the number of proposals accepted by a blocking partner within this segment and the weighted  average of the maximal number of disjoint blocking pairs in the market (weighted across time). The Figure reports the average of this ratio across all markets in our main treatments across the ten segments. The right panel plots an area covering the mass between the 25th and 75th percentiles across all markets in our main treatments.}
\end{figure}

\begin{table}
\centering\small
\caption{Description of first proposals: main and auxiliary treatments\label{first_proposal_all}\centering}
\begin{tabular}{lcccc}
 &  &  &  &  \\\hline\hline
 &  &  &  &  \\
 & 
\textit{Unilateral} & 
\textit{Main} & 
\textit{Large} & 
All \\\hline
 &  &  &  &  \\
\multicolumn{5}{l}{\textit{Proposer's and receiver's first proposal}}\\
 &  &  &  &  \\
\# \textit{proposals per mkt} & 5.4 & 5.0 & 8.3 & 5.5 \\
\% \textit{accepted} & 71.4 & 59.8 & 52.8 & 62.8 \\
\% \textit{to blocking pairs} & 100.0 & 100.0 & 100.0 & 100.0 \\
\% \textit{to stable partners} & 64.0 & 62.5 & 37.3 & 60.8 \\
\% \textit{to top choice} & 63.7 & 55.6 & 15.6 & 54.7 \\
 &  &  &  &  \\
\multicolumn{5}{l}{\textit{First proposal made by proposer}}\\
 &  &  &  &  \\
\# \textit{proposals per mkt} & 8.0 & 14.2 & 27.8 & 13.5 \\
\% \textit{accepted} & 67.2 & 52.2 & 40.9 & 55.8 \\
\% \textit{to blocking pairs} & 95.1 & 88.5 & 93.1 & 90.9 \\
\% \textit{to stable partners} & 62.5 & 60.0 & 37.9 & 58.9 \\
\% \textit{to top choice} & 64.2 & 58.8 & 19.9 & 57.1 \\
 &  &  &  &  \\\hline\hline
\multicolumn{5}{p{10cm}}{\scriptsize\textit{Notes.} The table reports average characteristics of first proposals across our main and auxiliary treatments, where we use two definition of a ``first proposal.'' First, we consider the proposals which are the first proposal made by the proposer \textit{and} the first proposal involving the receiver (as receiver or proposer). Second, we consider proposals that are the first proposal made by the proposer (but in which the receiver may have already been involved in a prior proposal, as receiver or proposer).}
\end{tabular}
\end{table}

\begin{table}
\centering\small
\caption{\centering\label{first_proposal_median} Description of first proposals in markets with five stable matchings and three stable partners (main treatment)}
\begin{tabular}{lccc}
 &  &  &  \\\hline\hline
 & 
\textit{\makecell[b]{ Stable \\ median }} & 
\textit{\makecell[b]{ Stable \\ non-median }} & 
\textit{Unstable} \\\hline
 &  &  &  \\
\textit{Mkts.} & 12 & 3 & 5 \\
 & (60\%) & (15\%) & (25\%) \\
 &  &  &  \\
\multicolumn{4}{l}{\textit{Proposer's and receiver's first proposal}}\\
 &  &  &  \\
\# \textit{proposals per mkt} & 5.1 & 4.7 & 5.0 \\
\% \textit{accepted} & 57.8 & 25.0 & 47.0 \\
\% \textit{to blocking pairs} & 100.0 & 100.0 & 100.0 \\
\% \textit{to stable partners} & 73.1 & 66.7 & 55.3 \\
\% \textit{to best stable partner} & 33.8 & 38.9 & 30.0 \\
\% \textit{to median stable partner} & 37.9 & 27.8 & 11.3 \\
\% \textit{to worst stable partner} & 1.4 & 0.0 & 14.0 \\
\% \textit{to top choice} & 33.7 & 47.2 & 55.3 \\
 &  &   &  \\
\multicolumn{4}{l}{\textit{First proposal made by proposer}}\\
 &  &  &   \\
\# \textit{proposals per mkt} & 14.8 & 15.3 & 14.4 \\
\% \textit{accepted} & 47.7 & 40.8 & 49.7 \\
\% \textit{to blocking pairs} & 88.7 & 91.1 & 87.2 \\
\% \textit{to stable partners} & 73.7 & 56.2 & 61.8 \\
\% \textit{to best stable partner} & 35.3 & 32.1 & 37.1 \\
\% \textit{to median stable partner} & 35.6 & 17.9 & 12.4 \\
\% \textit{to worst stable partner} & 2.8 & 6.2 & 12.3 \\
\% \textit{to top choice} & 38.3 & 60.7 & 44.4 \\
 &  &  &  \\\hline\hline
\multicolumn{4}{p{10.5cm}}{\scriptsize\textit{Notes.} The table reports average characteristics of first proposals in the markets of our main treatment with five stable matchings and three stable partners, differentiating them by whether they converged to the median stable matching, a non-median stable matching, or an unstable matching, where we use two definition of a ``first proposal.'' First, we consider the proposals which are the first proposal made by the proposer \textit{and} the first proposal involving the receiver (as receiver or proposer). Second, we consider proposals that are the first proposal made by the proposer (but in which the receiver may have already been involved in a prior proposal, as receiver or proposer).}
\end{tabular}
\end{table}

\begin{table}
\centering\small
\caption{Individual behavior by final match  in markets with five stable matchings and three stable partners (main treatment)\label{individual_median}\centering}
\begin{tabular}{lccccc}
 &  &  &  &  &  \\\hline\hline
 &  &  &  &  &  \\
 & 
{\small\textit{Stable}} & 
{\small\textit{Best}} & 
{\small\textit{Median}} & 
{\small\textit{Worst}} & 
{\small\textit{Unstable}} \\\hline
 &  &  &  &  &  \\
\textit{\# agents} & 15.5 & 1.7 & 12.1 & 1.7 & 0.5 \\
\textit{\% agents} & 96.9 & 10.6 & 75.6 & 10.6 & 3.1 \\
&  &  &  &  &  \\
\textit{\# proposals made} & 3.9 & 3.2 & 4.6 & 3.6 & 5.8 \\
\textit{\% accepted (of made)} & 49.1 & 52.3 & 47.6 & 54.0 & 35.8 \\
&  &  &  &  &  \\
\textit{\# proposals received} & 3.8 & 4.5 & 4.0 & 3.3 & 5.5 \\
\textit{\% accepted (of received)} & 45.8 & 40.5 & 47.8 & 48.4 & 62.5 \\
&  &  &  &  &  \\
\textit{\% proposals to blocking pairs} & 73.2 & 67.6 & 71.5 & 82.6 & 66.1 \\
\textit{\# proposals per minute} & 3.5 & 3.3 & 3.4 & 3.6 & 3.3 \\
 &  &  &  &  &  \\\hline\hline
\multicolumn{6}{p{12.25cm}}{\scriptsize\textit{Notes.} The table reports averages across all agents and proposals made in markets with five stable matchings and three stable partners in our main treatment, differentiating them by whether the agent or the proposer finalized the market matched to: as stable partner, their best stable partner, their median stable partner, their worst stable partner, or an unstable partner.}
\end{tabular}
\end{table}

\clearpage

\begin{figure}[tbp!]
    \centering
    
    \subfloat[Unique stable matching]{\includegraphics[width=\textwidth]{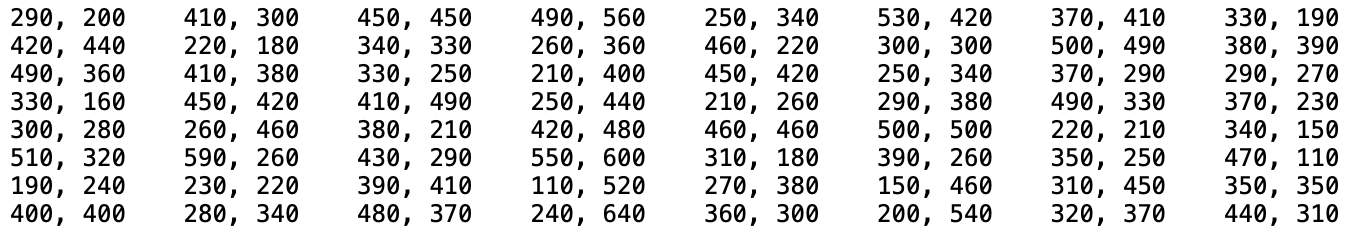}}  

    \vspace{0.5cm}

    \subfloat[Multiple stable matchings]{\includegraphics[width=\textwidth]{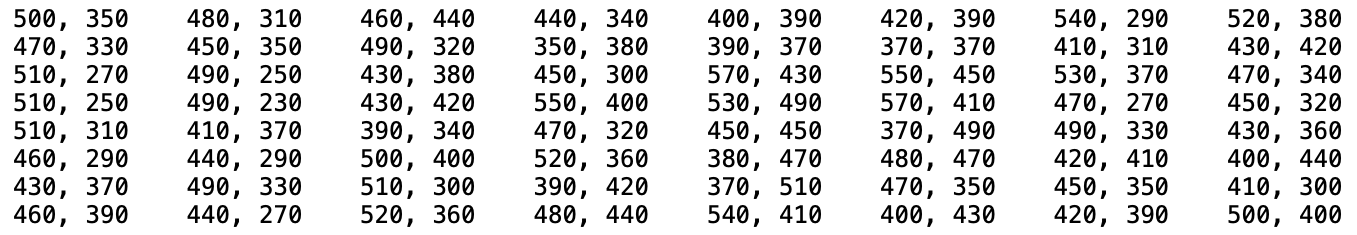}}  

    \vspace{0.5cm}
    
    \caption{Examples of payoff matrices\label{fig_payoffs}}
\end{figure}

\end{document}